\documentclass[fleqn,usenatbib]{mnras}

\usepackage{newtxtext,newtxmath}
\usepackage[T1]{fontenc}

\usepackage{graphicx}	% Including figure files
\usepackage{amsmath}	% Advanced maths commands

\usepackage[usenames]{xcolor}
\usepackage[normalem]{ulem}
\usepackage{amsmath}

\def\Mhalo{\rm M_{\rm h}}
\def\Mstar{\rm M_{\star}}

\def\Mbh{\rm M_{\rm BH}}
\newcommand{\msun}{{\rm M}_{\odot}}

\defcitealias{lupi2019}{L19}

\usepackage{verbatim}

\title[Hunting the first Cosmic Giants]{Hunting the first Cosmic Giants: formation and detectability of Direct Collapse Black Holes around high-redshift quasars}

\author[Trinca et al.]{
Alessandro Trinca$^{1,2,3,4}$\thanks{E-mail: atrinca@roe.ac.uk},
Alessandro Lupi$^{2,5}$,
Zolt\'an Haiman$^{6,7,8}$,
Marta Volonteri$^{9}$,
Rosa Valiante$^{3}$,
\newauthor Raffaella Schneider$^{4,10,11}$,
Roberto Decarli$^{12}$
\\\\
$^{1}$ Institute for Astronomy, University of Edinburgh, Royal Observatory, Blackford Hill, Edinburgh EH9 3HJ, UK\\
$^{2}$ Como Lake Center for Astrophysics, DiSAT, Università degli Studi dell'Insubria, via Valleggio 11, 22100 Como, Italy\\
$^{3}$ INAF/Osservatorio Astronomico di Roma, Via Frascati 33, 00040 Monte Porzio Catone, Italy\\
$^{4}$ INFN, Sezione Roma1, Dipartimento di Fisica, “Sapienza’’ Università di Roma, Piazzale Aldo Moro 2, 00185 Roma, Italy\\
$^{5}$ INFN, Sezione di Milano-Bicocca, Piazza della Scienza 3, I-20126 Milano, Italy\\
$^{6}$ Institute of Science and Technology Austria (ISTA), Am Campus 1, 3400 Klosterneuburg, Austria\\
$^{7}$ Department of Astronomy, Columbia University, New York, NY, 10027\\ 
$^{8}$ Department of Physics, Columbia University, New York, NY, 10027\\
$^{9}$ Institut d’Astrophysique de Paris, Sorbonne Université, CNRS, UMR 7095, 98 bis bd Arago, 75014 Paris, France, \\
$^{10}$ Dipartimento di Fisica, “Sapienza’’ Università di Roma, Piazzale Aldo Moro 2, 00185 Roma, Italy\\
$^{11}$ Sapienza School for Advanced Studies, Viale Regina Elena 291, 00161 Roma, Italy\\
$^{12}$ INAF Osservatorio di Astrofisica e Scienza dello Spazio di Bologna, via Gobetti 93/3, Bologna, 40129, Italy\\
}

\date{Accepted 2026 June 18. Received 2026 May 25; in original form 2026 January 27}

\pubyear{\the\year{}}

\begin{document}
\label{firstpage}
\pagerange{\pageref{firstpage}--\pageref{lastpage}}
\maketitle

% Abstract of the paper
\begin{abstract}
The rapid emergence of supermassive black holes (SMBHs) in the early Universe poses a challenge to current models of black hole growth. One promising formation pathway is the direct collapse black hole (DCBH) scenario, in which gas in pristine, low-metallicity halos forms supermassive (or quasi-) stars leading to massive black holes seeds under specific environmental conditions. In this work, we investigate the potential host environments of DCBHs by coupling a semi-analytic model tracing BH formation and galaxy co-evolution with high-resolution N-body dark matter merger trees. This allows us to trace the population of DCBHs formed during the hierarchical assembly of a $\sim 10^{12} ~\msun$ dark matter halo hosting a bright $10^9 \ \msun$ quasar at redshift $z \approx 7$. We find that, when accounting for local fluctuations in the UV radiation field within this early cosmic structure, massive BH seeds can form via direct collapse as early as $z \approx 22$. Even under more stringent conditions for heavy seed formation, tens of DCBHs are predicted to emerge within the simulated overdensity down to $z \sim 14$, at which point metal enrichment of the intergalactic medium inhibits further episodes of direct collapse. A significant fraction of the massive black hole population formed at $z > 14$ is expected to survive in satellite galaxies that do not merge with the central halo down to $z \approx 7$. We show that the existence of such a population of ungrown heavy BH seeds can be probed through deep JWST observations targeting regions surrounding bright high-redshift quasars, and we discuss tailored observational strategies to detect and identify these elusive systems.
\end{abstract}

\begin{keywords} 
quasars: supermassive black holes - galaxies: high-redshift - galaxies: evolution - galaxies: active 
\end{keywords}

\section{Introduction}

\label{sec:intro}
The presence of supermassive black holes (SMBHs) with masses up to $\sim 10^9~\msun$ at redshifts beyond $z > 7$ presents a key challenge to our understanding of black hole (BH) formation and growth in the early Universe. Recent observational campaigns, powered by the unprecedented sensitivity of the James Webb Space Telescope (JWST), have pushed the detection of candidate SMBHs out to even higher redshifts, with a few candidates now identified up to $z > 10$ \citep[e.g.][]{ bogdan2024, maiolino24, napolitano2025, kovacs2024}. They uncovered the presence of systems powered by SMBHs with $\rm \log(M_{\rm BH}/\msun) \sim 7-8$ already in place when the Universe was $\lesssim 500$ Myrs old \citep[e.g.][]{tripodi2025, taylor2025}.

Explaining the presence of such massive BHs at these early times is particularly difficult if one assumes they originated from stellar-mass remnants of the first (Population III) stars, even if formed as early as $z > 20$. In this case, sustained accretion close or above the Eddington limit would be required over several hundred million years, which represent an unlikely scenario due to the expected strong radiative feedback and the inefficient gas inflows that might characterize their early growth phases. Although recent studies suggest that rapid SMBH growth could be accomplished through repeated, short-lived phases of super- or even hyper-Eddington accretion \citep[e.g.][]{Lupi2016,pezzulli2016,Sassano2023,lupi2024SE,Husko2025,trinca2025,shi2023,Shi2024,Shi2024b,Shi2026,Gordon2024,Mehta2024,Metha2026}, the efficiency and sustainability of such episodes remain uncertain and actively debated \citep{pezzulli2017b,regan2019,massonneau2023,Zana2025}.

An alternative and promising pathway to form the earliest massive black holes is the direct collapse black hole (DCBH) scenario. In this framework, massive BH seeds ($M_{\rm seed} \sim 10^4$-$10^6~\msun$) form rapidly from the collapse of pristine, metal-poor gas in atomic cooling halos (ACH). For this to occur, very specific environmental conditions are required: the gas must remain hot and avoid fragmentation, which in turn suppresses early star formation. Such conditions may arise due to strong dynamical heating during rapid halo assembly \citep{Wise2019,lupi2021,Latif2022}, sustained gas inflows \citep{Lodato2006,Mayer2015, Mayer2024}, or the presence of an intense Lyman-Werner (LW) radiation background, which dissociates molecular hydrogen and inhibits efficient cooling in the first dark matter mini-halos \citep{Haiman1997,Bromm2003,Begelman2006formation,Volonteri2008,Wise2008,Regan2009,Chon2016}. In the latter scenario, once the halo grows past the atomic cooling threshold (with virial temperature $T_{\rm vir} \gtrsim 10^4$ K), the gas can undergo an isothermal collapse on a timescale of $\lesssim 10^6$ yr, possibly forming a supermassive star as an intermediate stage before collapsing into a BH \citep{Oh2002,Volonteri2010,Ferrara2014}.

While these rare conditions have been extensively explored in the literature through analytical studies and high-resolution simulations, large uncertainties remain in predicting the formation rate and characteristic mass distribution of DCBHs \citep[see e.g.][]{Obrennan2025,Wise2023}. However, it is generally agreed that the necessary conditions for direct collapse, including rapid halo assembly and strong LW backgrounds, are more readily met in highly overdense regions in the early Universe \citep[e.g.][]{dijkstra2008fluctuations}. Therefore, the most massive halos formed at high redshift, which are those usually observed to host luminous quasars \citep{Eilers2024,wang2024}, represent prime environments to search for signatures of these early population of \textit{heavy seeds}, and have thus been the target of tailored numerical simulations \citep{khandai2012,dimatteo2017,valiante2016,valiante2018statistics,valiante2018observability}.
The advent of JWST already enabled extraordinary improvements in the observation of high-redshift quasars. Several dedicated programs have begun to characterize quasars at $z > 6$ in unprecedented detail, providing rest-frame optical spectra, the first detections of starlight from quasar host galaxies \citep[e.g.][]{Ding2023,yue2024}, spatially resolved emission-line maps \citep{Marshall2023,Decarli2024}, and insights into their large-scale environments \citep{Wang2023,Eilers2024,Pizzati2024}. This opens an unprecedented opportunity for identifying companion systems and satellite galaxies in quasar fields. Not only does it allow tracing the potential remnants or descendants of early DCBHs formed within these rare, massive cosmological overdensities, but it also enables a detailed characterization of their environmental properties, such as gas metallicity and ionization state, which could provide distinctive signatures of the seed formation channel. If successful, this could pave the way for searches for possible remnants of this formation scenario at much later cosmic epochs, potentially extending to satellite systems of Milky Way-like galaxies in the local Universe \cite[see e.g. ][]{vanWassenhove2010, scoggins2025}.

In this work, we aim to investigate and quantify the role of massive cosmological overdensities, which give rise to bright high-redshift quasars, as favourable environments for the formation of early DCBHs. We focus on the assembly history of a massive dark matter halo, with final mass $\sim 10^{12}\rm\, M_\odot$ at $z\sim 6$, representative of bright quasar hosts.  By combining a detailed semi-analytic model (SAM) of early galaxy and black hole formation with high-resolution dark matter merger trees extracted from cosmological N-body simulations, we trace the abundance, spatial distribution, and environmental conditions of halos capable of hosting DCBHs over cosmic time. We further study how the emerging population of BH seeds depends on the specific physical requirements for direct collapse formation.
Building on this, we investigate the spectral emission of DCBH-host satellites in the vicinity of high-redshift quasars at $z \sim 7$, providing predictions for their detectability in current and future surveys and proposing tailored observational strategies. This approach offers a solid theoretical framework for identifying favourable environments for heavy seed formation and for predicting the characteristic signatures that future surveys should target to place stricter constraints on the early formation of SMBHs through the direct collapse scenario.

\section{Numerical setup}
\label{sec:model}
In this work, we investigate the evolution of high-redshift quasars by coupling dark matter merger trees obtained from zoom-in cosmological simulations with the modelling of the baryonic evolution provided by the Cosmic Archaeology Tool (\textsc{cat}) semi-analytical model.

\subsection{The N-body merger trees}
Since we are interested in investigating the early build up of the high-redshift quasar population, in this work we reconstruct the hierarchical merger history of massive dark matter halo of $M_{\rm vir} \approx 3 \times 10^{12} \rm ~\msun$ at $z = 6$, taking advantage of detailed N-body simulations. 
The halo merger history has been derived from the cosmological zoom-in simulation by \citet{lupi2019} (hereafter \citetalias{lupi2019}). The simulation is based on the \textsc{gizmo} particle-based code \citep{hopkins2015} and covers a $75 ~h^{-1} ~\rm cMpc$ box. Constrained initial conditions are imposed with \textsc{music} \citep{MUSIC} starting from $z = 100$ to follow the assembly of a DM halo with the target mass within the simulated box. Recursive levels of refinement allowed to reach a maximum mass resolution of $m_{\rm DM} = 9.19 \times 10^4 ~\msun$ within a sphere of radius $2.5 ~\rm R_{vir}$ around the centre of the massive halo. While in \citetalias{lupi2019} the evolution of the baryonic component is followed using \textsc{gizmo}, which resolves in detail the gas and stellar dynamics, as well as the co-evolution of massive BHs with their host galaxies, resolution limits prevent a proper characterization of the very early phase of evolution of cosmic structures, in particular in the mini-halo regime ($T_{\rm vir} < 10^4 ~\rm K$). Therefore, in \citetalias{lupi2019} a standard seeding prescription is assumed, where a massive BH of $\Mbh = 10^6 ~\msun$ is formed in galaxies above a threshold $\Mstar > 10^8 ~\msun$. 
Here, instead, we adopt a different approach. Starting from the companion DM-only simulation employed in \citetalias{lupi2019}, we extract the merger tree of the final target halo using the \textsc{rockstar} halo finder \citep{Behroozi2013}. The high DM resolution allows us to reach the mass-scale of the first star-forming minihalos ($\Mhalo \sim 10^6 ~\msun$), each resolved with $\gtrsim 10$ particles \footnote{Although halos resolved with $\sim 10$ particles lie at the resolution limit, this choice is necessary to avoid missing potential star-forming minihalos at early epochs, which is crucial for characterizing subsequent DCBH formation. In practice, only a subset of these halos satisfies the star-formation criteria (see e.g. Fig.~\ref{fig:MapBHz7}), and results are marginally sensitive to this threshold, as star-forming halos are typically resolved with significantly more particles, particularly at later epochs.}.
Building on this, we then follow the evolution of the baryonic component in all the progenitor halos using a tailored semi-analytical model for galaxy evolution, as described below. This approach ensures a more detailed characterization of the minihalo evolutionary phase and the formation of first BH \textit{seeds}, compared to the simplified treatment adopted in \citet{lupi2021}. 

\subsection{Baryonic evolution}
Once obtained the hierarchical merger tree of the target quasar-host halo, we follow the baryonic evolution in all its progenitors using the Cosmic Archaeology Tool \citep[\textsc{cat,}][]{trinca2022}. \textsc{cat} is a semi-analytical model developed to characterize the formation of early cosmic structures, resolving in detail the formation of the first populations of stars and the formation of the first BHs through different possible seeding scenarios, with the aim of following the co-evolution between galaxies and their nuclear SMBH across cosmic times. 
The model has already been used to investigate large statistical populations of galaxies across a wide dynamical range, from the first minihalos to the massive quasar hosts, with a limited computational cost. In particular, it has demonstrated its ability to reproduce simultaneously - and relying on a small number of free parameters - several key observables characterizing galaxy and AGN populations at high-redshift, including UV luminosity functions, cosmic star formation rate density, and BH mass functions.
Here we summarize the main features of the model, while we refer the reader to the original papers for a more detailed description \citep{trinca2022, trinca2023bh, trinca2024}.

\subsubsection{Modelling galaxies}
\textsc{cat} is tailored to track and reconstruct the evolution of gas, stars, metals, and dust in each progenitor galaxy along a DM merger tree.
Starting from halo virialization, the model follows the gas inflows from the intergalactic medium (IGM) onto each newly collapsed halo and the subsequent onset of star formation. The star formation rate (SFR) is computed from the available gas mass at a given timestep, $M_{\rm gas}$, as follows:
\begin{equation}
    SFR = f_{\rm cool} M_{\rm gas} \epsilon_{SF}/\tau_{\rm dyn},
    \label{eq:SFR}
\end{equation}
where $\epsilon_{\rm SF}$ is the star formation efficiency and $\rm \tau_{dyn} \equiv [R_{vir}^3 /(G M_{halo})]^{1/2}$ is the halo dynamical time. 
The factor $f_{\rm cool}$ quantifies the reduced cooling efficiency of molecular hydrogen.  Its value ranges between 0 and 1 and depends on the halo properties, being $=1$ only for atomic-cooling halos. In minihalos, below $T_{\rm vir}  = 10^4 ~\rm K$, where atomic cooling is not available, the presence of incident Lyman-Werner (LW) radiation, in the $[11.2-13.6]$ eV energy band, can photodissociate molecular hydrogen and suppress the gas cooling efficiency. Therefore, in these halos $f_{\rm cool} <1$ and depends on the halo virial temperature, redshift, gas metallicity and intensity of the illuminating LW radiation \citep[see detailed discussion and results in][]{valiante2016, debennassuti2017, sassano2021}.

The SF efficiency $\epsilon_{\rm SF}$ is a free parameter of the model and has been calibrated for different stellar populations based on empirical and numerical results, as detailed below. 
We assume that Population~III (PopIII) stars form in pristine/metal poor halos, where the metallicity is $ \rm Z < Z_{\rm crit} = 10^{-3.8} Z_\odot$, according to a top-heavy distribution in the mass range $10 \, \msun \leq m_* \leq 300 \, \msun$, parametrized by a Larson initial mass function \citep[IMF;][]{larson1998early}:
\begin{equation}
    \rm \Phi(m_*) = m_*^{\alpha-1} \, e^{-m_{ch}/m_{*}}
    \label{eq:larsonIMF}
\end{equation}
where $\alpha=-1.35$ and the characteristic mass is $\rm m_{ch} = 20 \, M_\odot$. The PopIII IMF is stochastically sampled as a function of the effective total stellar mass formed in each star formation episode \citep[][]{valiante2016, debennassuti2017}, enabling to extract information on the single stars formed, and in particular their subsequent remnants, as will be discussed below. As discussed in \citet{trinca2024}, the SF efficiency adopted for PopIII star formation is $\epsilon_{\rm SF,~ PopIII}=0.15$, to align with results of detailed hydrodynamical simulations of star formation in pristine clusters \citep{chon2022}.

In halos enriched above the metallicity threshold, $ \rm Z > Z_{\rm crit}$, less massive, Population~II (PopII) stars are assumed to form according to Eq. \ref{eq:larsonIMF} with masses in the range $0.1 \leq m_* \leq 100 \, \rm M_\odot$ and a characteristic mass $m_{ch}=0.35 \, \rm M_\odot$. The efficiency of star formation for PopII host halos is assumed to be $\epsilon_{\rm SF, ~PopII}=0.05$ \citep{trinca2022}, which has been shown to reproduce various properties of the high-redshift galaxy population, including the SFR density, stellar mass density, and galaxy UV luminosity function.
\textsc{cat} subsequently follows the evolution of stellar population, as well as the metal- and dust-enrichment of the interstellar medium (ISM) due to asymptotic giant branch stars and supernovae (SNe), based on mass- and metallicity-dependent yields. It also considers a two-phase ISM environment, where dust grains can both be destroyed by SN shocks expanding in the diffuse hot medium and grow in mass by accreting gas-phase metals in warm dense gas. 
The gas abundance inside each galaxy is regulated by mechanical feedback associated with both SN explosions and energy deposition from nuclear BH accretion. 
The energy released by these processes couples with the gas, driving galactic-scale winds, that can remove a substantial fraction of the galaxy baryonic reservoir.
The outflowing material is assumed to mix uniformly with the surrounding gas throughout the simulated overdensity, leading to a gradual enrichment of the IGM. The enriched IGM is then accreted onto newly virialized DM halos, as well as onto pre-existing star forming galaxies during subsequent inflow episodes. As a result, a homogeneous metallicity floor builds up across the overdensity, which evolves with redshift and affects the subsequent episodes of star and BH seed formation, particularly in newly formed halos, by preventing them from originating out of pristine, chemically unpolluted gas.

The effect of photo-heating feedback is also considered, leading to star formation suppression in halos with virial temperatures below the temperature of the intergalactic medium ($T_{\rm vir} < T_{\rm IGM}$), where the latter is modelled self-consistently with the reionization process and traced according to the ionizing photon production in the simulated cosmic overdensity.
Finally, we also account for dynamical heating related to rapid halo mass growth, as described in detail in \citet{ventura2023}, which might introduce a delay in the onset of star formation. In particular, we estimate the gas heating rate following \citet{wang2008}, and suppress star formation whenever this exceeds the gas cooling rate estimated within the galaxy \citep[see][for a detailed description]{valiante2016}.
We refer the interested reader to \citet{valiante2016,trinca2022}, and \citet{trinca2024}, where a more detailed description of how galaxy and AGN emission is modelled within the simulation is presented.

\subsubsection{Black Hole formation and growth}
The model follows simultaneously two different seed BH populations. \textit{Light} seeds, up to a few hundreds solar masses, are formed as remnants of the first generation of massive and metal-free Pop~III stars.
In our model, Pop~III stars forming in the mass interval $\rm 40 \, M_{\odot}< m_* < 140 \, M_\odot$ and $m_* > 260 \, M_\odot$ are assumed to directly collapse into BHs of comparable mass \citep{heger2002}, after a time delay consistent with their stellar progenitor lifetime. The more massive among the Pop~III remnants formed in a single stellar population is then considered as a \textit{light} seed settling in the nucleus of its host galaxy thereafter \citep[see][for more details]{trinca2022, trinca2024}.

\textit{Heavy} BH seeds are instead expected to form in the centre of galaxies through the rapid collapse of a pristine gas cloud under peculiar environmental conditions \citep[see][for a comprehensive review]{inayoshi2020}. The gas molecular and metal cooling channels must be strongly suppressed, and the monolithic collapse of a large gas cloud in a single massive BH can take place without triggering fragmentation, leading to a final mass up to $ \rm M_{BH} = 10^{5-6} ~ M_\odot$ \citep[e.g.][]{latif2013,Shlosman2016}. 
We parametrize these conditions, requiring the galaxy to be hosted in atomic-cooling halos ($T_{\rm vir}> 10^4 ~ \rm K$) of pristine gas composition ($Z < Z_{\rm crit}$), and illuminated by a Lyman-Werner flux, $\rm J_{LW}$, intense enough to dissociate H$_2$ molecules ($\rm J_{LW} > \, J_{\rm crit}$), thus preventing molecular cooling and subsequent star formation.

The critical value of the LW background intensity required to suppress $\rm H_2$ cooling in ACHs is $J_{\rm crit} \sim 1000 ~J_{21}$\footnote{$J_{21}$ is a reference LW flux of $10^{-21} \, \rm erg \, s^{-1} \, cm^{-2} \, Hz^{-1} \, sr^{-1}$.}, as found in both semi-analytic calculations \citep[e.g.][]{sugimura2014critical,wolcottgreen2017}, and hydrodynamical simulations (\citealt{shang2010supermassive}; see also the review by \citealt{inayoshi2020}). Note that this value is significantly higher than the critical LW intensity inferred for minihalos, where $J_{\rm crit} \sim 0.1-1 ~ J_{21}$. The difference arises because atomic cooling allows the gas to reach substantially higher densities than in minihalos. Since the $\rm H_2$ formation rate scales as $\propto \rho^2$, while the LW photodissociation rate is proportional to $J_{\rm LW} \times \rho$, the critical LW intensity increases with gas density.
Several previous studies have nonetheless adopted lower values of $J_{\rm crit}$ for ACHs, well below $\sim 1000$, including the original work used to calibrate the \textsc{cat} suite, which adopted $J_{\rm crit}=300 ~ J_{21}$ \citep{trinca2022}. As discussed in \citet{sugimura2014critical} and \citet{wolcottgreen2020suppression}, such a value would be appropriate only for very soft spectral energy distributions (SED), which reduce $J_{\rm crit}$ through the efficient photo-dissociation of $\rm H^-$ ions by infrared photons. While such soft SEDs may arise from metal-enriched post-starburst populations (with ages $\gtrsim 300$ Myr and metallicities $\gtrsim 0.2\,Z_\odot$), it appears unlikely that these sources would dominate the radiation fields illuminating DCBH host halos at $z \gtrsim 10$.

Despite these considerations, in the present paper we continue to adopt the fiducial value $J_{\rm crit}=300 ~ J_{21}$ for two reasons. First, adopting a higher threshold would require a re-calibration of the entire \textsc{cat} cosmological suite, as this value may be degenerate with other model's free parameters, which is beyond the scope of the present study. Second, there is growing evidence that DCBH formation may still be possible even when the LW radiation field is insufficient to completely suppress $\rm H_2$ cooling. For example, in the Renaissance simulation suite, \citet{Wise2019} identify two pristine ACHs exposed to LW fluxes as low as $\sim$ a few, which nevertheless experience rapid gas infall rates. Although $\rm H_2$ cooling is not suppressed in these halos, the large infalls appear to arise because of unusually strong dynamical heating. While in the core of the halo (within $\sim 1$ pc) the inflow rate drops below the critical value required for supermassive star (and hence DCBH) formation, high-resolution re-simulations of the halo's core suggest that this may still occur \citep{sakurai2020,toyouchi2023}. Similar conclusions are also reached by \citet{Regan2020b}. These results suggest therefore that DCBHs may form even when the LW flux remains below the critical value of $J_{\rm crit} \sim 1000 ~ J_{21}$ required for complete $\rm H_2$ cooling suppression.

The sensitivity of our results to the choice of $J_{\rm crit}$ and the implications of requiring higher incident LW fluxes will be explored in Section \ref{sec:Jcrit_dep}.
If the all the conditions described above are met, a heavy seed of $\rm M_{BH} = 10^5 \, \rm M_\odot$ is formed in the centre of the galaxy. 
The assumption of a single, monochromatic seed mass is primarily motivated by the current uncertainties surrounding the mass distribution of DCBHs. Although recent high-resolution simulations have started exploring how the mass spectrum of supermassive stars may depend on different environmental conditions \citep[see e.g.][]{Prole2024b, chon2025}, these constraints remain highly uncertain. We therefore adopt the simplified prescription originally implemented in the \textsc{cat} model calibration, which has been shown to successfully reproduce several observed properties of the high-redshift galaxy and AGN populations \citep{trinca2022, trinca2024}. The impact on our predictions of a lower initial seed mass, as well as the possibility of DCBH formation in mildly enriched environments, will be further discussed in Sections \ref{sec:results} and \ref{sec:discussion}.

Coupling the galaxy evolution predicted by \textsc{cat} with the N-body merger trees presented in Section 2.1, we have been able to track in detail the spatial distribution of emitting sources in our simulated overdensity, and describe with higher precision the incident LW flux on each halo depending on the spatial distribution of neighbouring galaxies. The details on how the LW flux is modelled within the simulation are provided in Appendix \ref{sec:AppendixA}.

After the formation of the initial seed, we follow the growth of BHs through gas accretion and mergers across cosmic time. In this work, we assumed the nuclear BH to accrete according to the Bondi-Hoyle-Lyttleton rate \citep{bondi1952,hoyle1941}, calculated as:
\begin{equation}
    \dot{M}_{\rm BHL} = \alpha \, \frac{4 \pi G^2 M_{\rm BH}^2 \rho_{\rm gas}(r_A)}{c_{s}^3}
\label{eq:bondi}
\end{equation}
where $c_s$ is the sound speed and $\rho_{\rm gas}(r_A)$ the gas mass density evaluated at the radius of gravitational influence of the BH, $r_{A}=2G M_{\rm BH} / c_s^{2}$. We assume here a gas density distribution described by a singular isothermal sphere profile with a flat core:
\begin{equation}
    \rho(R) = \frac{\rho_{\rm norm}}{1 +(R/R_{\rm core})^2}
\end{equation}
where $R_{\rm core}=0.012 \, R_{\rm vir}$ and the profile is normalized such that the total gas mass is always enclosed within the galaxy virial radius \citep[see][]{valiante2011}. 
The factor $\alpha$ is a boost parameter which is needed due to the inability of semi-analytical models, as well as low-resolution simulations, to accurately estimate the gas density within the Bondi radius, which can be significantly enhanced compared to the average density in the cored galactic profile \citep{dimatteo2012,schaye2015}. We set the value of this parameter to $\alpha=90$, which has been shown to reproduce the properties of the high redshift quasar population \citep{trinca2022,trinca2025}.
In this scenario, to be conservative we limit the BH mass growth rate ($\dot{M}_{\rm BH}$) to the Eddington limit, $ \dot{M}_{\rm Edd}=L_{\rm Edd}/(\epsilon_r c^2)$, where $\epsilon_r=0.1$ is the adopted radiative efficiency and $L_{\rm Edd}=4\pi G M_{BH} m_{\rm p}/\sigma_{\rm T}$ is the Eddington luminosity in which c is the speed of light, $m_{\rm p}$ is the proton mass, and $\rm \sigma_{\rm T}$ is the Thomson scattering cross-section. Therefore, the growth rate is computed as follows: 
\begin{equation}
  \dot{M}_{\rm BH}= (1-\epsilon_{\rm r}) \times \min(\dot{M}_{\rm BHL},\dot{M}_{\rm Edd}).
\end{equation} 

In addition, nuclear BHs can also grow though coalescence following the merger of their host galaxies.
In major mergers (defined as galaxy mergers where the stellar mass ratio $\mu$ is $> 1/10$), we assume the two nuclear BHs to efficiently sink into the centre of the newly formed galaxy and merge instantaneously, i.e. within the typical time step of the simulation, $\Delta t \sim 0.8-1.7 ~ \rm Myr$. Conversely, in minor galaxy mergers ($\mu < 1/10$), only the most massive of the two BHs is considered as the nuclear BH of the resulting galaxy, while the least massive one is considered a wandering system, that will sink with timescales $> t_{\rm Hubble} (z=6)$, and hence its evolution is no longer followed in the simulation.

\section{Results}
\label{sec:results}

\subsection{Quasar Evolution}
\begin{figure}
\centering
\includegraphics[width=\columnwidth]{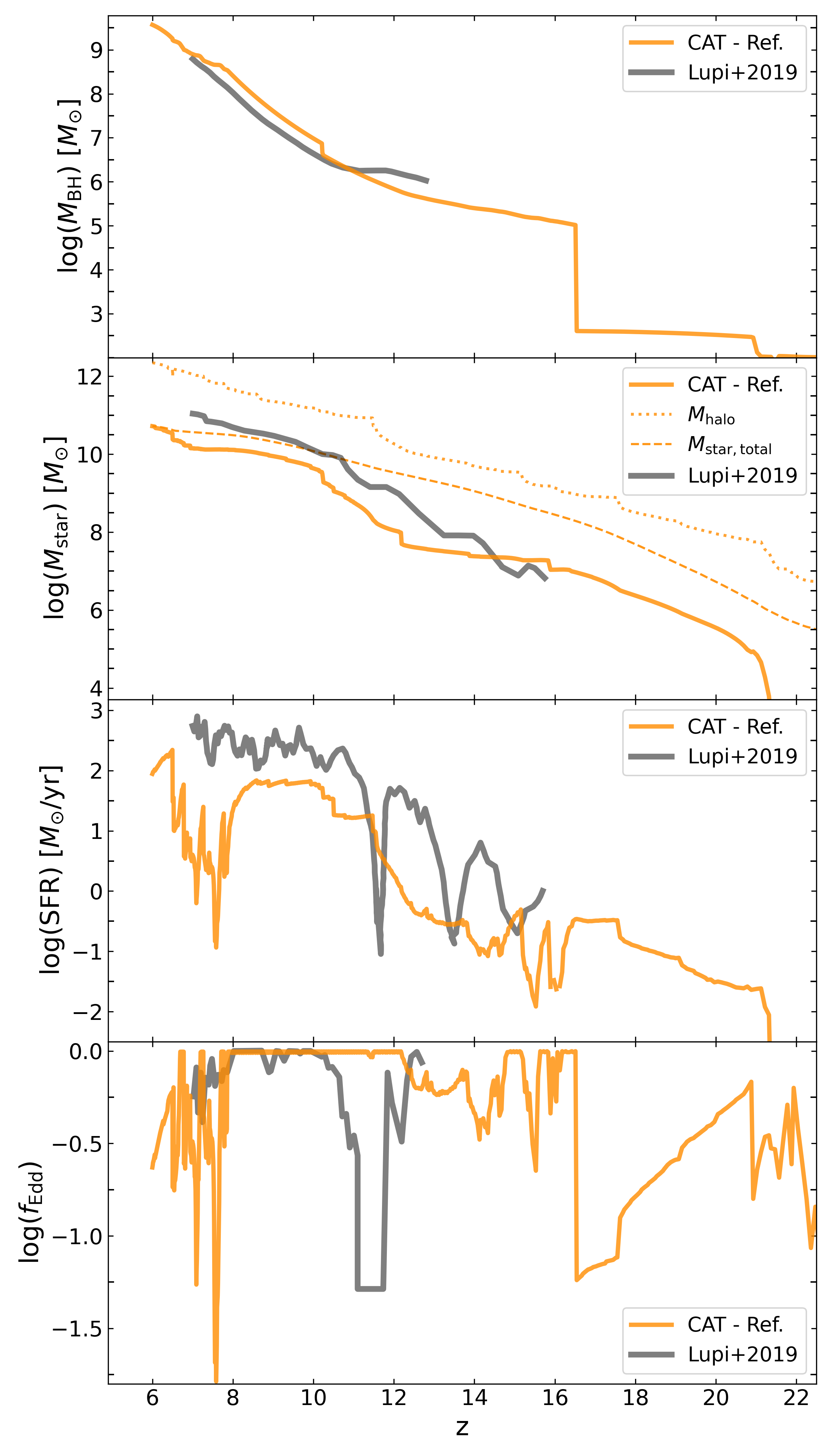}
\caption{Evolution of the properties of the main simulated galaxy. From top to bottom we show the BH mass, stellar mass, SFR, and BH accretion rate. \textsc{cat} predictions (orange lines) are compared with the results of the hydrodynamical simulation by \citetalias{lupi2019} (gray lines). For reference, the stellar mass panel also shows the DM halo mass growth and the total stellar mass assembled in the simulated overdensity with dotted and dashed lines, respectively.
\label{fig:GalEvo}}
\end{figure}

In Figure \ref{fig:GalEvo} we show the evolution in redshift of the main properties of our simulated massive quasar host halo at $z > 6$ as predicted by \textsc{cat}. We report the black hole mass, stellar mass, star formation rate and black hole accretion rate (BHAR) compared to the evolution obtained from the hydrodynamical simulation in \citetalias{lupi2019}, sharing the same underlying halo merger history \footnote{For a proper comparison, we track here the evolution of the most massive branch throughout the assembly history of the final quasar host galaxy.}.
We use this as a consistency check to see whether the model is able to provide consistent predictions for the properties of the final galaxy at $z \approx 6$. Note that in \citetalias{lupi2019} the baryonic simulations ends at $z \approx 7$, while the merger tree is extracted by the underlying DM run which extends to lower redshift.

The quasar SMBH mass evolution closely follows the result of the hydrodynamical simulation, with the  discrepancy remaining below $0.5 \ \rm dex$ across the entire evolution at $z<13$. The SMBH reaches $\log(\Mbh/\msun) = 8.9$ at $z=7$, within $0.1 \rm \, dex$ from the prediction of \citetalias{lupi2019}. In particular, the late phase of growth observed in the hydrodynamical simulation, below $z \sim 13$, is closely matched despite the simplified modelling of the accretion and gas distribution provided by the semi-analytical approach. This agreement is likely due to the growth occurring near the Eddington limit, which is set as the maximum accretion rate in both simulations. The sharp increase in BH mass at $z \sim 16$ marks the moment when the first DCBHs - formed in a neighbouring pristine ACH - enters the main galaxy branch and starts accreting gas efficiently. Following a phase of rapid growth, at $z \lesssim 8$ the increased BH feedback begins to deplete the galaxy's gas reservoir, suppressing star formation and stabilizing BH accretion at an average rate of $\sim 0.3 ~ f_{\rm Edd}$.

The final galaxy stellar mass at $z \approx 7$ predicted by \textsc{cat}, $\log(\Mstar/\msun) \approx  10.2$, appears lower by about $\sim 0.5$ dex compared to \citetalias{lupi2019}.
The difference is likely related to the different initial seed mass and early SMBH growth. In \textsc{cat}, as the SMBH grows from the initial heavy seed mass, $10^5~\rm M_\odot$, to $10^6~\rm M_\odot$, it significantly affects the host galaxy gas reservoir, slowing the stellar mass growth. In the original hydrodynamical simulation by \citetalias{lupi2019}, the SMBH is instead directly seeded at $M_{\rm seed} = 10^6~\rm M_\odot$. At later times ($z \sim 8 - 11$), when the simulations share comparable SMBH masses, the stellar mass shows a similar evolution, but maintaining the rigid offset built at earlier times. Another major difference is how the feedback is modelled in the simulation.
In \textsc{cat}, a fixed fraction of the energy released by the accreting SMBH is assumed to couple with the galaxy gas reservoir, generating large scale outflows that affect especially the late phase of galaxy evolution, as shown by the sharp drop in the SFR at $ z \lesssim 8$, following a prolonged phase of efficient BH accretion at the Eddington limit. In \citetalias{lupi2019}, the BH thermal feedback is instead consistently coupled with the gas on nuclear scales, and was shown to have only a minor impact on the overall galaxy. This translates in an observable difference in the BHAR and SFR at $z<8$. Conversely, stellar feedback in the early phases of evolution can affect the SMBH growth, as evidenced by the transient drop in BHAR at $z\sim 12$ observed in the hydrodynamical simulation and not captured by \textsc{cat}, coincident with a stellar feedback outburst that also temporarily suppresses the galaxy SFR.  In \citet{quadri2025}, a refined analysis of the same system is presented, incorporating a more detailed treatment of AGN feedback across different accretion regimes, including kinetic feedback from jets and radiatively-driven winds. They show how, although feedback from the accreting BH does not significantly impact the host galaxy during most of its evolution, episodes of high accretion rate combined with particularly favourable alignment between the jets and the galactic plane can lead to the formation of a central cavity of  a few 100s parsecs in size. This almost completely quenches SF, but represents only a transient phase of $\sim 50 ~\rm Myrs$, after which both the galaxy and the BH resume a rapid growth.
 
AGN-driven outflows may effectively regulate star formation only when the black hole has grown sufficiently massive and under specific configurations in which jets or radiatively driven winds interact strongly with the gas in the galactic disk plane. At other times, AGN feedback has a limited impact on star formation in these massive overdensities, where strong gas inflows prevent halo starvation.
It is also important to note that the star formation efficiency in hydrodynamical simulations is sensitive to the adopted subgrid physics \citep[e.g.,][]{Rosdahl2017}. In the updated simulations presented in \citet{quadri2025}, which were run from the same initial conditions as \citetalias{lupi2019}, SF is found to be less efficient due to a more accurate treatment of high-density cooling, as well as improved prescriptions for both stellar and SMBH feedback.

Despite these differences, the main properties and evolution of the simulated galaxy remain broadly consistent with previous results from hydrodynamical simulations, without requiring any fine-tuning of the model free parameter outside of the fiducial ranges previously adopted to reproduce the global galaxy population \citep{trinca2022}. In the following sections, we focus on characterizing the formation environments of the population of direct collapse black holes predicted by the \textsc{cat} model - representing the progenitors of the massive quasars observed at $z \approx 6$ - and how their abundance varies under different assumptions regarding the conditions that enable their formation.

\subsection{Tracing the local Lyman-Werner field}

The impact of nearby sources is crucial in the DCBH formation process, as they can provide a significant LW flux capable of suppressing $\rm H_2$ formation in pristine ACHs already at very early epochs.
Accurately resolving the spatial distribution of progenitor halos of a massive quasar host, and determining the irradiating LW flux that each potential DCBH host galaxy experiences, is essential for a correct characterization of the early population of massive BH seeds.

\begin{figure}
\centering
\includegraphics[width=1.05\linewidth]{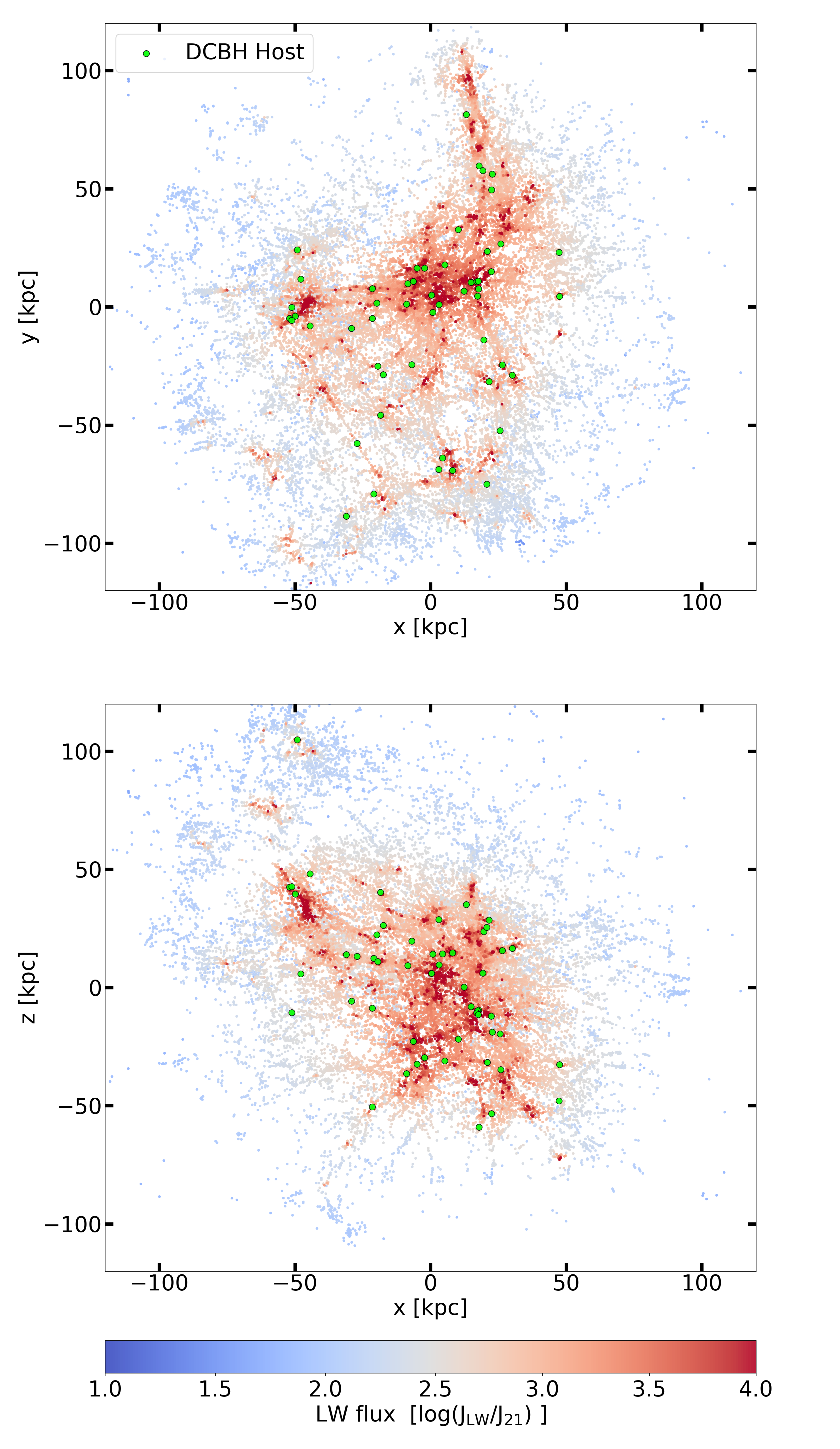}
\caption{Projections in the XY and XZ plane of the spatial distribution of all the quasar-host progenitor halos at $z=15$ (for a total number of $\rm N_{\rm halos} = 30649$), colour coded according to the LW flux illuminating each galaxy. Distances are reported in physical kpc. Regions where the LW intensity is above the threshold for DCBH seed formation, $J_{\rm crit} = 300 ~ J_{ \rm 21}$ are marked with red colours. Green points show the position of halos hosting a heavy seed descendant.
\label{fig:HseedsDistr}}
\end{figure}

In Figure \ref{fig:HseedsDistr}, we show the projected spatial distribution (in the XY and XZ planes) of all the progenitor DM halos of the final quasar host galaxy (for a total number of $\rm N_{halos} = 30649$) at $z = 15$, which marks the final epoch of DCBH formation predicted for the simulated overdensity (as will be discussed in Section \ref{sec:SeedDistribution}). Each halo is colour coded according to the external LW flux it receives, while those hosting DCBHs are highlighted in green. Red colours in the colourscale mark regions where the LW intensity exceeds the assumed threshold for DCBH formation, $J_{\rm crit} = 300 ~ J_{ \rm 21}$.

We find that a significant number of heavy seeds, $54$ in total, are expected to form within the simulated overdensity. As expected, the highest background LW fluxes arise in the central regions of the overdensity, naturally tracing the most massive progenitors and their associated high SF activity. Massive seeds formation occurs mainly in these central regions, where, however, rapid metal enrichment might play a competing role in suppressing the onset of direct collapse, as discussed in detail in Section \ref{sec:MetalPollution}.
Nonetheless, a non-negligible fraction of the population formed by $z\sim15$ is located in more external regions, extending up to $\sim 50-100$ physical kiloparsecs (pkpc) from the centre, predominantly associated with clustered substructures and filaments.
In the following sections, we analyse in detail the properties of this population of massive black hole seeds, focusing in particular on the distribution of their formation epochs and the intensity of the LW radiation incident on their host galaxies at the onset of the direct collapse process.

\subsection{DCBH formation: modelling and environmental dependencies}
\label{sec:SeedDistribution}
The predicted population of DCBHs depends on both the modelling of the LW background, and on the threshold flux assumed to suppress molecular hydrogen formation in pristine ACHs.
To this aim, we compare the \textsc{cat} predictions obtained running the model for two different scenarios:
\begin{enumerate}
    \item \textit{Homogeneous LW field}: the incident LW flux is estimated as a uniform background, averaging over the total emission by galaxies in the simulated overdensity. This approach represents a common approximation when relying on semi-analytical merger trees, which do not provide information on the spatial distribution of progenitor halos (e.g., in \citealt{trinca2022,trinca2023bh}).\\
    
    \item \textit{Local LW field}: for each galaxy, the incident LW flux is estimated self-consistently from the emission of all the neighbour systems, thanks to the halo spatial information provided by the N-body merger tree. This represents the reference used throughout this work.
\end{enumerate}

In Figure \ref{fig:Hist_LocalFlux} we compare the resulting distributions of newly formed heavy seed BHs for an homogeneous (panel a) and local (panel c) LW field. In panel (b) we show the corresponding cumulative distributions, while panel (d) shows the local LW flux received by their host halos at the time of BH formation, compared with the average background flux predicted across the overdensity. In both scenarios, we assume a critical LW threshold flux of $\rm J_{LW, crit} = 300 ~ J_{21}$.
Although the total number of DCBHs formed across cosmic time is similar in the two cases (54 for the local flux model and 47 for the homogeneous background), the redshift distribution of their formation times is significantly different. Massive BH seeds start to form at a much earlier cosmic epoch when we consistently follow the local LW field on DM halos, as the spatially resolved contribution from nearby sources can provide a sufficiently strong flux to trigger direct collapse even at very high-redshift. In this scenario, the first DCBH forms as early as $z \sim 22$, and a non-negligible fraction of the total population ($\sim 10/54$) forms at $z>17$. Before this epoch, as shown by the blue dashed line in panel (d), the average background flux is not high enough to trigger the formation of DCBHs in favourable pristine ACHs, hence no heavy seeds form in the first scenario.

Additionally, a large fraction of DCBH host halos are exposed to a LW flux from $J_{\rm LW} \approx  10^3$ up to $\gtrsim 10^5$ at the time of seed formation - significantly higher than the assumed critical threshold $J_{\rm crit} = 300 ~J_{21}$ - including systems where the SMBH forms at very early times. This highlights the importance of resolving the local contribution from nearby halos, in particular configurations usually known as ``synchronized galaxy pairs" \citep{dijkstra2008fluctuations,visbal2014synchrPairs,regan2017}. In such cases, one of the galaxies forms stars and provides a substantial LW flux which prevents efficient $\rm H_2$ cooling and therefore episodes of SF in the neighbouring minihalo. This suppresses $\rm H_2$ cooling and triggers the monolithic gas collapse as soon as the halo enters the atomic-cooling regime.

\subsubsection{Dependence on critical flux}
\label{sec:Jcrit_dep}
To test how much our predictions for the DCBH population depend on the assumed critical LW flux required to suppress $\rm H_2$ formation, we re-ran our simulation with the resolved local LW flux, adopting a higher threshold value of $\rm J_{LW, crit} = 1000$, which is likely more appropriate for realistic galaxy spectra at these early epochs, with the exception of rare metal-enriched post-starburst galaxy SED (see e.g., \citealt{sugimura2014critical,Wolcott-Green2017}).

\begin{figure*}
\centering
\includegraphics[width=0.8\textwidth]{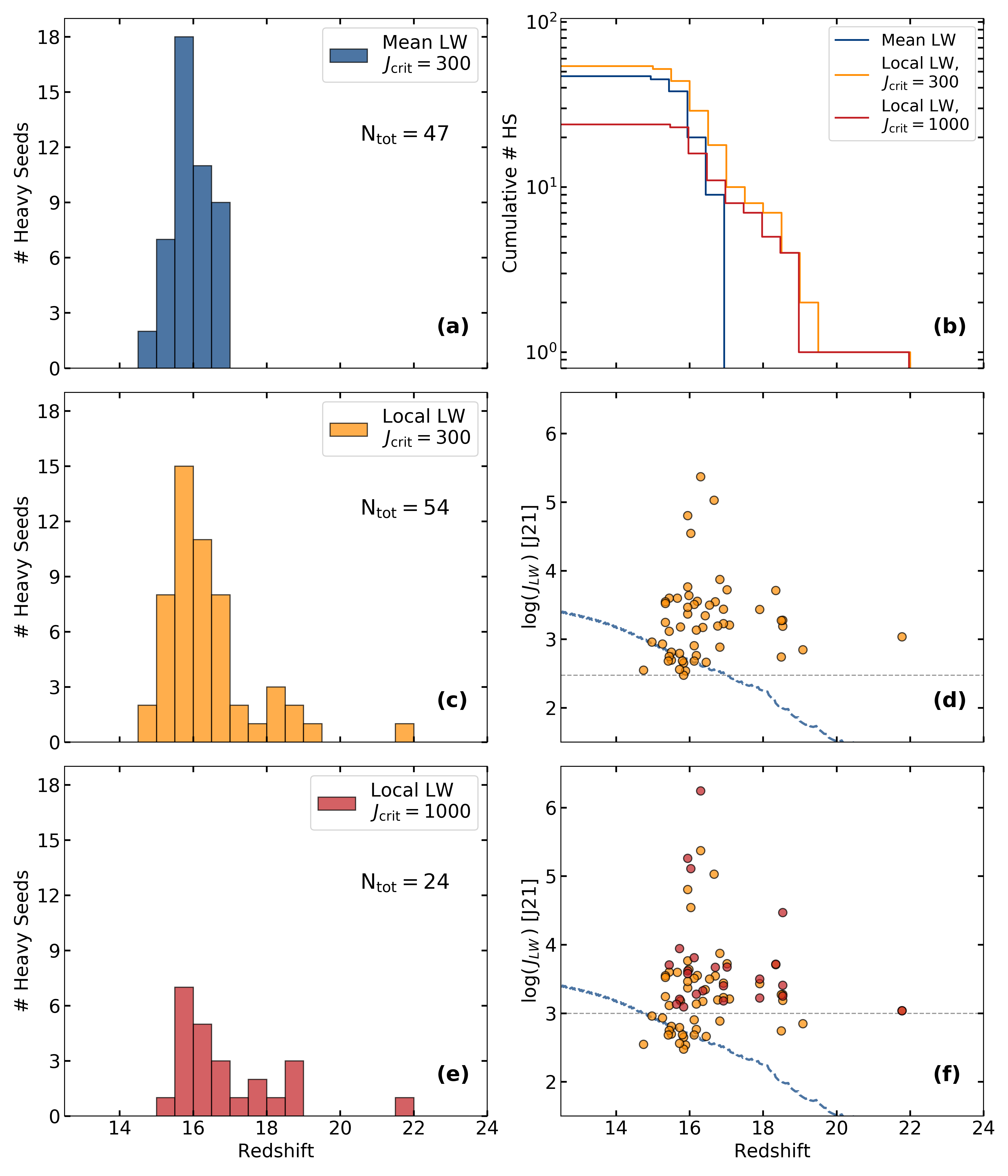}
\caption{
Redshift distribution of newly formed heavy BH seed for different model variants. Panel (a) shows the results for a homogeneous LW field (blue), while the results including the resolved local flux seen by each galaxy are shown in panels (c) and (e) for a critical threshold of, respectively, $\rm J_{LW, crit} = 300 ~J_{21}$ (orange) and $1000 ~J_{21}$ (red). Panel (b) shows the comparison between the corresponding cumulative seed distributions. The distribution of incident LW fluxes at DCBH formation vs redshift is shown in panels (d) and (f) for, respectively, local LW flux models with $\rm J_{LW,crit} = 300$ (orange points) and $1000$ (red points), where the blue-dashed line marks the background LW flux evolution within the simulated overdensity.
}
\label{fig:Hist_LocalFlux}
\end{figure*}

In Fig. \ref{fig:Hist_LocalFlux} we compare the results obtained for $\rm J_{LW, crit} = 300$ (our reference model) and $\rm J_{LW, crit} = 1000$ (in panels $(e)$ and $(f)$). As expected, increasing the critical flux reduces the total number of heavy seeds formed by $\sim 45 \%$, from 54 to 24. However, the largest difference occurs at redshift $z < 17$, where a significant fraction of DCBHs formed in the reference run are subject to LW fluxes below $10^3 \, \rm J_{21}$. In contrast, most of the seeds formed at earlier times are illuminated by higher LW fluxes, and that holds even for the earliest one formed at $z \approx 22$. This highlights once again how, at earlier cosmic epochs, the role of nearby halos is crucial to properly identify favourable environments for heavy seed formation, while at later times the combined contribution of multiple, more evolved surrounding galaxies becomes dominant in shaping the LW field, driven by the high clustering expected in overdense regions \citep[as shown in][]{lupi2021}.
Remarkably, in this scenario, even under a stricter assumption of a critical flux threshold of $J_{\rm crit} = 10^3 \, J_{21}$ - at the upper edge of the range commonly considered in literature - we still predict the formation of $24$ heavy seeds within a massive overdensity evolving into a $M_{\rm h} \sim 10^{12} \, \rm \msun$ quasar host halo by $z \approx 6$. 

Overall, these results suggest that:
\begin{itemize}
    \item Massive BH seeds from direct collapse can form at very early cosmic epochs within large overdensities. The LW flux produced by early star-forming structures might be high enough to photo-dissociate $H_{2}$ in nearby pristine halos, even in conditions where molecular clouds provides a significant self shielding of their internal regions. This might enable massive BHs to start growing already at early times, potentially relieving the tension with the detection of $M_{\rm BH} = 10^{6-7} ~\rm M_\odot$ systems already in place at $z>9$. However, this scenario would require that efficient gas accretion is already possible in these early structures, characterized by relatively shallow potential wells. \\
    \item The large number of heavy BH seeds predicted to form around the central galaxy would make the observability of multiple SMBHs in the vicinity of quasar hosts at high redshift more likely. This, in turn, could provide a promising pathway to constrain observationally the direct collapse scenario for SMBH formation, as we will further investigate in Section \ref{sec:HseedObservability}.
\end{itemize}

\subsubsection{Impact of external metal pollution}
\label{sec:MetalPollution}
An important factor that might affect our results is the role of potential external metal pollution of otherwise pristine DM halos, due to supernova-driven ejecta from nearby star forming galaxies. Such external enrichment could enable efficient gas cooling through fine-structure metal lines even in the presence of intense LW radiation. This enhanced cooling might favour rapid fragmentation and subsequent episodes of star formation, preventing the gas from collapsing isothermally and forming a massive BH \citep{omukai2001,omukai2008}. However, the actual impact of external metal pollution in the early Universe remains largely uncertain. 
On the one hand, several studies have shown that synchronized galaxy pairs, where one galaxy starts forming stars while the other remains pristine, can lead to a metal enrichment of the latter through SN-driven winds before it enters the atomic-cooling regime, potentially suppressing DCBH formation \citep[see e.g. discussions in][]{habouzit2016a,habouzit2016, regan2017,lupi2021,Obrennan2025}. On the other hand, the efficiency of SN outflows in low-mass systems is limited, and the ability of enriched gas to penetrate and mix with the dense, pristine cores of neighbouring halos is still debated, and highly dependent on the velocity and mass-loading of SN winds in these early systems \citep{smith2015,habouzit2016,Mead2025}, as well as on how quickly the pristine halo grows above the atomic cooling threshold, triggering rapid gas collapse.

In the \textsc{cat} model, due to this complex interplay of mechanisms that are beyond the descriptive capability of the SAM, we evolve the IGM metallicity as a volume-averaged quantity across the simulated overdense region. 
This is done by tracking the metal-enriched outflows driven by AGN and SN feedback from each galaxy and diluting this material into the surrounding medium, which represents a localized patch of the IGM within the simulated cosmological volume that then serves as a reservoir for subsequent episodes of gas accretion onto the main halo and its progenitors.
This approach introduces two opposite effects. On the one hand, an averaged IGM metallicity could result in an accelerated enrichment for halos in the outskirts, which will be subject to enriched inflows as soon as the first strong SN winds start to take place. This might lead to underestimating the number of potential DCBH host halos, since some of them might be evolving within metal-poor patches of the IGM down to later cosmic epochs. On the other hand, in case of synchronized-pairs, the outflows and SN-winds originated from the star forming galaxy might reach rapidly the pristine companion, before the metals gets diluted in the metal-poor IGM gas. Neglecting the role of these localized early enrichment events could therefore lead to overestimating the number of potential DCBH formation sites \citep{lupi2021}.

\begin{figure}
\includegraphics[width=\columnwidth]{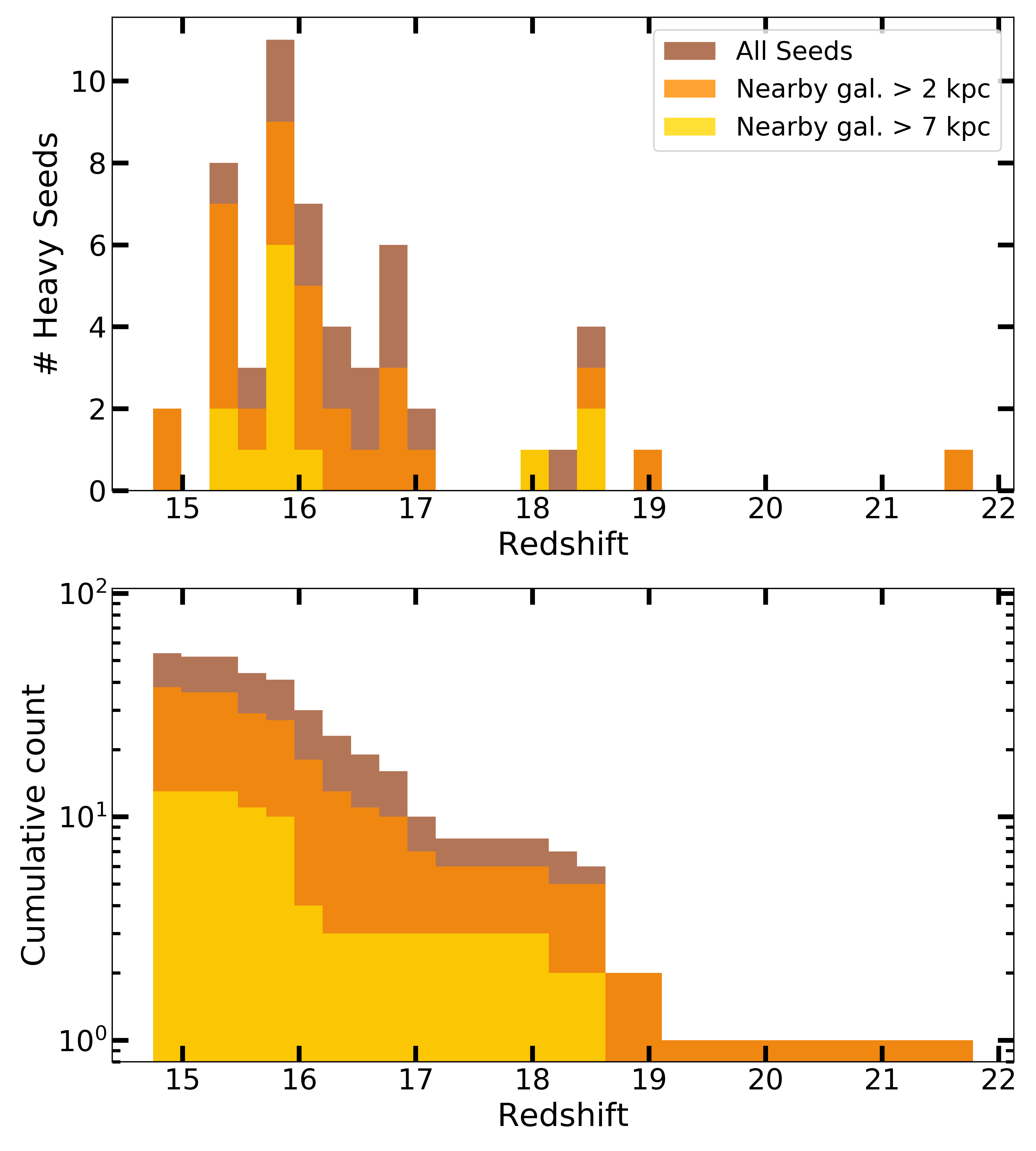}
\caption{
Number of heavy seeds formed in the simulation as a function of redshift (upper panel) and related cumulative distribution (lower panel). Orange and yellow histograms show, respectively, the fraction of DCBH host halos having - at the time of seed formation - the closest star forming galaxy at a distance larger than 2 and 7 physical kpc.}
\label{fig:PollutionImpact}
\end{figure}

Previous semi-analytical studies based on large cosmological volumes have shown that at $z \sim 20$ - when the first heavy seeds form in our simulation - the typical extent of metal-enriched bubbles around star-forming galaxies reaches up to $\sim 40-50~\rm ckpc$. By the end of the DCBH formation epoch ($z \sim 15$), this enrichment can extend to $\sim 80~\rm ckpc$ \citep{ventura2024}.
Building on these estimates, we provide a conservative assessment of the potential impact of metal enrichment from nearby galaxies on our results. In Figure \ref{fig:PollutionImpact} we show again the distribution of direct-collapse host halos formed in the reference simulation ($J_{\rm crit} = 300\, J_{21}$), highlighting the fraction of systems for which, at the time of heavy seed formation, the nearest star-forming galaxy lies at a distance larger than 2 and 7 physical kpc, corresponding to, respectively, $ \sim 40~ \rm ckpc$ at $z = 20$ and $ \sim 80~ \rm ckpc$ at $z = 15$.
We find that, at $z > 20$, even under the assumption of relatively extended metal-enriched bubbles (up to $\sim 50~\rm ckpc$), the first heavy seeds are still expected to form in pristine environments. At later times, external pollution from surrounding galaxies may start to reduce the number of massive BH seeds formed. However, even under the very conservative assumption that all star-forming galaxies enrich their surroundings up to $7~\rm pkpc$, $13$ DCBHs are still predicted to form in metal-free regions within the simulated overdensity.
For completeness, we repeated the same analysis for the \textsc{cat} run where we assumed $J_{\rm LW, crit} = 1000$, since a higher LW threshold would require favourable DCBH host systems to have a nearby irradiating companion, especially at early times. This, in turns, could make them more sensitive to the effects of local external enrichment. However, even in this more stringent case, 13 and 4 (out of a total of 24) heavy seeds are predicted to form in metal-free regions, assuming enriched bubbles of, respectively, 2 and 7 pkpc around each star forming system.
This suggests that while external metal enrichment likely plays an important role in regulating the formation of massive seeds within the first cosmic structures, it is unlikely to completely suppress their formation in highly clustered regions of the early Universe.

It is worth noting, in addition, that detailed hydrodynamical simulations \citep{omukai2008,chon2020,chon2025} have shown that massive BH seeds can still form in massive, strongly irradiated clouds even in the presence of moderate metal-enrichment, up to $\rm Z \sim 10^{-3} ~Z_\odot$, due to strong gas inflows overcoming small scale fragmentation. If that is the case, in such non-pristine environments also internal radiation - produced by the first stellar population formed within the halo itself - has been proposed to play a role in suppressing $\rm H_2$ and promoting DCBH formation \citep{Chiaki2023}, although recent simulations suggest this effect is unlikely to play a significant role \citep{sullivan2025}.
These results suggest that the impact of moderate gas enrichment on DCBH formation might be less restrictive than previously thought.

\subsubsection{DCBHs number density}
As a benchmark, it is instructive to compare our estimates with the results presented by \citet{lupi2021}, who adopted the same halo evolutionary history but employed a simplified semi-analytical framework to identify potential DCBH host halos among the quasar progenitors. Their analysis focused on a more restrictive formation pathway, searching for synchronized halo pairs, in which the smaller halo crosses the atomic-cooling threshold while being irradiated by a nearby source within a maximum separation of $\sim 0.2 - 1 ~\rm pkpc$. Within this framework, in their fiducial model they predict between 9 and 94 synchronized pairs, depending on the assumed maximum separation of the irradiating source ($0.5-1 ~\rm pkpc$), of which only $\sim 1-19$ systems are expected to remain pristine and thus capable of hosting DCBH formation.
Despite the substantial differences in methodology and physical assumptions, these estimates are broadly consistent in order of magnitude with our results. In the \textsc{cat} model, we predict approximately $\sim 24-54$ DCBH host candidates, depending on the adopted value of $J_{\rm LW,crit}$, with this number decreasing to $\sim 4-13$ under the most stringent assumptions regarding external metal enrichment. Taken together, this comparison suggests two key points. First, both approaches point toward the possibility of multiple DCBH formation sites within the same overdense region. Second, while synchronized pairs with very small separations provide a favourable channel for DCBH formation, our results highlight the importance of the cumulative LW radiation field produced by multiple sources at larger distances. This collective contribution can be sufficient to exceed the critical LW irradiation on pristine halos, thereby increasing the number of viable DCBH candidates and partially mitigating the suppressive effects of external metal pollution, as discussed also in \citet{lupi2021}. This effect naturally explains the larger number of DCBH host candidates predicted in the \textsc{cat} model.

Based on our predictions, we can also infer an approximate lower limit on the cosmological number density of DCBHs formed in similar environments. Given that our analysis focuses on a highly biased overdensity at $z \approx 6$, it is reasonable to assume that comparable evolutionary pathways may also characterize more massive halos. Adopting a Sheth-Tormen mass function \citep{sheth2001}, such systems are expected to have a number density of $n_{\rm halo}(>3 \times 10^{12} ~\msun) \simeq 7.2 \times 10^{-9} ~\rm{cMpc^{-3}}$. Under these assumptions, the corresponding number density of DCBHs formed in similar or more massive overdensities is estimated to lie in the range $\approx [3 \times 10^{-8} - 4 \times 10^{-7}] ~ \rm cMpc^{-3}$.  

Note that these estimates are approximately three orders of magnitude lower than recent constraints on the number density of MBHs ($\Mbh \gtrsim 10^6 ~\msun$) detected by JWST at $z > 5$, primarily in the form of broad-line AGNs \citep{taylor2025b,Madau2026}. If DCBH formation represents a dominant seeding channel for the high-redshift MBH population, this comparison suggests that the same mechanism must also take place efficiently in significantly less massive and less biased environments. This possibility has been recently proposed by \citet{Baggen2026}, motivated by the large fraction of MBH candidates observed in close proximity to bright star-forming companion galaxies with modest stellar masses of $10^8-10^9 ~\msun$, potentially supporting a `synchronised-pair' formation scenario.

Alternatively, a comparable number of DCBH formation sites, but associated with less massive central halos than those considered in this work, has also been reported in detailed hydrodynamical simulations by \citet{Dunn2018}, where internal radiation fields play a dominant role. In particular, they predict the formation of tens to hundreds of heavy seeds over the evolutionary history of a Milky Way-like halo, whose most massive progenitor reaches $\approx 10^{10} ~\msun$ at $z=5$, depending on the adopted LW threshold in the range $J_{\rm LW, ~crit}=30-10^3 ~J_{21}$. In their model, however, the majority of DCBHs form within pristine gas pockets embedded in the same halo hosting the star-forming regions responsible for the dissociating LW radiation, i.e. within environments that we would classify in CAT as “enriched galaxies”. It should be noted, however, that subsequent higher-resolution simulations, capable of resolving the internal structure of collapsing gas clouds and the formation of protostellar cores, generally do not support a significant role for internal LW radiation from nearby clumps in triggering the formation of supermassive stars \citep{sullivan2025}.

\section{Observability of direct collapse black holes}
\label{sec:HseedObservability}
We have shown that the conditions required for the formation of a significant number of heavy seeds - on the order of several tens - via a direct collapse scenario are naturally met within cosmic overdensities that eventually host the massive quasars observed at redshift $z \sim 6-7$.
This raises the compelling question of whether this clustered population of DCBHs might be observable with current facilities. 
Since we predict the formation window of these DCBHs to occur at much earlier epochs compared to typical high-redshift quasar observations, their subsequent evolution will be significantly affected by the growth history of the main galaxy. In particular, a fraction of these heavy seeds will enter the main galaxy as it undergoes a rapid assembly. Once merged in the system, they may either contribute to the growth of the central supermassive black hole by efficiently migrating toward the nucleus and merging, or wander in the galaxy outskirts and sink toward the centre on much larger timescales. In the latter scenario, the BH will reside in low-density environments, making any electromagnetic signature coming from its accretion very unlikely to be detected. At the same time, the rapid migration of ex-situ massive seeds into the nuclear region, driven by efficient dynamical friction, could lead to the formation of a massive BH binary, potentially giving rise to several detectable electromagnetic (EM) signatures that are expected from a dual AGN system \citep{dotti2023,Haiman2023,bertassi2025,Xin2025}. However, the spectral and variability features expected from dual/binary AGNs are not unique to these systems, and can be challenging to distinguish from intrinsic emission features in single AGN systems \citep{rigamonti2025}.
At later times, the resulting SMBH binary would emit gravitational-wave signals well within the sensitivity range of the next generation of GW observatories, and especially of the future Laser Interferometer Space Antenna (LISA) \citep{valiante2021,LISA2023}, although accurately localizing the host galaxy and robustly determining the redshift of such binary systems will remain challenging.

The remaining fraction of the DCBH population formed at $z>10$ might instead survive outside of the main galaxy down to the cosmic epoch where massive BHs are typically observed, residing within less massive and less evolved satellite halos. In the following, we investigate this population by testing the possibility of observing such systems around bright quasars detected at $z \simeq 7.5$, and characterizing their spectral features. 

\begin{figure*}
\includegraphics[width=\textwidth]{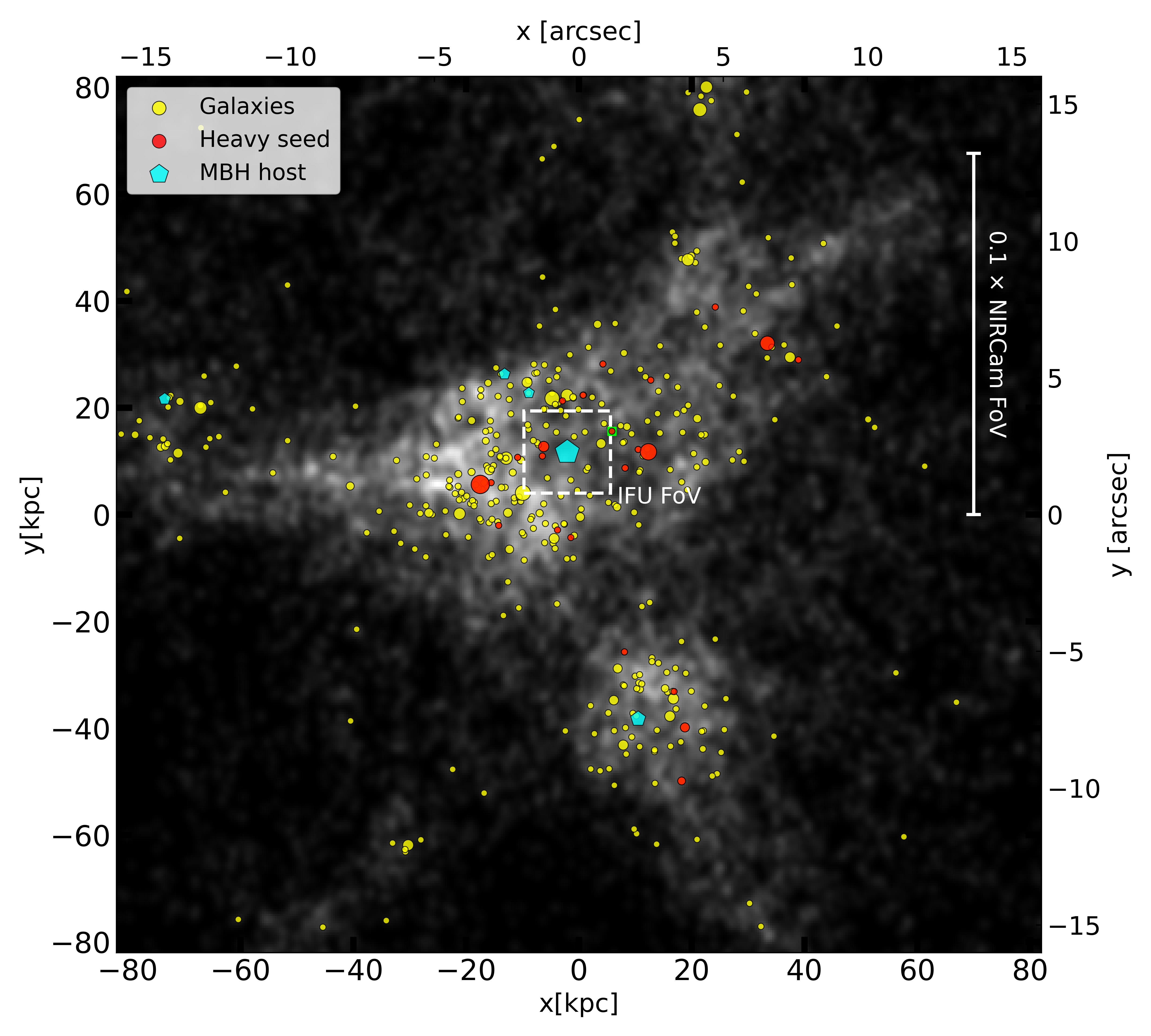}
\caption{Projection of the simulated overdensity at $z \approx 7.5$, with distance from the centre of the structure marked in physical kpc and arcseconds. Star forming galaxies are marked with yellow points, while halos hosting heavy seed descendants or SMBHs ($\log(\Mbh/M_\odot) > 6.5$) are highlighted as red circles and cyan pentagons, respectively. For illustrative purposes, the size of each point is scaled according to the stellar mass of the corresponding galaxy. The white smooth background points represent the underlying distribution of dark matter halos, most of which do not meet the conditions for triggering star formation. The green square highlights the system \textsc{ID48871}, for which we analysed the potential AGN detectability in the following sections. For reference, we overplot as a white box the $3'' \times 3''$ field of view of IFU observations centred on the central quasar, and we also report the scale corresponding to $0.1 \times$ the NIRCam FoV, to highlight how the entire simulated overdensity would be encompassed within the $2' \times 2'$ coverage of a single NIRCam module.}
\label{fig:MapBHz7}
\end{figure*}

\subsection{SED of massive seeds around high-z quasars}

In Figure~\ref{fig:MapBHz7}, we show a projection of the simulated overdensity at redshift $z \approx 7.5$. The plot shows the underlying distribution of dark matter halos within $\sim 80 ~\rm kpc$ (physical) of the overdensity centre, with star forming galaxies highlighted in yellow. Halos hosting heavy BH seeds that have grown only modestly from their birth mass of $10^5 \ \msun$ are shown as red points, while halos containing SMBHs that have grown above the threshold of $\rm \log(M_{\rm BH}/M_\odot) = 6.5$ are marked with cyan pentagons. As shown, $22$ heavy BH seeds out of $54$ survived outside the main halo down to redshift $z \approx 7.5$, while only $5$ of them were able to grow in mass efficiently above the assumed threshold.
A natural question that arises is the observability of this population of poorly grown heavy seeds clustering around the main quasar host.
To investigate this aspect, we used the  \textsc{Synthesizer} tool \citep{Roper2025} to generate synthetic spectra for our galaxy population. In particular, we focus on retrieving the expected spectral energy distribution (SED) of heavy seed-host galaxies surrounding the bright quasar at $z \sim 7$, when the central SMBH has grown to $\rm M_{SMBH} \, \simeq 10^{8.9} ~\rm \msun$. The aim is to determine whether these faint systems might be detectable with current observational facilities and whether they show characteristic spectral signatures that can be uniquely related to the presence of an accreting BH. 

We first examined the BH mass distribution of all heavy-seed descendants that appear in Figure \ref{fig:MapBHz7}. While a few seeds undergo substantial growth between $z \approx 7-8$, we find that within this redshift interval - largely accessible with forthcoming JWST surveys - the vast majority of the population ($>50 \%$) remains relatively low-mass, with $\log(\Mbh/\msun) \lesssim 5.5$. 
At the same time, their host stellar masses span several orders of magnitude, ranging from $\approx10^3 - 10^9 ~\msun$. We refer the interested reader to Appendix~\ref{sec:AppendixB} for the full BH mass distribution of the sample, its evolution across the redshift range considered in this study, as well as the corresponding stellar mass distribution of their host galaxies.
Examining the evolution of this satellite MBHs, we find that they are characterized by a strong variability in their AGN accretion rate, which will directly affect their observability.
Motivated by this, we selected one representative system within the sample, characterized by typical BH and host galaxy stellar masses, to illustrate its expected spectral features and investigate the feasibility of detecting a similar source. This therefore represents a conservative estimate, since more massive (though rarer) companion BHs hosted in brighter galaxies would naturally provide more optimistic observational chances. A statistical analysis of detectability for the entire SMBH satellite population will be presented in Section \ref{sec:population_detectability}.

\begin{figure*}
\centering
\includegraphics[width=0.8\linewidth]{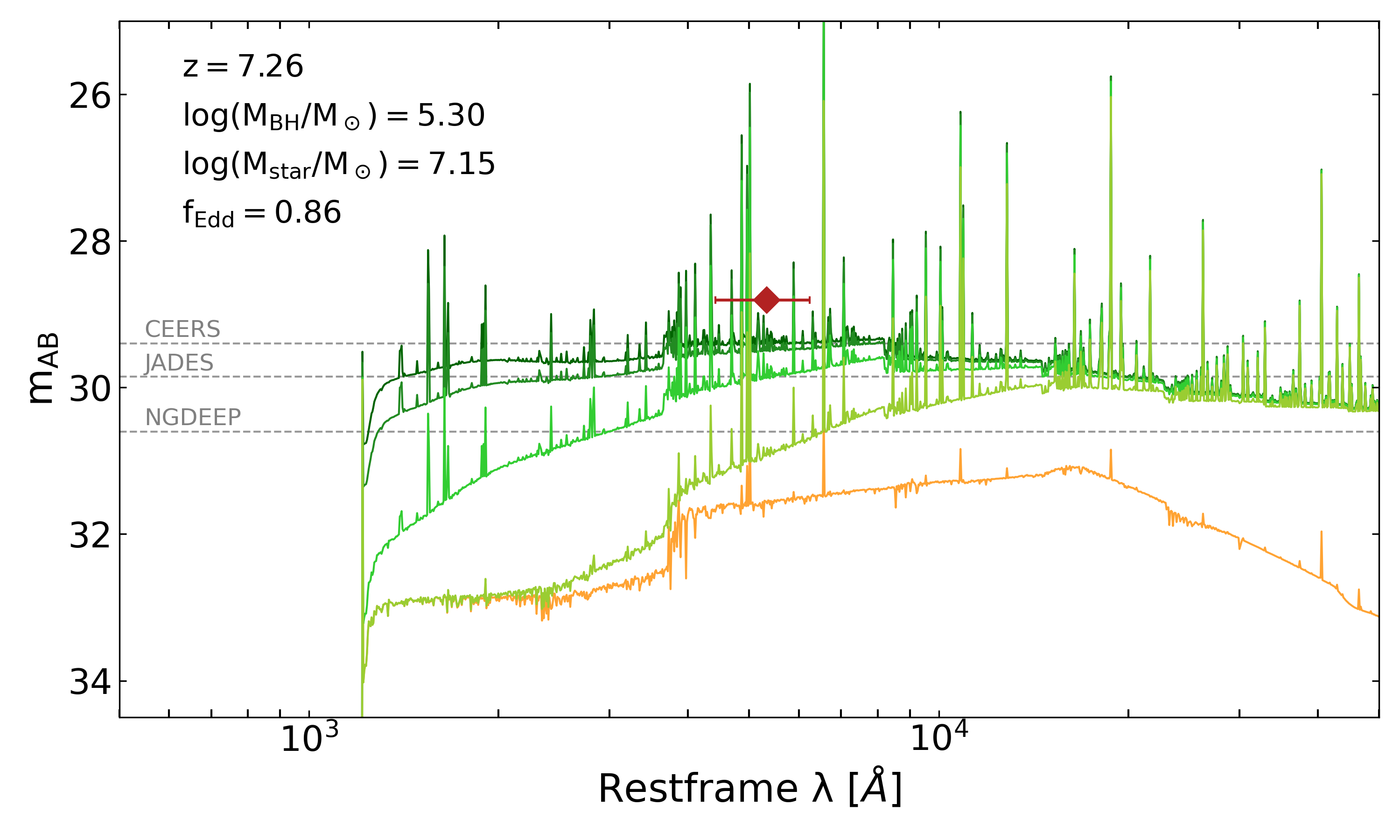}
\includegraphics[width=0.8\linewidth]{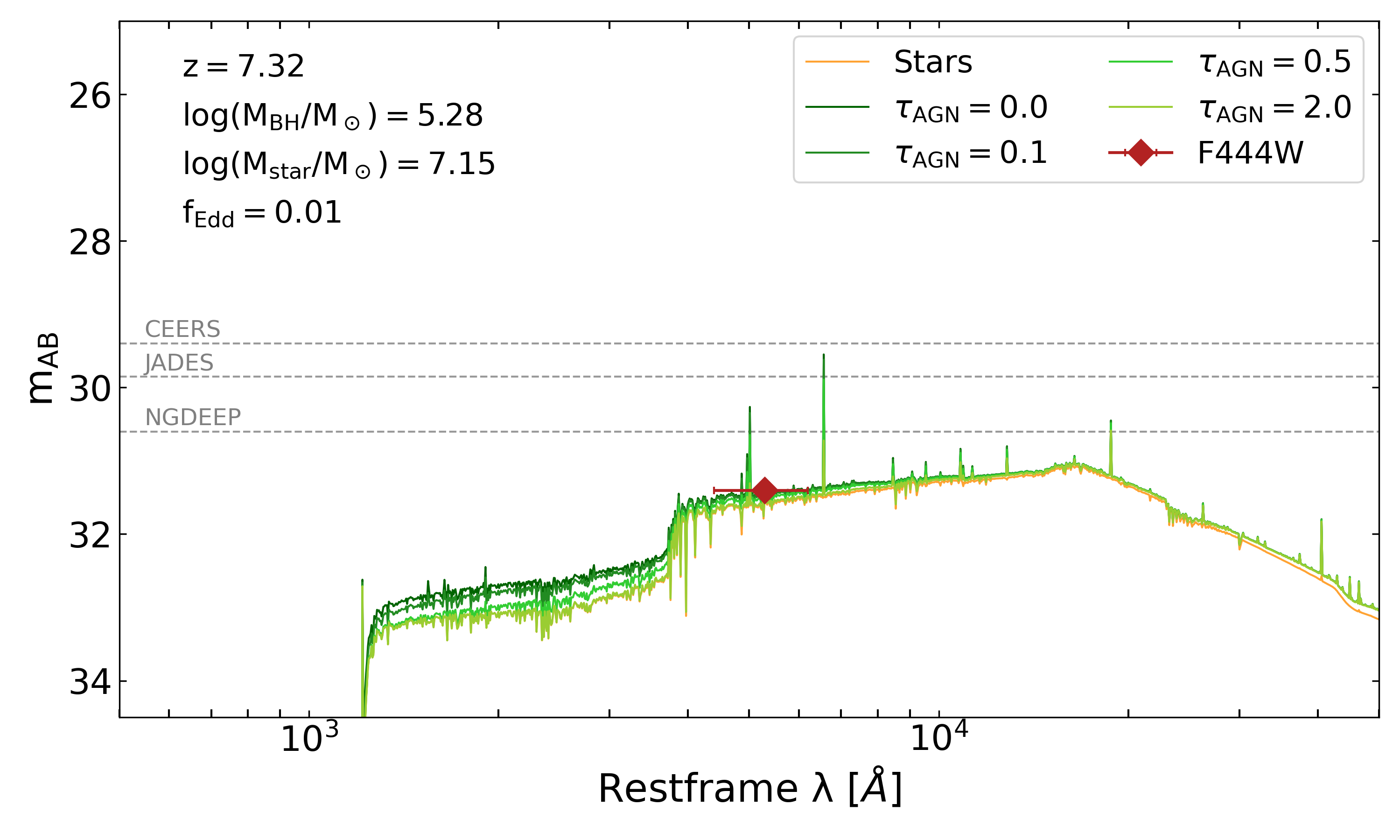}
\caption{Spectrum of the quasar satellite galaxy $\rm ID48871$, highlighted as a green square in Fig.~\ref{fig:MapBHz7}. This system hosts a representative heavy BH-seed descendant at $z \approx 7-8$, which exhibits strong AGN variability.
The spectrum is shown at two different snapshots around $z \sim 7$ 
(in an active phase in the top panel and in a dormant phase in the bottom panel), separated by $\sim 8 ~\rm Myr$. The stellar contribution to the SED is shown by the solid orange line, while the total galaxy emission is marked by the green lines, exploring different values of V-band optical depth $\tau_{\rm V, \,  AGN}$ for the AGN attenuation. The red data point displays the predicted system photometry in the F444W JWST filter assuming no AGN attenuation. We mark, for comparison, the limiting magnitudes at which different deep JWST surveys reach a completeness $ >50\% $, shown as dashed gray lines.}
\label{fig:GalaxySpectrum}
\end{figure*}

In Figure \ref{fig:GalaxySpectrum} we report the spectrum of such system, \textsc{ID48871}, highlighted as a green square in Figure \ref{fig:MapBHz7}. At $z \sim 7.5$, this galaxy is located at $ \sim 5 ~\rm$ projected physical kpc from the central quasar (corresponding to $ \sim 1.5''a$), with a stellar mass of $\log(\Mstar/\msun) \approx 7.16$ and a BH mass of $\log(\Mbh/\msun) = 5.3$. Figure \ref{fig:GalaxySpectrum} displays the galaxy's synthetic spectrum at two different snapshots, $z = 7.32$ and $z=7.26$, separated by a temporal interval of $\approx 8 ~\rm Myr$, highlighting the substantial changes in the emission due to AGN variability.
At $z = 7.26$ (upper panel) we clearly see that the accreting BH dominates the intrinsic galaxy emission compared to the stellar component. To test the potential impact of gas attenuation, often invoked to explain the properties of several faint AGN candidates recently identified at high-z \citep{Naidu2025,degraaff2025}, we overplot the total emission assuming increasing levels of AGN attenuation. This is sampled using different values of optical depth in the rest-frame V band, ranging between $\tau_{\rm V, \,  AGN} = 0 - 2$, and assuming an SMC attenuation law. The resulting spectra are shown for $\tau_{\rm V, \,  AGN} = [0,0.1,0.5,2]$, which, given the galaxy metallicity of $\log(Z/Z_\odot)  \simeq -1.3$, correspond\footnote{We assume here a dust-to-metal ratio of 0.3, representative of early, low-mass galaxies.}
to gas column densities of $N_{\rm H}\approx [0,0.39,1.9,7.7] \times 10^{23} \, \rm cm^{-2}$, respectively. The AGN emission remains dominant even under significant attenuation, in which case it contributes as a steeply rising red optical continuum. This behaviour is consistent with the spectral features observed in a significant fraction of high-redshift AGN candidates detected by JWST, commonly referred to as \textit{``little red dots"} \citep[LRDs,][]{Matthee2024,Greene2024,kokorev2025}. However, the nature of such strong attenuation remains debated, either due to moderate but compact ($\lesssim10^{5-6} \, \msun$) dust reservoirs \citep{Akins2025,Casey2025} or to the presence of extremely dense, metal-poor gas in the nuclear region enshrouding the central BH \citep{inayoshi2024,Ji2025}.
During this phase of efficient accretion, characterized by an Eddington ratio $f_{\rm Edd} = 0.86$, the system is predicted to reach observed magnitudes in JWST rest-frame optical bands (e.g., F444W) in the range $m_{\rm F444W} \sim 29-31$ \footnote{All magnitudes presented in this work are in the AB magnitude system.}, depending on the level of attenuation. These values place the source close or just below the detection threshold for both current wide-field JWST photometric surveys - such as CEERS, JADES and PRIMER - and deep spectroscopic surveys like NGDEEP \citep[see for instance the large compilation of LRDs presented by][]{Kocevski2024}.
For comparison, Fig. \ref{fig:GalaxySpectrum} shows the limiting magnitudes corresponding to a detection completeness greater than $> 50 \%$ for several of these wide-field and deep JWST surveys \citep{Merlin2024}.

However, less than $10 ~\rm Myr$ before this active phase of BH growth, at $z=7.32$ (lower panel in Figure \ref{fig:GalaxySpectrum}), the AGN was largely quiescent, with an Eddington ratio of $f_{\rm Edd} = 0.01$. At this time, even for moderate AGN attenuation, the relatively faint stellar emission dominates the galaxy SED. The system would show a rest-frame optical magnitude of $m_{\rm F444W} > 31$, placing it well below the typical detection threshold of JWST, unless targeted with extremely long exposures or in the presence of gravitational lensing \citep[e.g.][]{maiolino2025}. Therefore, this representative heavy-seed host galaxy is expected to remain completely undetected during its `dormant' AGN phase in current deep wide-field surveys. More massive BH companions, hosted in more luminous satellite galaxies, may instead lie above the detection threshold even during periods of inefficient BH accretion. In such cases, however, it would become considerably more challenging to identify unambiguous signatures of AGN activity in such faint, but detectable, systems, as discussed in the following sections.

\subsection{Detectability and observational strategies}
\label{sec:population_detectability}

Given the high variability of BH accretion we predict in the population of galaxies hosting heavy seed descendants around a massive quasar, it is crucial to explore the probability of detecting at least one system in an active phase within a realistic observable redshift window. For this reason, we focused again on the range $7 < z < 8$, and evaluated the average number of accreting BHs that would be detectable for different sensitivity limits in the JWST F444W filter band, tracing their optical rest-frame emission. Since our goal is to assess the potential observability of sources that can be identified as descendants of DCBHs, we focus our analysis on systems with $\log(\Mbh/\msun) < 6.5$ (marked in red in Fig. \ref{fig:MapBHz7}, see also additional details in Appendix \ref{sec:AppendixB}).

The results are shown in Figure \ref{fig:AvgObservedSyst}, under two different assumptions: a) relying only on the AGN emission to evaluate the system detectability; b) including the total AGN + stellar emission, but counting only systems where the intrinsic UV emission is dominated by the AGN. 
The first case provides a conservative lower limit on the number of detectable galaxies hosting a heavy seed, while the second provides an estimate of how many systems could be unambiguously identified as AGNs. We find that the observational estimates differ significantly between these two scenarios.
If we consider only the intrinsic AGN emission, there is a substantial probability of detecting at least one system hosting an accreting DCBH in the vicinity of the bright quasar. The average number of detectable systems ranges from $\rm \bar{N}_{det}\sim 3$ for a threshold of $m_{\rm F444W} < 29$, up to $\rm \bar{N}_{det}\sim 5$ for $m_{\rm F444W} < 31$. Assuming detections are independent and follow Poissonian statistics, this translates in a probability of detecting at least one system, $P_{\rm det}(\geq1) = 1-e^{\rm -\bar{N}_{det}}$, of $95 \%$ ($\rm \bar{N}_{det}=3$) and $99 \%$ ($\rm \bar{N}_{det}=5$), respectively. 
However, when limiting our selection to systems with an AGN-dominated UV budget, these estimates drop noticeably. In this case, a detection threshold of $m_{\rm F444W} < 29$ corresponds to $\rm \bar{N}_{det} < 1$, implying a reduced probability of observing any active satellite DCBH ($P_{\rm det}\approx 60 \%$). Increasing the limiting magnitude to $m_{\rm F444W} < 30$ raises the average number to $\rm \bar{N}_{det}\sim 1.4$, while pushing it to $m_{\rm F444W} < 31$ yields $\rm \bar{N}_{det} \sim 1.8$ ($P_{\rm det}\approx 83 \%$).

Interestingly, the predicted average number of detectable systems is higher when considering only the AGN emission than when also including the host galaxy contribution, but counting only AGN-dominated systems. This reflects the fact that, although many DCBH hosts present detectable AGN component, their total emission is often dominated by stellar light. In particular, less massive DCBH descendants, hosted in smaller galaxies (see Appendix \ref{sec:AppendixB}), are likely to show clear AGN signatures during active accretion phases (as shown in Fig. \ref{fig:GalaxySpectrum}), but these episodes will be rarer due to the limited gas availability. Conversely, DCBHs descendants hosted in more massive galaxies will experience more steady accretion flows, allowing them to remain above the detection threshold for longer periods of time, despite their total UV emission being dominated by the host galaxy, which makes their characterization more challenging. As a result, photometric observations alone may struggle to confirm the presence of an AGN without additional spectral diagnostics, which could instead be provided by dedicated spectroscopic surveys, as discussed in the following.

\section{Discussion}
\label{sec:discussion}
The unprecedented capabilities of JWST have already enabled major breakthroughs in the characterization of high-redshift quasar environments and their satellites. Recent studies have focused on detecting quasar hosts and constraining galaxy properties \citep{Marshall2023,Stone2024}, as well as on the remarkable discovery of quasar satellite galaxies caught in the process of merging \citep{loiacono2024,Decarli2024}.
The detection of SMBH companions around luminous quasars at high redshift, however, remains tentative. 
In the less massive AGN regime, probed by extensive JWST wide field surveys, \citet{Maiolino2024bhs} reported indications of dual AGN candidates in up to $25\%$ of the analysed high-redshift sample. These detections, however, rely on resolving multiple broad Gaussian components in the fit to the $H\alpha$ emission line, which could alternatively be explained by different physical processes, such as small scale outflows/inflows or non-virialised BLRs. Instead, a spatially resolved candidate dual system has been identified through deep NIRSpec-Integral Field Units (IFU) spectroscopy by \citet{ubler2024}, where the off-centre SMBH lies at a physical distance of $\sim 680~\rm pc$. More recent is instead the discovery of a potential triple system at $z \sim 5$, where a central $10^8 ~\msun$ BH is surrounded by two accreting companions with $\log(\Mbh/M_\odot) \sim 6$ at projected separations of $\sim 200$ pc and $1.7$ kpc, respectively \citep{Ubler2025}.
These detections are promising and suggest that SMBH companions could be relatively common in the early Universe, even in less biased cosmological overdensities. 
However, identifying potential heavy-seed descendants in the vicinity of high-redshift quasars may require tailored observational strategies, as our simulations indicate that, even during phases of efficient accretion, their expected luminosities remain modest. This could help explain the observational challenges in detecting such systems, even in the deepest JWST datasets, and why larger, statistically significant samples of confirmed companions are still lacking.

In this work, we do not focus on the role of the few MBH satellites that undergo efficient growth down to redshift $z \approx 7.5$, two of which reach masses above $10^8 ~\msun $, as shown in Appendix~\ref{sec:AppendixB}, since they highly depend on the presence of massive satellite halos in our single merger-tree realization, one of which eventually undergoes a major merger with the quasar host by $z \sim 5$.
Nevertheless, the presence of a small number of massive DCBH descendants is an intriguing outcome, suggesting that environments hosting multiple nearby quasars could be particularly promising targets for deeper observations aimed at testing this formation scenario. A notable example is provided by the merging twin quasars reported by \citet{Matsuoka2024} at $z \approx 6$, consisting of two $\sim 10^8 ~\msun $ quasars separated by a projected distance of $\sim 12 ~\rm pkpc$, not too different from what is found in our simulated overdensity (see Fig.~\ref{fig:MapBHz7}).

\begin{figure}
\centering
\includegraphics[width=1.0\linewidth]{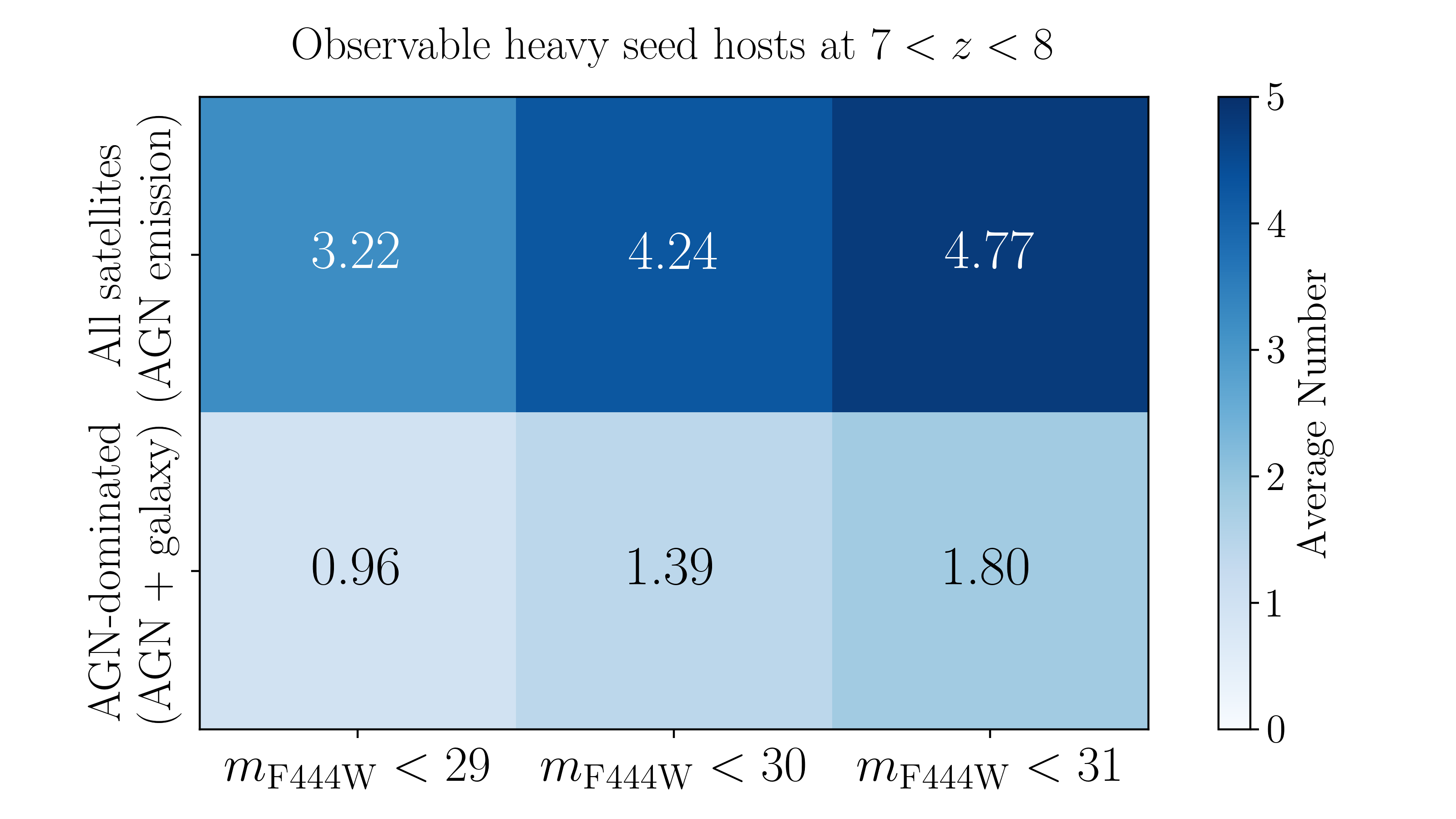}
\caption{Average number of detectable galaxies hosting heavy BH seeds descendants ($5 < \log(\Mbh) < 6.5$) in the redshift range $7 < z < 8 $, assuming different magnitude threshold limits $m_{\rm F444W} < 29, 30, 31$. The galaxy detectability has been estimated: a) considering only the - intrinsic - emission from the AGN (top row), and b) considering both AGN and stellar emission, but only counting systems that are AGN dominated in the rest-frame UV continuum (bottom row). All the systems within the simulated overdensity, shown in Fig. \ref{fig:MapBHz7}, are considered here, independently of their distance from the central quasar.}
\label{fig:AvgObservedSyst}
\end{figure}

It is important to note that the average number of observable active massive seeds shown in Fig. \ref{fig:AvgObservedSyst} is derived from the evolutionary history of a single simulated quasar. Dedicated surveys targeting multiple quasars at similar redshifts, and focusing on sources as luminous as those hosting $\sim10^{9} \ \msun$ SMBHs, are therefore expected to significantly increase the likelihood of detecting at least one of such companions. This will be particularly crucial for detecting, among their satellites, rarer systems that can be unambiguously identified as accreting AGNs, and that may also offer deeper insights into the original seed formation environment. Such campaigns could therefore place much more stringent constraints on the existence and properties of this potential population of heavy-seed descendants.
NIRCam imaging across multiple broadband filters offers a powerful way to explore this scenario. Systems similar to the simulated ones, located around bright quasars at $z \sim 7$ and reaching luminosities of $m_{\rm AB} \approx 29$, could be detected with $S/N > 3$ in the F444W filter with a total integration time of roughly one hour.\footnote{estimated through the JWST Exposure Time Calculator (https://jwst.etc.stsci.edu/) assuming DEEP8 NIRCam readout mode} Such observations would enable the search for luminous companions on scales of tens of kiloparsecs, thereby increasing the likelihood of identifying systems caught in an active accretion phase. Naturally, complementary observations in multiple filters would be required to mitigate contamination from lower-redshift interlopers and improve the reliability of candidate selection.
Photometry alone, though, is unlikely to reveal whether these satellite galaxies host accreting massive black holes. For the most promising cases, where multiple candidate companions are found, follow-up observations with deep IFU spectroscopy become essential to the identification of unambiguous AGN signatures, particularly the detection of broad Balmer emission lines. IFU data would also provide kinematic information, enabling a dynamical characterization of how these AGN companions interact with the primary halo.
Even under relatively efficient accretion, confirming the presence of heavy-seed descendants requires substantial integration time to properly resolve strong emission lines. Nevertheless, the H$\alpha$ flux predicted for promising simulated systems such as \textsc{ID48871} ($f_{\rm H\alpha} \sim 10^{-18}~\rm erg / s / cm^2 /\AA$, see Fig.~\ref{fig:GalaxySpectrum}) should be sufficient to detect the broad component within $1$--$2~\rm h$ of integration, as recently shown in \citet{Ubler2025}.

The complementary use of photometry and spectroscopy will also be required in order to survey a larger region around the central quasar. As shown in Fig.~\ref{fig:MapBHz7}, we predict satellite systems hosting heavy-seed descendants out to $\sim 70$ physical kpc from the central source at $z \approx 7$. These systems would fall fully within the NIRCam field of view (FoV), whereas only a minority ($\sim 3/22$, averaging over different projections) would be enclosed within the $3'' \times 3''$ area probed by IFU observations centred on the bright quasar (which we marked for reference as a white box in Figure \ref{fig:MapBHz7}). On the other hand, if the satellite BHs are already in an advanced phase of merging with the quasar host galaxy, their emission may be indistinguishable from that of the central quasar, being embedded within its PSF ( $\lesssim 1~\rm kpc$ at $z=7$ and $\lambda \sim 4.4 \rm ~\mu m$ both for NIRCam and NIRSpec IFU). Nevertheless, even in this case, the companion emission could still be inferred from asymmetries in the H$\alpha$ profile through careful decomposition of the broad-line components, provided the signal-to-noise ratio is sufficient \citep{Maiolino2024bhs}. 

An alternative observational strategy is to exploit the capability of the NIRCam Wide Field Slitless Spectroscopy (WFSS) mode, which provides simultaneous spectroscopic coverage of all sources within the field of view. This approach has been successfully employed in the EIGER \citep{kashino2023} and ASPIRE \citep{Wang2023,Champagne2025} programs, which surveyed more than 30 quasar fields at $5 \lesssim z \lesssim 7$ and identified tens of [O III] emitters in overdense environments. These campaigns have offered new insights into the role of environment in early galaxy evolution and provided the most robust clustering-based estimates of the DM halo masses of typical $z>6$ quasars to date \citep{Eilers2024}. In this context, NIRCam WFSS could represent an efficient tool to search for heavy-seed descendants around $z \sim 7$ quasars, as a single pointing can fully cover the entire simulated overdensity and identify strong rest-optical emission lines potentially associated with AGN activity in satellite systems.

Systematic searches for broad $\rm H\alpha$ emitters using WFSS have also been performed relying on the EIGER, FRESCO, and ASPIRE surveys \citep{Matthee2024, lin2024}. These efforts have proven very effective in detecting AGNs and LRDs at $z \sim 4-6$, reaching luminosities as faint as $\rm m_{F356W} \gtrsim 25.5$, and have also identified candidates in overdense regions with multiple $\rm H\alpha$ emitting companions. It is important to note, however, that even with deep, targeted observations of a limited set of promising quasar fields, the sensitivity achievable with WFSS might be significantly lower than that obtained through NIRCam imaging combined with targeted NIRSpec IFU follow-up. As a result, detections will likely be restricted to the brightest companion AGNs, either the ones hosting the most massive BHs or those undergoing transient phases of enhanced accretion.
In this context, a valuable aid could be provided by observations of strongly lensed quasar fields \citep[see e.g.][]{fan2019}, as magnification and/or multiple images have proven to enable the characterization of faint sources at flux levels otherwise inaccessible in blank fields at these high redshifts, reaching magnitudes far below the typical detection limits (see, e.g., \citealt{furtak2024, Juodzbalis2025, Yanagisawa2026}).

A final caveat to consider is that, in our modelling, we assumed an initial seed mass for DCBHs of $M_{\rm seed} = 10^5 ~\msun$. Recent detailed hydrodynamical simulations suggest that, even in the presence of strong LW fluxes, $\rm H_2$ formation might still occur in the inner regions of the collapsing cloud, leading to fragmentation and a maximum BH mass of $\sim 10^4 ~\msun$ \citep{Prole2024b}. 
In this scenario, the detection of heavy-seed descendants could become more challenging, as their luminosities during the early growth phase would be significantly lower. Nevertheless, the possible formation of supermassive stars with masses of $\sim 10^4~\msun$ even in more enriched environments, as recently suggested by \citet{chon2025}, could partially compensate for the lower seed masses by increasing the overall number of favourable formation sites accessible to future surveys. Alternatively, even if forming with lower initial mass, DCBHs may still undergo episodes of highly efficient, possibly super-Eddington, accretion, enhancing their detectability.

On this point, it is worth stressing that such episodes of accretion may not be uncommon among the investigated population. In this work, we have assumed that BH accretion is always capped at the Eddington limit. Over the evolution of the heavy-seed descendants examined between redshifts $z = 8$ and $7$, we find that they accrete at the Eddington limit for approximately $\sim 60 \%$ of the time spent in an active accretion phase, defined as $\rm \dot{M}/\dot{M}_{\rm Edd} > 0.1$. Although a detailed characterization of potential super-critical accretion episodes, and of their impact on the host galaxy, is beyond the scope of this work and would require dedicated spectral modelling, these findings nonetheless suggest that the estimated number of detectable systems could increase when accounting for such accretion phases.

\subsection{Toward empirical constraints of massive seed formation}

The predictions presented in this work offer a quantitative framework against which future targeted observational campaigns can be interpreted, providing important insight into the physical conditions that could enable massive seed formation in the early Universe.

The detection of a significantly larger number of satellite galaxies hosting candidate DCBHs than predicted by our reference models would suggest that the conditions enabling direct collapse are more permissive than currently assumed. This could imply that efficient supermassive-star formation and collapse may occur even in moderately metal-enriched environments, or that additional mechanisms suppressing fragmentation during the collapse phase, such as turbulence, dynamical heating driven by gas inflows, or the effect of internal radiative feedback within the host galaxy, play a more important role than typically considered.
A larger number of detectable companions could also alternatively reflect faster growth and/or prolonged active phases of the massive seeds expected to form in these regions. However, widespread sustained accretion across the majority of the satellite population may be difficult to achieve, given the relatively low gas masses and modest stellar components characterizing these systems.

In this context, rapid growth of lighter BH seeds could still represent a competitive scenario and potentially contaminate the interpretation of these observations. Dedicated spectroscopic observations targeting the properties of the companion AGNs and their host galaxies, including constraints on BH-to-stellar mass ratios, line widths, and metal enrichment, could provide partial leverage in disentangling heavy-seed populations from rapidly evolved lighter seeds, though this remains a non-trivial challenge even with high-quality data. Nevertheless, the simultaneous super-Eddington growth of multiple low- to intermediate-mass BHs in nearby environments is unlikely, as there is no particular reason to expect it to be favoured in large cosmological overdensities, and therefore to cluster preferentially around the most massive high-redshift quasars. In this regard, increased statistical samples may ultimately represent the most powerful discriminant.

Conversely, even the lack of MBH companion detections in dedicated surveys would provide valuable, and potentially even tighter, constraints on the mechanisms of early BH seeding. In fact, an underabundance relative to our predictions would suggest either that the environmental conditions for DCBH formation are more restrictive than currently assumed, or that their initial masses, even in the most favourable sites, remain well below $\sim 10^5 ~\msun$. This would also suggest that alternative pathways, such as rapid accretion onto smaller seeds or merger-driven formation triggered by extreme gas inflows, could play a more dominant role in the building up of the first massive quasars.

Finally, extending this analysis to more average cosmological environments would provide an important complementary benchmark to the predictions derived here for extreme quasar overdensities. Exploring less massive and more representative regions of the early Universe, aiming at larger observational statistics, would help determine whether efficient massive seed formation through the DCBH channel is restricted to the most extreme environments, or may also occur under more common conditions, potentially implying higher MBH number densities at early times.
Such investigations would be particularly relevant to assess a possible connection between DCBHs and the recently identified population of high-redshift MBH candidates, especially LRDs \citep[see e.g.][]{pacucci2026}, for a significant fraction of which bright UV companion sources has recently been reported, producing the required LW intensity  \citep{Baggen2026}. We defer this analysis to future work.

\section{Conclusions}
In this work, following up on \cite{lupi2021}, we presented a detailed investigation of the role of high-redshift quasar environments as potential hosts of early DCBHs, and provided theoretical predictions for the detectability of their descendants with current and future JWST surveys. Focusing on the assembly history of a massive dark matter halo with a final mass of $\sim 10^{12} ~\msun$ at $z \sim 6$, representative of bright quasar hosts, we combined high-resolution dark matter merger trees extracted from a cosmological N-body simulation to the \textsc{Cosmic Archaeology Tool} semi-analytic model to track the early galaxy evolution and BH seed formation. This allowed us to trace the abundance, spatial distribution, and environmental conditions of halos capable of forming DCBHs over cosmic time, exploring how these trends depend on the specific physical requirements for heavy seed formation.
The analysis revealed that:
\begin{itemize}

    \item When the impact of local LW emitters is spatially resolved, a large number of seeds can be formed ($> 50$), starting at very early epochs (up to $z \sim 22$), assuming a critical Lyman-Werner flux threshold of $J_{\rm LW, crit} = 300 \, J_{21}$ for triggering DCBH formation. Even imposing a stricter threshold of $J_{\rm LW, crit} = 1000 \, J_{21}$, massive seeds continue to form in a large number ($> 20$) and at early epochs, since they are formed thanks to nearby irradiating systems.\\
    
    \item Although efficient external pollution from nearby star forming galaxies may inhibit the formation of heavy seeds, we find that a large fraction of DCBHs formed in the simulation at $z > 15$ originate in isolated regions, with $70\%$ ($24\%$) of them forming more than 2 kpc (7 kpc) away from the nearest enriched system.\\

    \item A significant fraction ($22/54 \sim 41 \%$) of the predicted heavy seed population survives outside of the main galaxy down to $z \approx 7$, on scales between $\sim 5 -70$ kpc from the central bright quasar. Only a few systems are able to grow efficiently above $\log(\Mbh/\msun) = 6.5$, while most of the population remains in a mass range close to the initial seed mass.
\end{itemize}

Building on this, we focused on predicting the observational features of the population of DCBH descendants expected to reside near high-redshift quasars at $z \sim 7$. We coupled the \textsc{cat} predictions with the \textsc{Synthesizer} spectral synthesis tool to retrieve the galaxy and AGN emission of the candidate population, and test their detectability within current or future JWST observations.

We find that:

\begin{itemize}
    \item Despite the large number of predicted quasar companion systems, the AGN emission from galaxies hosting heavy seed descendants is highly variable, with accretion rates evolving from $f_{\rm Edd} \sim 1$ down to $\sim 0.01$ on timescales of a few Myr. Nevertheless, during active phases, individual companions may be clearly visible even in current wide-field JWST surveys, reaching magnitudes brighter than $m_{F444W} < 29$.\\

    \item Performing a statistical study of the observability of this population, we find that the average number of detectable systems with clearly AGN-dominated emission around a single quasar in the redshift range $7 < z < 8$ increases from $<1$ at a magnitude limit of $m_{F444W} = 29$ to $\sim 1.8$ for deeper observations reaching $m_{F444W} = 31$. If we instead consider all satellites hosting a DCBH descendant in an active phase, we predict an average of $>3$ detectable systems already for $m_{F444W} < 29$, where although disentangling the nuclear emission from that of the host galaxy may be more challenging.

\end{itemize}

These results demonstrate that forthcoming JWST observations could already be used to test the heavy seed formation scenario and to search for signatures of the extreme conditions required for direct collapse. However, a targeted observational strategy would be required to robustly identify such systems.
In particular, a synergistic approach combining JWST photometry and spectroscopy will be crucial. NIRCam imaging, with its large field of view, can be used to probe promising candidates located up to several tens of kiloparsecs from the central bright quasar. Subsequently, deep NIRSpec IFU follow-up observations focusing on the most promising regions would be essential to identify clear spectral signatures confirming the AGN nature of these companions and to characterize the properties and dynamics of their central BH through the detection and spatially resolved analysis of bright, broad Balmer emission lines.

Alternatively, given the substantial number of detectable DCBH hosts predicted even at moderately deep sensitivity limits ($m_{\rm F444W} \sim 29-30$), a multi-target observational strategy may also prove effective, reducing the need for extremely deep exposures and prohibitively long integration times.
This could be achieved by exploiting the capabilities of NIRCam WFSS, which would allow the spectroscopic characterization of all satellite galaxies within several hundred physical kiloparsecs of the central bright source in a single observation for multiple target quasars. In this case, however, robust detections would likely be limited to the brighter satellites, corresponding to the more massive or highly accreting SMBH companions. \newline 

Our results provide a theoretical framework to test the conditions that favour heavy seed formation and to assess the role of direct collapse as a pathway to the SMBH population observed at very high redshift. The possible identification of a large population of quasar-companion AGN candidates in future surveys would represent a strong indication that early massive BH formation preferentially occurs in highly clustered, overdense environments, and the number of detectable systems predicted here will serve as a concrete benchmark for upcoming observational campaigns.

\section*{Acknowledgements}
AT, AL and RV acknowledge support by the PRIN MUR “2022935STW" funded by European Union-Next Generation EU, Missione 4 Componente 2, CUP C53D23000950006. AT acknowledges financial support from the Bando Ricerca Fondamentale INAF 2023, Mini-grant ``Cosmic Archaeology with the first black hole seeds" (Ob.Fu. RSN1 1.05.23.04.01). RV acknowledge support from Bando Ricerca Fondamentale INAF 2023, Theory Grant "Theoretical models for Black Holes Archaeology". ZH acknowledges support by NASA grants
80NSSC22K0822 and 80NSSC24K0440.

%%%%%%%%%%%%%%%%%%%%%%%%%%%%%%%%%%%%%%%%%%%%%%%%%%
\section*{Data Availability}
The simulated data underlying this article will be shared on reasonable request to the corresponding author.

\vspace{5mm}
%\facilities{HST(STIS), Swift(XRT and UVOT), AAVSO, CTIO:1.3m,CTIO:1.5m,CXO}

%%%%%%%%%%%%%%%%%%%% REFERENCES %%%%%%%%%%%%%%%%%%

\bibliographystyle{mnras}
\bibliography{DCBH_environment} 

@ARTICLE{kovacs2024,
       author = {{Kov{\'a}cs}, Orsolya E. and {Bogd{\'a}n}, {\'A}kos and {Natarajan}, Priyamvada and {Werner}, Norbert and {Azadi}, Mojegan and {Volonteri}, Marta and {Tremblay}, Grant R. and {Chadayammuri}, Urmila and {Forman}, William R. and {Jones}, Christine and {Kraft}, Ralph P.},
        title = "{A Candidate Supermassive Black Hole in a Gravitationally Lensed Galaxy at Z {\ensuremath{\approx}} 10}",
      journal = {\apjl},
         year = 2024,
        month = apr,
       volume = {965},
       number = {2},
          eid = {L21},
        pages = {L21},
          doi = {10.3847/2041-8213/ad391f},
archivePrefix = {arXiv},
       eprint = {2403.14745},
 primaryClass = {astro-ph.GA},
       adsurl = {https://ui.adsabs.harvard.edu/abs/2024ApJ...965L..21K}
}

@ARTICLE{habouzit2016a,
       author = {{Habouzit}, M{\'e}lanie and {Volonteri}, Marta and {Latif}, Muhammad and {Nishimichi}, Takahiro and {Peirani}, S{\'e}bastien and {Dubois}, Yohan and {Mamon}, Gary A. and {Silk}, Joseph and {Chevallard}, Jacopo},
        title = "{Black hole formation and growth with non-Gaussian primordial density perturbations}",
      journal = {\mnras},
         year = 2016,
        month = feb,
       volume = {456},
       number = {2},
        pages = {1901-1912},
          doi = {10.1093/mnras/stv2740},
archivePrefix = {arXiv},
       eprint = {1507.05971},
 primaryClass = {astro-ph.GA},
       adsurl = {https://ui.adsabs.harvard.edu/abs/2016MNRAS.456.1901H}
}

@ARTICLE{Decarli2024,
       author = {{Decarli}, Roberto and {Loiacono}, Federica and {Farina}, Emanuele Paolo and {Dotti}, Massimo and {Lupi}, Alessandro and {Meyer}, Romain A. and {Mignoli}, Marco and {Pensabene}, Antonio and {Strauss}, Michael A. and {Venemans}, Bram and {Yang}, Jinyi and {Walter}, Fabian and {Wolf}, Julien and {Ba{\~n}ados}, Eduardo and {Blecha}, Laura and {Bosman}, Sarah and {Carilli}, Chris L. and {Comastri}, Andrea and {Connor}, Thomas and {Costa}, Tiago and {Eilers}, Anna-Christina and {Fan}, Xiaohui and {Gilli}, Roberto and {Jun}, Hyunsung D. and {Liu}, Weizhe and {Marshall}, Madeline A. and {Mazzucchelli}, Chiara and {Neeleman}, Marcel and {Onoue}, Masafusa and {Overzier}, Roderik and {Pudoka}, Maria Anne and {Riechers}, Dominik A. and {Rix}, Hans-Walter and {Schindler}, Jan-Torge and {Trakhtenbrot}, Benny and {Trebitsch}, Maxime and {Vestergaard}, Marianne and {Volonteri}, Marta and {Wang}, Feige and {Zhang}, Huanian and {Zou}, Siwei},
        title = "{A quasar-galaxy merger at z {\ensuremath{\sim}} 6.2: Rapid host growth via the accretion of two massive satellite galaxies}",
      journal = {\aap},
         year = 2024,
        month = sep,
       volume = {689},
          eid = {A219},
        pages = {A219},
          doi = {10.1051/0004-6361/202449239},
archivePrefix = {arXiv},
       eprint = {2406.06697},
 primaryClass = {astro-ph.GA},
       adsurl = {https://ui.adsabs.harvard.edu/abs/2024A&A...689A.219D}
}

@ARTICLE{visbal2014synchrPairs,
       author = {{Visbal}, Eli and {Haiman}, Zolt{\'a}n and {Bryan}, Greg L.},
        title = "{Direct collapse black hole formation from synchronized pairs of atomic cooling haloes}",
      journal = {\mnras},
         year = 2014,
        month = nov,
       volume = {445},
       number = {1},
        pages = {1056-1063},
          doi = {10.1093/mnras/stu1794},
archivePrefix = {arXiv},
       eprint = {1406.7020},
 primaryClass = {astro-ph.GA},
       adsurl = {https://ui.adsabs.harvard.edu/abs/2014MNRAS.445.1056V}
}

@ARTICLE{Oh2002,
       author = {{Oh}, S. Peng and {Haiman}, Zolt{\'a}n},
        title = "{Second-Generation Objects in the Universe: Radiative Cooling and Collapse of Halos with Virial Temperatures above {}10$^{4}$ K}",
      journal = {\apj},
         year = 2002,
        month = apr,
       volume = {569},
       number = {2},
        pages = {558-572},
          doi = {10.1086/339393},
archivePrefix = {arXiv},
       eprint = {astro-ph/0108071},
 primaryClass = {astro-ph},
       adsurl = {https://ui.adsabs.harvard.edu/abs/2002ApJ...569..558O}
}

@ARTICLE{Haiman1997,
       author = {{Haiman}, Zolt{\'a}n and {Rees}, Martin J. and {Loeb}, Abraham},
        title = "{Destruction of Molecular Hydrogen during Cosmological Reionization}",
      journal = {\apj},
         year = 1997,
        month = feb,
       volume = {476},
       number = {2},
        pages = {458-463},
          doi = {10.1086/303647},
archivePrefix = {arXiv},
       eprint = {astro-ph/9608130},
 primaryClass = {astro-ph},
       adsurl = {https://ui.adsabs.harvard.edu/abs/1997ApJ...476..458H}
}

@ARTICLE{Latif2022,
       author = {{Latif}, M.~A. and {Whalen}, D.~J. and {Khochfar}, S. and {Herrington}, N.~P. and {Woods}, T.~E.},
        title = "{Turbulent cold flows gave birth to the first quasars}",
      journal = {\nat},
         year = 2022,
        month = jul,
       volume = {607},
       number = {7917},
        pages = {48-51},
          doi = {10.1038/s41586-022-04813-y},
archivePrefix = {arXiv},
       eprint = {2207.05093},
 primaryClass = {astro-ph.GA},
       adsurl = {https://ui.adsabs.harvard.edu/abs/2022Natur.607...48L}
}

@ARTICLE{Shlosman2016,
       author = {{Shlosman}, Isaac and {Choi}, Jun-Hwan and {Begelman}, Mitchell C. and {Nagamine}, Kentaro},
        title = "{Supermassive black hole seed formation at high redshifts: long-term evolution of the direct collapse}",
      journal = {\mnras},
         year = 2016,
        month = feb,
       volume = {456},
       number = {1},
        pages = {500-511},
          doi = {10.1093/mnras/stv2700},
archivePrefix = {arXiv},
       eprint = {1508.05098},
 primaryClass = {astro-ph.GA},
       adsurl = {https://ui.adsabs.harvard.edu/abs/2016MNRAS.456..500S}
}

@ARTICLE{Behroozi2013,
       author = {{Behroozi}, Peter S. and {Wechsler}, Risa H. and {Wu}, Hao-Yi},
        title = "{The ROCKSTAR Phase-space Temporal Halo Finder and the Velocity Offsets of Cluster Cores}",
      journal = {\apj},
         year = 2013,
        month = jan,
       volume = {762},
       number = {2},
          eid = {109},
        pages = {109},
          doi = {10.1088/0004-637X/762/2/109},
archivePrefix = {arXiv},
       eprint = {1110.4372},
 primaryClass = {astro-ph.CO},
       adsurl = {https://ui.adsabs.harvard.edu/abs/2013ApJ...762..109B}
}

@ARTICLE{Rosdahl2017,
       author = {{Rosdahl}, Joakim and {Schaye}, Joop and {Dubois}, Yohan and {Kimm}, Taysun and {Teyssier}, Romain},
        title = "{Snap, crackle, pop: sub-grid supernova feedback in AMR simulations of disc galaxies}",
      journal = {\mnras},
         year = 2017,
        month = apr,
       volume = {466},
       number = {1},
        pages = {11-33},
          doi = {10.1093/mnras/stw3034},
archivePrefix = {arXiv},
       eprint = {1609.01296},
 primaryClass = {astro-ph.GA},
       adsurl = {https://ui.adsabs.harvard.edu/abs/2017MNRAS.466...11R}
}

@ARTICLE{Stone2024,
       author = {{Stone}, Meredith A. and {Lyu}, Jianwei and {Rieke}, George H. and {Alberts}, Stacey and {Hainline}, Kevin N.},
        title = "{Undermassive Host Galaxies of Five z {\ensuremath{\sim}} 6 Luminous Quasars Detected with JWST}",
      journal = {\apj},
         year = 2024,
        month = mar,
       volume = {964},
       number = {1},
          eid = {90},
        pages = {90},
          doi = {10.3847/1538-4357/ad2a57},
archivePrefix = {arXiv},
       eprint = {2310.18395},
 primaryClass = {astro-ph.GA},
       adsurl = {https://ui.adsabs.harvard.edu/abs/2024ApJ...964...90S}
}

@ARTICLE{Haiman2000,
       author = {{Haiman}, Zolt{\'a}n and {Abel}, Tom and {Rees}, Martin J.},
        title = "{The Radiative Feedback of the First Cosmological Objects}",
      journal = {\apj},
         year = 2000,
        month = may,
       volume = {534},
       number = {1},
        pages = {11-24},
          doi = {10.1086/308723},
archivePrefix = {arXiv},
       eprint = {astro-ph/9903336},
 primaryClass = {astro-ph},
       adsurl = {https://ui.adsabs.harvard.edu/abs/2000ApJ...534...11H}
}

@ARTICLE{Mayer2024,
       author = {{Mayer}, Lucio and {Capelo}, Pedro R. and {Zwick}, Lorenz and {Di Matteo}, Tiziana},
        title = "{Direct Formation of Massive Black Holes via Dynamical Collapse in Metal-enriched Merging Galaxies at z   10: Fully Cosmological Simulations}",
      journal = {\apj},
         year = 2024,
        month = jan,
       volume = {961},
       number = {1},
          eid = {76},
        pages = {76},
          doi = {10.3847/1538-4357/ad11cf},
archivePrefix = {arXiv},
       eprint = {2304.02066},
 primaryClass = {astro-ph.GA},
       adsurl = {https://ui.adsabs.harvard.edu/abs/2024ApJ...961...76M}
}

@ARTICLE{loiacono2024,
       author = {{Loiacono}, Federica and {Decarli}, Roberto and {Mignoli}, Marco and {Farina}, Emanuele Paolo and {Ba{\~n}ados}, Eduardo and {Bosman}, Sarah and {Eilers}, Anna-Christina and {Schindler}, Jan-Torge and {Strauss}, Michael A. and {Vestergaard}, Marianne and {Wang}, Feige and {Blecha}, Laura and {Carilli}, Chris L. and {Comastri}, Andrea and {Connor}, Thomas and {Costa}, Tiago and {Dotti}, Massimo and {Fan}, Xiaohui and {Gilli}, Roberto and {Jun}, Hyunsung D. and {Liu}, Weizhe and {Lupi}, Alessandro and {Marshall}, Madeline A. and {Mazzucchelli}, Chiara and {Meyer}, Romain A. and {Neeleman}, Marcel and {Overzier}, Roderik and {Pensabene}, Antonio and {Riechers}, Dominik A. and {Trakhtenbrot}, Benny and {Trebitsch}, Maxime and {Venemans}, Bram and {Walter}, Fabian and {Yang}, Jinyi},
        title = "{A quasar-galaxy merger at z {\ensuremath{\sim}} 6.2: Black hole mass and quasar properties from the NIRSpec spectrum}",
      journal = {\aap},
         year = 2024,
        month = may,
       volume = {685},
          eid = {A121},
        pages = {A121},
          doi = {10.1051/0004-6361/202348535},
archivePrefix = {arXiv},
       eprint = {2402.13319},
 primaryClass = {astro-ph.GA},
       adsurl = {https://ui.adsabs.harvard.edu/abs/2024A&A...685A.121L}
}

@ARTICLE{napolitano2025,
       author = {{Napolitano}, Lorenzo and {Castellano}, Marco and {Pentericci}, Laura and {Vignali}, Cristian and {Gilli}, Roberto and {Fontana}, Adriano and {Santini}, Paola and {Treu}, Tommaso and {Calabr{\`o}}, Antonello and {Llerena}, Mario and {Piconcelli}, Enrico and {Zappacosta}, Luca and {Mascia}, Sara and {Tripodi}, Roberta and {Arrabal Haro}, Pablo and {Bergamini}, Pietro and {Bakx}, Tom J.~L.~C. and {Dickinson}, Mark and {Glazebrook}, Karl and {Henry}, Alaina and {Leethochawalit}, Nicha and {Mazzolari}, Giovanni and {Merlin}, Emiliano and {Morishita}, Takahiro and {Nanayakkara}, Themiya and {Paris}, Diego and {Puccetti}, Simonetta and {Roberts-Borsani}, Guido and {Rojas Ruiz}, Sofia and {Rosati}, Piero and {Vanzella}, Eros and {Vito}, Fabio and {Vulcani}, Benedetta and {Wang}, Xin and {Yoon}, Ilsang and {Zavala}, Jorge A.},
        title = "{The Dual Nature of GHZ9: Coexisting Active Galactic Nuclei and Star Formation Activity in a Remote X-Ray Source at z = 10.145}",
      journal = {\apj},
         year = 2025,
        month = aug,
       volume = {989},
       number = {1},
          eid = {75},
        pages = {75},
          doi = {10.3847/1538-4357/ade706},
archivePrefix = {arXiv},
       eprint = {2410.18763},
 primaryClass = {astro-ph.GA},
       adsurl = {https://ui.adsabs.harvard.edu/abs/2025ApJ...989...75N}
}

@ARTICLE{tripodi2025,
       author = {{Tripodi}, Roberta and {Martis}, Nicholas and {Markov}, Vladan and {Brada{\v{c}}}, Maru{\v{s}}a and {Di Mascia}, Fabio and {Cammelli}, Vieri and {D'Eugenio}, Francesco and {Willott}, Chris and {Curti}, Mirko and {Bhatt}, Maulik and {Gallerani}, Simona and {Rihtar{\v{s}}i{\v{c}}}, Gregor and {Singh}, Jasbir and {Gaspar}, Gaia and {Harshan}, Anishya and {Jude{\v{z}}}, Jon and {Merida}, Rosa M. and {Desprez}, Guillaume and {Sawicki}, Marcin and {Goovaerts}, Ilias and {Muzzin}, Adam and {Noirot}, Ga{\"e}l and {Sarrouh}, Ghassan T.~E. and {Abraham}, Roberto and {Asada}, Yoshihisa and {Brammer}, Gabriel and {Estrada Carpenter}, Vicente and {Felicioni}, Giordano and {Fujimoto}, Seiji and {Iyer}, Kartheik and {Mowla}, Lamiya and {Strait}, Victoria},
        title = "{Red, hot, and very metal poor: extreme properties of a massive accreting black hole in the first 500 Myr}",
      journal = {arXiv e-prints},
         year = 2024,
        month = dec,
          eid = {arXiv:2412.04983},
        pages = {arXiv:2412.04983},
          doi = {10.48550/arXiv.2412.04983},
archivePrefix = {arXiv},
       eprint = {2412.04983},
 primaryClass = {astro-ph.GA},
       adsurl = {https://ui.adsabs.harvard.edu/abs/2024arXiv241204983T}
}

@ARTICLE{Marshall2023,
       author = {{Marshall}, Madeline A. and {Perna}, Michele and {Willott}, Chris J. and {Maiolino}, Roberto and {Scholtz}, Jan and {{\"U}bler}, Hannah and {Carniani}, Stefano and {Arribas}, Santiago and {L{\"u}tzgendorf}, Nora and {Bunker}, Andrew J. and {Charlot}, Stephane and {Ferruit}, Pierre and {Jakobsen}, Peter and {Rix}, Hans-Walter and {Rodr{\'\i}guez Del Pino}, Bruno and {B{\"o}ker}, Torsten and {Cameron}, Alex J. and {Cresci}, Giovanni and {Curtis-Lake}, Emma and {Jones}, Gareth C. and {Kumari}, Nimisha and {P{\'e}rez-Gonz{\'a}lez}, Pablo G. and {Reed}, Sophie L.},
        title = "{GA-NIFS: Black hole and host galaxy properties of two z ≃ 6.8 quasars from the NIRSpec IFU}",
      journal = {\aap},
         year = 2023,
        month = oct,
       volume = {678},
          eid = {A191},
        pages = {A191},
          doi = {10.1051/0004-6361/202346113},
archivePrefix = {arXiv},
       eprint = {2302.04795},
 primaryClass = {astro-ph.GA},
       adsurl = {https://ui.adsabs.harvard.edu/abs/2023A&A...678A.191M}
}

@ARTICLE{Wise2023,
       author = {{Wise}, John H. and {Regan}, John A. and {O'Shea}, Brian W. and {Norman}, Michael L. and {Downes}, Turlough P. and {Xu}, Hao},
        title = "{Formation of massive black holes in rapidly growing pre-galactic gas clouds}",
      journal = {\nat},
         year = 2019,
        month = jan,
       volume = {566},
       number = {7742},
        pages = {85-88},
          doi = {10.1038/s41586-019-0873-4},
archivePrefix = {arXiv},
       eprint = {1901.07563},
 primaryClass = {astro-ph.GA},
       adsurl = {https://ui.adsabs.harvard.edu/abs/2019Natur.566...85W}
}

@ARTICLE{Ding2023,
       author = {{Ding}, Xuheng and {Onoue}, Masafusa and {Silverman}, John D. and {Matsuoka}, Yoshiki and {Izumi}, Takuma and {Strauss}, Michael A. and {Jahnke}, Knud and {Phillips}, Camryn L. and {Li}, Junyao and {Volonteri}, Marta and {Haiman}, Zoltan and {Andika}, Irham Taufik and {Aoki}, Kentaro and {Baba}, Shunsuke and {Bieri}, Rebekka and {Bosman}, Sarah E.~I. and {Bottrell}, Connor and {Eilers}, Anna-Christina and {Fujimoto}, Seiji and {Habouzit}, Melanie and {Imanishi}, Masatoshi and {Inayoshi}, Kohei and {Iwasawa}, Kazushi and {Kashikawa}, Nobunari and {Kawaguchi}, Toshihiro and {Kohno}, Kotaro and {Lee}, Chien-Hsiu and {Lupi}, Alessandro and {Lyu}, Jianwei and {Nagao}, Tohru and {Overzier}, Roderik and {Schindler}, Jan-Torge and {Schramm}, Malte and {Shimasaku}, Kazuhiro and {Toba}, Yoshiki and {Trakhtenbrot}, Benny and {Trebitsch}, Maxime and {Treu}, Tommaso and {Umehata}, Hideki and {Venemans}, Bram P. and {Vestergaard}, Marianne and {Walter}, Fabian and {Wang}, Feige and {Yang}, Jinyi},
        title = "{Detection of stellar light from quasar host galaxies at redshifts above 6}",
      journal = {\nat},
         year = 2023,
        month = sep,
       volume = {621},
       number = {7977},
        pages = {51-55},
          doi = {10.1038/s41586-023-06345-5},
archivePrefix = {arXiv},
       eprint = {2211.14329},
 primaryClass = {astro-ph.GA},
       adsurl = {https://ui.adsabs.harvard.edu/abs/2023Natur.621...51D}
}

@ARTICLE{Wang2023,
       author = {{Wang}, Feige and {Yang}, Jinyi and {Hennawi}, Joseph F. and {Fan}, Xiaohui and {Sun}, Fengwu and {Champagne}, Jaclyn B. and {Costa}, Tiago and {Habouzit}, Melanie and {Endsley}, Ryan and {Li}, Zihao and {Lin}, Xiaojing and {Meyer}, Romain A. and {Schindler}, Jan-Torge and {Wu}, Yunjing and {Ba{\~n}ados}, Eduardo and {Barth}, Aaron J. and {Bhowmick}, Aklant K. and {Bieri}, Rebekka and {Blecha}, Laura and {Bosman}, Sarah and {Cai}, Zheng and {Colina}, Luis and {Connor}, Thomas and {Davies}, Frederick B. and {Decarli}, Roberto and {De Rosa}, Gisella and {Drake}, Alyssa B. and {Egami}, Eiichi and {Eilers}, Anna-Christina and {Evans}, Analis E. and {Farina}, Emanuele Paolo and {Haiman}, Zoltan and {Jiang}, Linhua and {Jin}, Xiangyu and {Jun}, Hyunsung D. and {Kakiichi}, Koki and {Khusanova}, Yana and {Kulkarni}, Girish and {Li}, Mingyu and {Liu}, Weizhe and {Loiacono}, Federica and {Lupi}, Alessandro and {Mazzucchelli}, Chiara and {Onoue}, Masafusa and {Pudoka}, Maria A. and {Rojas-Ruiz}, Sof{\'\i}a and {Shen}, Yue and {Strauss}, Michael A. and {Tee}, Wei Leong and {Trakhtenbrot}, Benny and {Trebitsch}, Maxime and {Venemans}, Bram and {Volonteri}, Marta and {Walter}, Fabian and {Xie}, Zhang-Liang and {Yue}, Minghao and {Zhang}, Haowen and {Zhang}, Huanian and {Zou}, Siwei},
        title = "{A SPectroscopic Survey of Biased Halos in the Reionization Era (ASPIRE): JWST Reveals a Filamentary Structure around a z = 6.61 Quasar}",
      journal = {\apjl},
         year = 2023,
        month = jul,
       volume = {951},
       number = {1},
          eid = {L4},
        pages = {L4},
          doi = {10.3847/2041-8213/accd6f},
archivePrefix = {arXiv},
       eprint = {2304.09894},
 primaryClass = {astro-ph.GA},
       adsurl = {https://ui.adsabs.harvard.edu/abs/2023ApJ...951L...4W}
}

@ARTICLE{taylor2025,
       author = {{Taylor}, Anthony J. and {Kokorev}, Vasily and {Kocevski}, Dale D. and {Akins}, Hollis B. and {Cullen}, Fergus and {Dickinson}, Mark and {Finkelstein}, Steven L. and {Arrabal Haro}, Pablo and {Bromm}, Volker and {Giavalisco}, Mauro and {Inayoshi}, Kohei and {Juneau}, Stephanie and {Leung}, Gene C.~K. and {Perez-Gonzalez}, Pablo G. and {Somerville}, Rachel S. and {Trump}, Jonathan R. and {Amorin}, Ricardo O. and {Barro}, Guillermo and {Burgarella}, Denis and {Brooks}, Madisyn and {Carnall}, Adam and {Casey}, Caitlin M. and {Cheng}, Yingjie and {Chisholm}, John and {Chworowsky}, Katherine and {Davis}, Kelcey and {Donnan}, Callum T. and {Dunlop}, James S. and {Ellis}, Richard S. and {Fernandez}, Vital and {Fujimoto}, Seiji and {Grogin}, Norman A. and {Gupta}, Ansh R. and {Hathi}, Nimish P. and {Jung}, Intae and {Hirschmann}, Michaela and {Kartaltepe}, Jeyhan S. and {Koekemoer}, Anton M. and {Larson}, Rebecca L. and {Leung}, Ho-Hin and {Llerena}, Mario and {Lucas}, Ray A. and {McLeod}, Derek J. and {McLure}, Ross and {Napolitano}, Lorenzo and {Papovich}, Casey and {Stanton}, Thomas M. and {Tripodi}, Roberta and {Wang}, Xin and {Wilkins}, Stephen M. and {Yung}, L.~Y. Aaron and {Zavala}, Jorge A.},
        title = "{CAPERS-LRD-z9: A Gas Enshrouded Little Red Dot Hosting a Broad-line AGN at z=9.288}",
      journal = {arXiv e-prints},
         year = 2025,
        month = may,
          eid = {arXiv:2505.04609},
        pages = {arXiv:2505.04609},
          doi = {10.48550/arXiv.2505.04609},
archivePrefix = {arXiv},
       eprint = {2505.04609},
 primaryClass = {astro-ph.GA},
       adsurl = {https://ui.adsabs.harvard.edu/abs/2025arXiv250504609T}
}

@ARTICLE{taylor2025b,
       author = {{Taylor}, Anthony J. and {Finkelstein}, Steven L. and {Kocevski}, Dale D. and {Jeon}, Junehyoung and {Bromm}, Volker and {Amor{\'\i}n}, Ricardo O. and {Arrabal Haro}, Pablo and {Backhaus}, Bren E. and {Bagley}, Micaela B. and {Banados}, Eduardo and {Bhatawdekar}, Rachana and {Brooks}, Madisyn and {Calabr{\`o}}, Antonello and {Ch{\'a}vez Ortiz}, {\'O}scar A. and {Cheng}, Yingjie and {Cleri}, Nikko J. and {Cole}, Justin W. and {Davis}, Kelcey and {Dickinson}, Mark and {Donnan}, Callum and {Dunlop}, James S. and {Ellis}, Richard S. and {Fern{\'a}ndez}, Vital and {Fontana}, Adriano and {Fujimoto}, Seiji and {Giavalisco}, Mauro and {Grazian}, Andrea and {Guo}, Jingsong and {Hathi}, Nimish P. and {Holwerda}, Benne W. and {Hirschmann}, Michaela and {Inayoshi}, Kohei and {Kartaltepe}, Jeyhan S. and {Khusanova}, Yana and {Koekemoer}, Anton M. and {Kokorev}, Vasily and {Larson}, Rebecca L. and {Leung}, Gene C.~K. and {Lucas}, Ray A. and {McLeod}, Derek J. and {Napolitano}, Lorenzo and {Onoue}, Masafusa and {Pacucci}, Fabio and {Papovich}, Casey and {P{\'e}rez-Gonz{\'a}lez}, Pablo G. and {Pirzkal}, Nor and {Somerville}, Rachel S. and {Trump}, Jonathan R. and {Wilkins}, Stephen M. and {Yung}, L.~Y. Aaron and {Zhang}, Haowen},
        title = "{Broad-line AGNs at 3.5 < z < 6: The Black Hole Mass Function and a Connection with Little Red Dots}",
      journal = {\apj},
         year = 2025,
        month = jun,
       volume = {986},
       number = {2},
          eid = {165},
        pages = {165},
          doi = {10.3847/1538-4357/add15b},
archivePrefix = {arXiv},
       eprint = {2409.06772},
 primaryClass = {astro-ph.GA},
       adsurl = {https://ui.adsabs.harvard.edu/abs/2025ApJ...986..165T}
}

@ARTICLE{Pacucci2026,
       author = {{Pacucci}, Fabio and {Ferrara}, Andrea and {Kocevski}, Dale D.},
        title = "{The Little Red Dots Are Direct Collapse Black Holes}",
      journal = {arXiv e-prints},
         year = 2026,
        month = jan,
          eid = {arXiv:2601.14368},
        pages = {arXiv:2601.14368},
          doi = {10.48550/arXiv.2601.14368},
archivePrefix = {arXiv},
       eprint = {2601.14368},
 primaryClass = {astro-ph.GA},
       adsurl = {https://ui.adsabs.harvard.edu/abs/2026arXiv260114368P}
}

@ARTICLE{Dunn2018,
       author = {{Dunn}, Glenna and {Bellovary}, Jillian and {Holley-Bockelmann}, Kelly and {Christensen}, Charlotte and {Quinn}, Thomas},
        title = "{Sowing Black Hole Seeds: Direct Collapse Black Hole Formation with Realistic Lyman-Werner Radiation in Cosmological Simulations}",
      journal = {\apj},
         year = 2018,
        month = jul,
       volume = {861},
       number = {1},
          eid = {39},
        pages = {39},
          doi = {10.3847/1538-4357/aac7c2},
archivePrefix = {arXiv},
       eprint = {1803.01007},
 primaryClass = {astro-ph.GA},
       adsurl = {https://ui.adsabs.harvard.edu/abs/2018ApJ...861...39D}
}

@ARTICLE{Baggen2026,
       author = {{Baggen}, Josephine F.~W. and {Scoggins}, Matthew T. and {van Dokkum}, Pieter and {Haiman}, Zolt{\'a}n and {Torralba}, Alberto and {Matthee}, Jorryt},
        title = "{Connecting the Dots: UV-bright Companions of Little Red Dots as Lyman─Werner Sources Enabling Direct-collapse Black Hole Formation}",
      journal = {\apjl},
         year = 2026,
        month = may,
       volume = {1002},
       number = {1},
          eid = {L4},
        pages = {L4},
          doi = {10.3847/2041-8213/ae58a5},
archivePrefix = {arXiv},
       eprint = {2602.02702},
 primaryClass = {astro-ph.GA},
       adsurl = {https://ui.adsabs.harvard.edu/abs/2026ApJ..1002L...4B}
}

@ARTICLE{Madau2026,
       author = {{Madau}, Piero and {Maiolino}, Roberto},
        title = "{Little red dots as obscured little blue dots: relative abundances, luminosities, and black-hole masses}",
      journal = {arXiv e-prints},
         year = 2026,
        month = may,
          eid = {arXiv:2605.05074},
        pages = {arXiv:2605.05074},
archivePrefix = {arXiv},
       eprint = {2605.05074},
 primaryClass = {astro-ph.GA},
       adsurl = {https://ui.adsabs.harvard.edu/abs/2026arXiv260505074M}
}

@ARTICLE{inayoshi2024,
       author = {{Inayoshi}, Kohei and {Maiolino}, Roberto},
        title = "{Extremely Dense Gas around Little Red Dots and High-redshift Active Galactic Nuclei: A Nonstellar Origin of the Balmer Break and Absorption Features}",
      journal = {\apjl},
         year = 2025,
        month = feb,
       volume = {980},
       number = {2},
          eid = {L27},
        pages = {L27},
          doi = {10.3847/2041-8213/adaebd},
archivePrefix = {arXiv},
       eprint = {2409.07805},
 primaryClass = {astro-ph.GA},
       adsurl = {https://ui.adsabs.harvard.edu/abs/2025ApJ...980L..27I}
}

@ARTICLE{Casey2025,
       author = {{Casey}, Caitlin M. and {Akins}, Hollis B. and {Finkelstein}, Steven L. and {Franco}, Maximilien and {Fujimoto}, Seiji and {Liu}, Daizhong and {Long}, Arianna S. and {Magdis}, Georgios and {Manning}, Sinclaire M. and {McKinney}, Jed and {Shuntov}, Marko and {Tanaka}, Takumi S.},
        title = "{An Upper Limit of {}10$^{6}$ M$_{{\ensuremath{\odot}}}$ in Dust from ALMA Observations in 60 Little Red Dots}",
      journal = {\apjl},
         year = 2025,
        month = sep,
       volume = {990},
       number = {2},
          eid = {L61},
        pages = {L61},
          doi = {10.3847/2041-8213/adfa91},
archivePrefix = {arXiv},
       eprint = {2505.18873},
 primaryClass = {astro-ph.GA},
       adsurl = {https://ui.adsabs.harvard.edu/abs/2025ApJ...990L..61C}
}

@ARTICLE{wang2024,
       author = {{Wang}, Feige and {Yang}, Jinyi and {Hennawi}, Joseph F. and {Fan}, Xiaohui and {Yue}, Minghao and {Ba{\~n}ados}, Eduardo and {Bechtel}, Shane and {Bian}, Fuyan and {Bosman}, Sarah and {Champagne}, Jaclyn B. and {Davies}, Frederick B. and {Decarli}, Roberto and {Farina}, Emanuele Paolo and {Mazzucchelli}, Chiara and {Venemans}, Bram and {Walter}, Fabian},
        title = "{A Massive Protocluster Anchored by a Luminous Quasar at z = 6.63}",
      journal = {\apjl},
         year = 2024,
        month = feb,
       volume = {962},
       number = {1},
          eid = {L11},
        pages = {L11},
          doi = {10.3847/2041-8213/ad20ef},
archivePrefix = {arXiv},
       eprint = {2402.01844},
 primaryClass = {astro-ph.GA},
       adsurl = {https://ui.adsabs.harvard.edu/abs/2024ApJ...962L..11W}
}

@ARTICLE{chon2025,
       author = {{Chon}, Sunmyon and {Omukai}, Kazuyuki},
        title = "{Formation of supermassive stars and dense star clusters in metal-poor clouds exposed to strong FUV radiation}",
      journal = {\mnras},
         year = 2025,
        month = may,
       volume = {539},
       number = {3},
        pages = {2561-2582},
          doi = {10.1093/mnras/staf598},
archivePrefix = {arXiv},
       eprint = {2412.14900},
 primaryClass = {astro-ph.GA},
       adsurl = {https://ui.adsabs.harvard.edu/abs/2025MNRAS.539.2561C}
}

@ARTICLE{Zana2025,
       author = {{Zana}, Tommaso and {Capelo}, Pedro R. and {Boresta}, Mairo and {Schneider}, Raffaella and {Lupi}, Alessandro and {Trinca}, Alessandro and {Mayer}, Lucio and {Valiante}, Rosa and {Graziani}, Luca},
        title = "{Super-Eddington accretion in protogalactic cores}",
      journal = {arXiv e-prints},
         year = 2025,
        month = aug,
          eid = {arXiv:2508.21114},
        pages = {arXiv:2508.21114},
          doi = {10.48550/arXiv.2508.21114},
archivePrefix = {arXiv},
       eprint = {2508.21114},
 primaryClass = {astro-ph.GA},
       adsurl = {https://ui.adsabs.harvard.edu/abs/2025arXiv250821114Z}
}

@ARTICLE{trinca2025,
       author = {{Trinca}, Alessandro and {Valiante}, Rosa and {Schneider}, Raffaella and {Juod{\v{z}}balis}, Ignas and {Maiolino}, Roberto and {Graziani}, Luca and {Lupi}, Alessandro and {Natarajan}, Priyamvada and {Volonteri}, Marta and {Zana}, Tommaso},
        title = "{Episodic super-Eddington accretion as a clue to Overmassive Black Holes in the early Universe}",
      journal = {arXiv e-prints},
         year = 2024,
        month = dec,
          eid = {arXiv:2412.14248},
        pages = {arXiv:2412.14248},
          doi = {10.48550/arXiv.2412.14248},
archivePrefix = {arXiv},
       eprint = {2412.14248},
 primaryClass = {astro-ph.GA},
       adsurl = {https://ui.adsabs.harvard.edu/abs/2024arXiv241214248T}
}

@software{MUSIC,
       author = {{Hahn}, Oliver and {Abel}, Tom},
        title = "{MUSIC: MUlti-Scale Initial Conditions}",
 howpublished = {Astrophysics Source Code Library, record ascl:1311.011},
         year = 2013,
        month = nov,
          eid = {ascl:1311.011},
       adsurl = {https://ui.adsabs.harvard.edu/abs/2013ascl.soft11011H}
}

@ARTICLE{quadri2025,
       author = {{Quadri}, Giada and {Trinca}, Alessandro and {Lupi}, Alessandro and {Colpi}, Monica and {Volonteri}, Marta},
        title = "{Super-Eddington accretion in high-redshift quasar hosts: black-hole driven outflows, galaxy quenching, and the nature of Little Red Dots}",
      journal = {arXiv e-prints},
         year = 2025,
        month = may,
          eid = {arXiv:2505.05556},
        pages = {arXiv:2505.05556},
          doi = {10.48550/arXiv.2505.05556},
archivePrefix = {arXiv},
       eprint = {2505.05556},
 primaryClass = {astro-ph.GA},
       adsurl = {https://ui.adsabs.harvard.edu/abs/2025arXiv250505556Q}
}

@article{smith2015,
   title={The first Population II stars formed in externally enriched mini-haloes},
   volume={452},
   ISSN={1365-2966},
   url={http://dx.doi.org/10.1093/mnras/stv1509},
   DOI={10.1093/mnras/stv1509},
   number={3},
   journal={Monthly Notices of the Royal Astronomical Society},
   publisher={Oxford University Press (OUP)},
   author={Smith, Britton D. and Wise, John H. and O’Shea, Brian W. and Norman, Michael L. and Khochfar, Sadegh},
   year={2015},
   month=jul, pages={2822–2836} }

@ARTICLE{Husko2025,
       author = {{Hu{\v{s}}ko}, Filip and {Lacey}, Cedric G. and {Roper}, William J. and {Schaye}, Joop and {Briggs}, Jemima Mae and {Schaller}, Matthieu},
        title = "{The effects of super-Eddington accretion and feedback on the growth of early supermassive black holes and galaxies}",
      journal = {\mnras},
         year = 2025,
        month = mar,
       volume = {537},
       number = {3},
        pages = {2559-2578},
          doi = {10.1093/mnras/staf146},
archivePrefix = {arXiv},
       eprint = {2410.09450},
 primaryClass = {astro-ph.GA},
       adsurl = {https://ui.adsabs.harvard.edu/abs/2025MNRAS.537.2559H}
}

@ARTICLE{hopkins2015,
       author = {{Hopkins}, Philip F.},
        title = "{A new class of accurate, mesh-free hydrodynamic simulation methods}",
      journal = {\mnras},
         year = 2015,
        month = jun,
       volume = {450},
       number = {1},
        pages = {53-110},
          doi = {10.1093/mnras/stv195},
archivePrefix = {arXiv},
       eprint = {1409.7395},
 primaryClass = {astro-ph.CO},
       adsurl = {https://ui.adsabs.harvard.edu/abs/2015MNRAS.450...53H}
}

@ARTICLE{Eilers2024,
       author = {{Eilers}, Anna-Christina and {Mackenzie}, Ruari and {Pizzati}, Elia and {Matthee}, Jorryt and {Hennawi}, Joseph F. and {Zhang}, Haowen and {Bordoloi}, Rongmon and {Kashino}, Daichi and {Lilly}, Simon J. and {Naidu}, Rohan P. and {Simcoe}, Robert A. and {Yue}, Minghao and {Frenk}, Carlos S. and {Helly}, John C. and {Schaller}, Matthieu and {Schaye}, Joop},
        title = "{EIGER. VI. The Correlation Function, Host Halo Mass, and Duty Cycle of Luminous Quasars at z {\ensuremath{\gtrsim}} 6}",
      journal = {\apj},
         year = 2024,
        month = oct,
       volume = {974},
       number = {2},
          eid = {275},
        pages = {275},
          doi = {10.3847/1538-4357/ad778b},
       adsurl = {https://ui.adsabs.harvard.edu/abs/2024ApJ...974..275E}
}

@ARTICLE{Lupi2016,
       author = {{Lupi}, A. and {Haardt}, F. and {Dotti}, M. and {Fiacconi}, D. and {Mayer}, L. and {Madau}, P.},
        title = "{Growing massive black holes through supercritical accretion of stellar-mass seeds}",
      journal = {\mnras},
         year = 2016,
        month = mar,
       volume = {456},
       number = {3},
        pages = {2993-3003},
          doi = {10.1093/mnras/stv2877},
archivePrefix = {arXiv},
       eprint = {1512.02651},
 primaryClass = {astro-ph.GA},
       adsurl = {https://ui.adsabs.harvard.edu/abs/2016MNRAS.456.2993L}
}

@ARTICLE{lupi2024SE,
       author = {{Lupi}, Alessandro and {Quadri}, Giada and {Volonteri}, Marta and {Colpi}, Monica and {Regan}, John A.},
        title = "{Sustained super-Eddington accretion in high-redshift quasars}",
      journal = {\aap},
         year = 2024,
        month = jun,
       volume = {686},
          eid = {A256},
        pages = {A256},
          doi = {10.1051/0004-6361/202348788},
archivePrefix = {arXiv},
       eprint = {2312.08422},
 primaryClass = {astro-ph.GA},
       adsurl = {https://ui.adsabs.harvard.edu/abs/2024A&A...686A.256L}
}

@ARTICLE{Shi2024,
       author = {{Shi}, Yanlong and {Kremer}, Kyle and {Hopkins}, Philip F.},
        title = "{From Seeds to Supermassive Black Holes: Capture, Growth, Migration, and Pairing in Dense Protobulge Environments}",
      journal = {\apjl},
         year = 2024,
        month = jul,
       volume = {969},
       number = {2},
          eid = {L31},
        pages = {L31},
          doi = {10.3847/2041-8213/ad5a95},
archivePrefix = {arXiv},
       eprint = {2405.17338},
 primaryClass = {astro-ph.GA},
       adsurl = {https://ui.adsabs.harvard.edu/abs/2024ApJ...969L..31S}
}

@ARTICLE{Akins2025,
       author = {{Akins}, Hollis B. and {Casey}, Caitlin M. and {Lambrides}, Erini and {Allen}, Natalie and {Andika}, Irham T. and {Brinch}, Malte and {Champagne}, Jaclyn B. and {Cooper}, Olivia and {Ding}, Xuheng and {Drakos}, Nicole E. and {Faisst}, Andreas and {Finkelstein}, Steven L. and {Franco}, Maximilien and {Fujimoto}, Seiji and {Gentile}, Fabrizio and {Gillman}, Steven and {Gozaliasl}, Ghassem and {Harish}, Santosh and {Hayward}, Christopher C. and {Hirschmann}, Michaela and {Ilbert}, Olivier and {Kartaltepe}, Jeyhan S. and {Kocevski}, Dale D. and {Koekemoer}, Anton M. and {Kokorev}, Vasily and {Liu}, Daizhong and {Long}, Arianna S. and {McCracken}, Henry Joy and {McKinney}, Jed and {Onoue}, Masafusa and {Paquereau}, Louise and {Renzini}, Alvio and {Rhodes}, Jason and {Robertson}, Brant E. and {Shuntov}, Marko and {Silverman}, John D. and {Tanaka}, Takumi S. and {Toft}, Sune and {Trakhtenbrot}, Benny and {Valentino}, Francesco and {Zavala}, Jorge},
        title = "{COSMOS-Web: The Overabundance and Physical Nature of ``Little Red Dots''{\textemdash}Implications for Early Galaxy and SMBH Assembly}",
      journal = {\apj},
         year = 2025,
        month = sep,
       volume = {991},
       number = {1},
          eid = {37},
        pages = {37},
          doi = {10.3847/1538-4357/ade984},
archivePrefix = {arXiv},
       eprint = {2406.10341},
 primaryClass = {astro-ph.GA},
       adsurl = {https://ui.adsabs.harvard.edu/abs/2025ApJ...991...37A}
}

@ARTICLE{Gordon2024,
       author = {{Gordon}, Simone T. and {Smith}, Britton D. and {Khochfar}, Sadegh and {Beckmann}, Ricarda S.},
        title = "{Conditions for super-Eddington accretion onto the first black holes}",
      journal = {\mnras},
         year = 2025,
        month = feb,
       volume = {537},
       number = {2},
        pages = {674-690},
          doi = {10.1093/mnras/staf054},
archivePrefix = {arXiv},
       eprint = {2412.06888},
 primaryClass = {astro-ph.GA},
       adsurl = {https://ui.adsabs.harvard.edu/abs/2025MNRAS.537..674G}
}

@ARTICLE{Bogdan2024,
       author = {{Bogd{\'a}n}, {\'A}kos and {Goulding}, Andy D. and {Natarajan}, Priyamvada and {Kov{\'a}cs}, Orsolya E. and {Tremblay}, Grant R. and {Chadayammuri}, Urmila and {Volonteri}, Marta and {Kraft}, Ralph P. and {Forman}, William R. and {Jones}, Christine and {Churazov}, Eugene and {Zhuravleva}, Irina},
        title = "{Evidence for heavy-seed origin of early supermassive black holes from a z {\ensuremath{\approx}} 10 X-ray quasar}",
      journal = {Nature Astronomy},
         year = 2024,
        month = jan,
       volume = {8},
        pages = {126-133},
          doi = {10.1038/s41550-023-02111-9},
archivePrefix = {arXiv},
       eprint = {2305.15458},
 primaryClass = {astro-ph.GA},
       adsurl = {https://ui.adsabs.harvard.edu/abs/2024NatAs...8..126B}
}

@ARTICLE{maiolino24,
       author = {{Maiolino}, Roberto and {Scholtz}, Jan and {Witstok}, Joris and {Carniani}, Stefano and {D'Eugenio}, Francesco and {de Graaff}, Anna and {{\"U}bler}, Hannah and {Tacchella}, Sandro and {Curtis-Lake}, Emma and {Arribas}, Santiago and {Bunker}, Andrew and {Charlot}, St{\'e}phane and {Chevallard}, Jacopo and {Curti}, Mirko and {Looser}, Tobias J. and {Maseda}, Michael V. and {Rawle}, Timothy D. and {Rodr{\'\i}guez del Pino}, Bruno and {Willott}, Chris J. and {Egami}, Eiichi and {Eisenstein}, Daniel J. and {Hainline}, Kevin N. and {Robertson}, Brant and {Williams}, Christina C. and {Willmer}, Christopher N.~A. and {Baker}, William M. and {Boyett}, Kristan and {DeCoursey}, Christa and {Fabian}, Andrew C. and {Helton}, Jakob M. and {Ji}, Zhiyuan and {Jones}, Gareth C. and {Kumari}, Nimisha and {Laporte}, Nicolas and {Nelson}, Erica J. and {Perna}, Michele and {Sandles}, Lester and {Shivaei}, Irene and {Sun}, Fengwu},
        title = "{A small and vigorous black hole in the early Universe}",
      journal = {\nat},
         year = 2024,
        month = mar,
       volume = {627},
       number = {8002},
        pages = {59-63},
          doi = {10.1038/s41586-024-07052-5},
archivePrefix = {arXiv},
       eprint = {2305.12492},
 primaryClass = {astro-ph.GA},
       adsurl = {https://ui.adsabs.harvard.edu/abs/2024Natur.627...59M}
}

@ARTICLE{ferrara2014,
       author = {{Ferrara}, A. and {Salvadori}, S. and {Yue}, B. and {Schleicher}, D.},
        title = "{Initial mass function of intermediate-mass black hole seeds}",
      journal = {\mnras},
         year = 2014,
        month = sep,
       volume = {443},
       number = {3},
        pages = {2410-2425},
          doi = {10.1093/mnras/stu1280},
archivePrefix = {arXiv},
       eprint = {1406.6685},
 primaryClass = {astro-ph.GA},
       adsurl = {https://ui.adsabs.harvard.edu/abs/2014MNRAS.443.2410F}
}

@ARTICLE{kashino2023,
       author = {{Kashino}, Daichi and {Lilly}, Simon J. and {Matthee}, Jorryt and {Eilers}, Anna-Christina and {Mackenzie}, Ruari and {Bordoloi}, Rongmon and {Simcoe}, Robert A.},
        title = "{EIGER. I. A Large Sample of [O III]-emitting Galaxies at 5.3 < z < 6.9 and Direct Evidence for Local Reionization by Galaxies}",
      journal = {\apj},
         year = 2023,
        month = jun,
       volume = {950},
       number = {1},
          eid = {66},
        pages = {66},
          doi = {10.3847/1538-4357/acc588},
archivePrefix = {arXiv},
       eprint = {2211.08254},
 primaryClass = {astro-ph.GA},
       adsurl = {https://ui.adsabs.harvard.edu/abs/2023ApJ...950...66K}
}

@ARTICLE{Matthee2024,
       author = {{Matthee}, Jorryt and {Naidu}, Rohan P. and {Brammer}, Gabriel and {Chisholm}, John and {Eilers}, Anna-Christina and {Goulding}, Andy and {Greene}, Jenny and {Kashino}, Daichi and {Labbe}, Ivo and {Lilly}, Simon J. and {Mackenzie}, Ruari and {Oesch}, Pascal A. and {Weibel}, Andrea and {Wuyts}, Stijn and {Xiao}, Mengyuan and {Bordoloi}, Rongmon and {Bouwens}, Rychard and {van Dokkum}, Pieter and {Illingworth}, Garth and {Kramarenko}, Ivan and {Maseda}, Michael V. and {Mason}, Charlotte and {Meyer}, Romain A. and {Nelson}, Erica J. and {Reddy}, Naveen A. and {Shivaei}, Irene and {Simcoe}, Robert A. and {Yue}, Minghao},
        title = "{Little Red Dots: An Abundant Population of Faint Active Galactic Nuclei at z {\ensuremath{\sim}} 5 Revealed by the EIGER and FRESCO JWST Surveys}",
      journal = {\apj},
         year = 2024,
        month = mar,
       volume = {963},
       number = {2},
          eid = {129},
        pages = {129},
          doi = {10.3847/1538-4357/ad2345},
archivePrefix = {arXiv},
       eprint = {2306.05448},
 primaryClass = {astro-ph.GA},
       adsurl = {https://ui.adsabs.harvard.edu/abs/2024ApJ...963..129M}
}

@ARTICLE{trinca2024,
       author = {{Trinca}, Alessandro and {Schneider}, Raffaella and {Valiante}, Rosa and {Graziani}, Luca and {Ferrotti}, Arianna and {Omukai}, Kazuyuki and {Chon}, Sunmyon},
        title = "{Exploring the nature of UV-bright z {\ensuremath{\gtrsim}} 10 galaxies detected by JWST: star formation, black hole accretion, or a non-universal IMF?}",
      journal = {\mnras},
         year = 2024,
        month = apr,
       volume = {529},
       number = {4},
        pages = {3563-3581},
          doi = {10.1093/mnras/stae651},
archivePrefix = {arXiv},
       eprint = {2305.04944},
 primaryClass = {astro-ph.GA},
       adsurl = {https://ui.adsabs.harvard.edu/abs/2024MNRAS.529.3563T}
}

@ARTICLE{ubler2024,
       author = {{{\"U}bler}, Hannah and {Maiolino}, Roberto and {P{\'e}rez-Gonz{\'a}lez}, Pablo G. and {D'Eugenio}, Francesco and {Perna}, Michele and {Curti}, Mirko and {Arribas}, Santiago and {Bunker}, Andrew and {Carniani}, Stefano and {Charlot}, St{\'e}phane and {Rodr{\'\i}guez Del Pino}, Bruno and {Baker}, William and {B{\"o}ker}, Torsten and {Cresci}, Giovanni and {Dunlop}, James and {Grogin}, Norman A. and {Jones}, Gareth C. and {Kumari}, Nimisha and {Lamperti}, Isabella and {Laporte}, Nicolas and {Marshall}, Madeline A. and {Mazzolari}, Giovanni and {Parlanti}, Eleonora and {Rawle}, Tim and {Scholtz}, Jan and {Venturi}, Giacomo and {Witstok}, Joris},
        title = "{GA-NIFS: JWST discovers an offset AGN 740 million years after the big bang}",
      journal = {\mnras},
         year = 2024,
        month = jun,
       volume = {531},
       number = {1},
        pages = {355-365},
          doi = {10.1093/mnras/stae943},
archivePrefix = {arXiv},
       eprint = {2312.03589},
 primaryClass = {astro-ph.GA},
       adsurl = {https://ui.adsabs.harvard.edu/abs/2024MNRAS.531..355U}
}

@ARTICLE{rigamonti2025,
       author = {{Rigamonti}, Fabio and {Bertassi}, Lorenzo and {Buscicchio}, Riccardo and {Cocchiararo}, Fabiola and {Covino}, Stefano and {Dotti}, Massimo and {Sesana}, Alberto and {Severgnini}, Paola},
        title = "{Variability in the supermassive black hole binary candidate SDSS J2320+0024: No evidence of periodic modulation}",
      journal = {arXiv e-prints},
         year = 2025,
        month = may,
          eid = {arXiv:2505.22706},
        pages = {arXiv:2505.22706},
          doi = {10.48550/arXiv.2505.22706},
archivePrefix = {arXiv},
       eprint = {2505.22706},
 primaryClass = {astro-ph.GA},
       adsurl = {https://ui.adsabs.harvard.edu/abs/2025arXiv250522706R}
}

@ARTICLE{sullivan2025,
       author = {{Sullivan}, James and {Haiman}, Zolt{\'a}n and {Kulkarni}, Mihir and {Visbal}, Eli},
        title = "{Can supermassive stars form in protogalaxies due to internal Lyman{\textendash}Werner feedback?}",
      journal = {\mnras},
         year = 2025,
        month = sep,
       volume = {542},
       number = {2},
        pages = {822-838},
          doi = {10.1093/mnras/staf1269},
archivePrefix = {arXiv},
       eprint = {2501.12986},
 primaryClass = {astro-ph.GA},
       adsurl = {https://ui.adsabs.harvard.edu/abs/2025MNRAS.542..822S}
}

@ARTICLE{Haiman2023,
       author = {{Haiman}, Zolt{\'a}n and {Xin}, Chengcheng and {Bogdanovi{\'c}}, Tamara and {Amaro Seoane}, Pau and {Bonetti}, Matteo and {Casey-Clyde}, J. Andrew and {Charisi}, Maria and {Colpi}, Monica and {Davelaar}, Jordy and {De Rosa}, Alessandra and {D'Orazio}, Daniel J. and {Futrowsky}, Kate and {Gandhi}, Poshak and {Graham}, Alister W. and {Greene}, Jenny E. and {Habouzit}, Melanie and {Haggard}, Daryl and {Holley-Bockelmann}, Kelly and {Liu}, Xin and {Mangiagli}, Alberto and {Mastrobuono-Battisti}, Alessandra and {McGee}, Sean and {Mingarelli}, Chiara M.~F. and {Nemmen}, Rodrigo and {Palmese}, Antonella and {Porquet}, Delphine and {Sesana}, Alberto and {Stemo}, Aaron and {Torres-Orjuela}, Alejandro and {Zrake}, Jonathan},
        title = "{Massive Black Hole Binaries as LISA Precursors in the Roman High Latitude Time Domain Survey}",
      journal = {arXiv e-prints},
         year = 2023,
        month = jun,
          eid = {arXiv:2306.14990},
        pages = {arXiv:2306.14990},
          doi = {10.48550/arXiv.2306.14990},
archivePrefix = {arXiv},
       eprint = {2306.14990},
 primaryClass = {astro-ph.HE},
       adsurl = {https://ui.adsabs.harvard.edu/abs/2023arXiv230614990H}
}

@ARTICLE{bertassi2025,
       author = {{Bertassi}, Lorenzo and {Sottocorno}, Erika and {Rigamonti}, Fabio and {D'Orazio}, Daniel J. and {Eracleous}, Michael and {Haiman}, Zolt{\'a}n and {Dotti}, Massimo},
        title = "{Testing compact massive black hole binary candidates through multi-epoch spectroscopy}",
      journal = {arXiv e-prints},
         year = 2025,
        month = apr,
          eid = {arXiv:2504.06349},
        pages = {arXiv:2504.06349},
          doi = {10.48550/arXiv.2504.06349},
archivePrefix = {arXiv},
       eprint = {2504.06349},
 primaryClass = {astro-ph.GA},
       adsurl = {https://ui.adsabs.harvard.edu/abs/2025arXiv250406349B}
}

@ARTICLE{dotti2023,
       author = {{Dotti}, Massimo and {Rigamonti}, Fabio and {Rinaldi}, Stefano and {Del Pozzo}, Walter and {Decarli}, Roberto and {Buscicchio}, Riccardo},
        title = "{A fast test for the identification and confirmation of massive black hole binaries}",
      journal = {\aap},
         year = 2023,
        month = dec,
       volume = {680},
          eid = {A69},
        pages = {A69},
          doi = {10.1051/0004-6361/202346916},
archivePrefix = {arXiv},
       eprint = {2310.06896},
 primaryClass = {astro-ph.HE},
       adsurl = {https://ui.adsabs.harvard.edu/abs/2023A&A...680A..69D}
}

@ARTICLE{Ji2025,
       author = {{Ji}, Xihan and {Maiolino}, Roberto and {{\"U}bler}, Hannah and {Scholtz}, Jan and {D'Eugenio}, Francesco and {Sun}, Fengwu and {Perna}, Michele and {Turner}, Hannah and {Arribas}, Santiago and {Bennett}, Jake S. and {Bunker}, Andrew and {Carniani}, Stefano and {Charlot}, St{\'e}phane and {Cresci}, Giovanni and {Curti}, Mirko and {Egami}, Eiichi and {Fabian}, Andy and {Inayoshi}, Kohei and {Isobe}, Yuki and {Jones}, Gareth and {Juod{\v{z}}balis}, Ignas and {Kumari}, Nimisha and {Lyu}, Jianwei and {Mazzolari}, Giovanni and {Parlanti}, Eleonora and {Robertson}, Brant and {Rodr{\'\i}guez Del Pino}, Bruno and {Schneider}, Raffaella and {Sijacki}, Debora and {Tacchella}, Sandro and {Trinca}, Alessandro and {Valiante}, Rosa and {Venturi}, Giacomo and {Volonteri}, Marta and {Willott}, Chris and {Witten}, Callum and {Witstok}, Joris},
        title = "{BlackTHUNDER -- A non-stellar Balmer break in a black hole-dominated little red dot at $z=7.04$}",
      journal = {arXiv e-prints},
         year = 2025,
        month = jan,
          eid = {arXiv:2501.13082},
        pages = {arXiv:2501.13082},
          doi = {10.48550/arXiv.2501.13082},
archivePrefix = {arXiv},
       eprint = {2501.13082},
 primaryClass = {astro-ph.GA},
       adsurl = {https://ui.adsabs.harvard.edu/abs/2025arXiv250113082J}
}

@ARTICLE{Ubler2025,
       author = {{{\"U}bler}, Hannah and {Mazzolari}, Giovanni and {Maiolino}, Roberto and {D'Eugenio}, Francesco and {Davari}, Nazanin and {Juod{\v{z}}balis}, Ignas and {Schneider}, Raffaella and {Valiante}, Rosa and {Arribas}, Santiago and {Bertola}, Elena and {Bunker}, Andrew J. and {Bromm}, Volker and {Carniani}, Stefano and {Charlot}, St{\'e}phane and {Cresci}, Giovanni and {Curti}, Mirko and {Davies}, Richard and {Eisenhauer}, Frank and {Fabian}, Andrew and {F{\"o}rster Schreiber}, Natascha M. and {Genzel}, Reinhard and {Inayoshi}, Kohei and {Ivey}, Lucy R. and {Jones}, Gareth C. and {Liu}, Boyuan and {Lutz}, Dieter and {Mackenzie}, Ruari and {Matthee}, Jorryt and {Parlanti}, Eleonora and {Perna}, Michele and {Robertson}, Brant and {Rodr{\'\i}guez del Pino}, Bruno and {Shimizu}, T. Taro and {Sijacki}, Debora and {Sturm}, Eckhard and {Tacchella}, Sandro and {Tacconi}, Linda and {Tozzi}, Giulia and {Trinca}, Alessandro and {Venturi}, Giacomo and {Volonteri}, Marta and {Willot}, Chris and {Zhang}, Saiyang},
        title = "{BlackTHUNDER: evidence for three massive black holes in a z\raisebox{-0.5ex}\textasciitilde5 galaxy}",
      journal = {arXiv e-prints},
         year = 2025,
        month = sep,
          eid = {arXiv:2509.21575},
        pages = {arXiv:2509.21575},
archivePrefix = {arXiv},
       eprint = {2509.21575},
 primaryClass = {astro-ph.GA},
       adsurl = {https://ui.adsabs.harvard.edu/abs/2025arXiv250921575U}
}

@ARTICLE{Merlin2024,
       author = {{Merlin}, E. and {Santini}, P. and {Paris}, D. and {Castellano}, M. and {Fontana}, A. and {Treu}, T. and {Finkelstein}, S.~L. and {Dunlop}, J.~S. and {Arrabal Haro}, P. and {Bagley}, M. and {Boyett}, K. and {Calabr{\`o}}, A. and {Correnti}, M. and {Davis}, K. and {Dickinson}, M. and {Donnan}, C.~T. and {Ferguson}, H.~C. and {Fortuni}, F. and {Giavalisco}, M. and {Glazebrook}, K. and {Grazian}, A. and {Grogin}, N.~A. and {Hathi}, N. and {Hirschmann}, M. and {Kartaltepe}, J.~S. and {Kewley}, L.~J. and {Kirkpatrick}, A. and {Kocevski}, D.~D. and {Koekemoer}, A.~M. and {Leung}, G. and {Lotz}, J.~M. and {Lucas}, R.~A. and {Magee}, D.~K. and {Marchesini}, D. and {Mascia}, S. and {McLeod}, D.~J. and {McLure}, R.~J. and {Nanayakkara}, T. and {Napolitano}, L. and {Nonino}, M. and {Papovich}, C. and {Pentericci}, L. and {P{\'e}rez-Gonz{\'a}lez}, P.~G. and {Pirzkal}, N. and {Ravindranath}, S. and {Roberts-Borsani}, G. and {Somerville}, R.~S. and {Trenti}, M. and {Trump}, J.~R. and {Vulcani}, B. and {Wang}, X. and {Watson}, P.~J. and {Wilkins}, S.~M. and {Yang}, G. and {Yung}, L.~Y.~A.},
        title = "{ASTRODEEP-JWST: NIRCam-HST multi-band photometry and redshifts for half a million sources in six extragalactic deep fields}",
      journal = {\aap},
         year = 2024,
        month = nov,
       volume = {691},
          eid = {A240},
        pages = {A240},
          doi = {10.1051/0004-6361/202451409},
archivePrefix = {arXiv},
       eprint = {2409.00169},
 primaryClass = {astro-ph.GA},
       adsurl = {https://ui.adsabs.harvard.edu/abs/2024A&A...691A.240M}
}

@ARTICLE{regan2017,
       author = {{Regan}, John A. and {Visbal}, Eli and {Wise}, John H. and {Haiman}, Zolt{\'a}n and {Johansson}, Peter H. and {Bryan}, Greg L.},
        title = "{Rapid formation of massive black holes in close proximity to embryonic protogalaxies}",
      journal = {Nature Astronomy},
         year = 2017,
        month = mar,
       volume = {1},
          eid = {0075},
        pages = {0075},
          doi = {10.1038/s41550-017-0075},
archivePrefix = {arXiv},
       eprint = {1703.03805},
 primaryClass = {astro-ph.GA},
       adsurl = {https://ui.adsabs.harvard.edu/abs/2017NatAs...1E..75R}
}

@ARTICLE{ventura2023,
       author = {{Ventura}, Emanuele M. and {Trinca}, Alessandro and {Schneider}, Raffaella and {Graziani}, Luca and {Valiante}, Rosa and {Wyithe}, J. Stuart B.},
        title = "{The role of Pop III stars and early black holes in the 21cm signal from Cosmic Dawn}",
      journal = {\mnras},
         year = 2023,
        month = jan,
          doi = {10.1093/mnras/stad237},
archivePrefix = {arXiv},
       eprint = {2210.10281},
 primaryClass = {astro-ph.CO},
       adsurl = {https://ui.adsabs.harvard.edu/abs/2023MNRAS.tmp..252V}
}

@ARTICLE{trinca2023bh,
       author = {{Trinca}, Alessandro and {Schneider}, Raffaella and {Maiolino}, Roberto and {Valiante}, Rosa and {Graziani}, Luca and {Volonteri}, Marta},
        title = "{Seeking the growth of the first black hole seeds with JWST}",
      journal = {\mnras},
         year = 2023,
        month = mar,
       volume = {519},
       number = {3},
        pages = {4753-4764},
          doi = {10.1093/mnras/stac3768},
archivePrefix = {arXiv},
       eprint = {2211.01389},
 primaryClass = {astro-ph.GA},
       adsurl = {https://ui.adsabs.harvard.edu/abs/2023MNRAS.519.4753T}
}

@ARTICLE{lupi2019,
       author = {{Lupi}, Alessandro and {Volonteri}, Marta and {Decarli}, Roberto and {Bovino}, Stefano and {Silk}, Joseph and {Bergeron}, Jacqueline},
        title = "{High-redshift quasars and their host galaxies - I. Kinematical and dynamical properties and their tracers}",
      journal = {\mnras},
         year = 2019,
        month = sep,
       volume = {488},
       number = {3},
        pages = {4004-4022},
          doi = {10.1093/mnras/stz1959},
archivePrefix = {arXiv},
       eprint = {1901.02464},
 primaryClass = {astro-ph.GA},
       adsurl = {https://ui.adsabs.harvard.edu/abs/2019MNRAS.488.4004L}
}

@ARTICLE{regan2019,
       author = {{Regan}, John A. and {Downes}, Turlough P. and {Volonteri}, Marta and {Beckmann}, Ricarda and {Lupi}, Alessandro and {Trebitsch}, Maxime and {Dubois}, Yohan},
        title = "{Super-Eddington accretion and feedback from the first massive seed black holes}",
      journal = {\mnras},
         year = 2019,
        month = jul,
       volume = {486},
       number = {3},
        pages = {3892-3906},
          doi = {10.1093/mnras/stz1045},
archivePrefix = {arXiv},
       eprint = {1811.04953},
 primaryClass = {astro-ph.GA},
       adsurl = {https://ui.adsabs.harvard.edu/abs/2019MNRAS.486.3892R}
}

@ARTICLE{volonteri2008,
       author = {{Volonteri}, Marta and {Lodato}, Giuseppe and {Natarajan}, Priyamvada},
        title = "{The evolution of massive black hole seeds}",
      journal = {\mnras},
         year = 2008,
        month = jan,
       volume = {383},
       number = {3},
        pages = {1079-1088},
          doi = {10.1111/j.1365-2966.2007.12589.x},
archivePrefix = {arXiv},
       eprint = {0709.0529},
 primaryClass = {astro-ph},
       adsurl = {https://ui.adsabs.harvard.edu/abs/2008MNRAS.383.1079V}
}

@ARTICLE{dimatteo2017,
       author = {{Di Matteo}, Tiziana and {Croft}, Rupert A.~C. and {Feng}, Yu and
         {Waters}, Dacen and {Wilkins}, Stephen},
        title = "{The origin of the most massive black holes at high-z: BlueTides and the next quasar frontier}",
      journal = {\mnras},
         year = 2017,
        month = jun,
       volume = {467},
       number = {4},
        pages = {4243-4251},
          doi = {10.1093/mnras/stx319},
archivePrefix = {arXiv},
       eprint = {1606.08871},
 primaryClass = {astro-ph.GA},
       adsurl = {https://ui.adsabs.harvard.edu/abs/2017MNRAS.467.4243D}
}

@article{volonteri2010,
  title={Formation of supermassive black holes},
  author={Volonteri, Marta},
  journal={The Astronomy and Astrophysics Review},
  volume={18},
  number={3},
  pages={279--315},
  year={2010},
  publisher={Springer}
}

@ARTICLE{inayoshi2020,
       author = {{Inayoshi}, Kohei and {Visbal}, Eli and {Haiman}, Zolt{\'a}n},
        title = "{The Assembly of the First Massive Black Holes}",
      journal = {\araa},
         year = 2020,
        month = aug,
       volume = {58},
        pages = {27-97},
          doi = {10.1146/annurev-astro-120419-014455},
archivePrefix = {arXiv},
       eprint = {1911.05791},
 primaryClass = {astro-ph.GA},
       adsurl = {https://ui.adsabs.harvard.edu/abs/2020ARA&A..58...27I}
}

@article{valiante2016,
  title={From the first stars to the first black holes},
  author={Valiante, Rosa and Schneider, Raffaella and Volonteri, Marta and Omukai, Kazuyuki},
  journal={Monthly Notices of the Royal Astronomical Society},
  volume={457},
  number={3},
  pages={3356--3371},
  year={2016},
  publisher={Oxford University Press}
}

@article{omukai2008,
  title={Can supermassive black holes form in metal-enriched high-redshift protogalaxies?},
  author={Omukai, K and Schneider, R and Haiman, Z},
  journal={The Astrophysical Journal},
  volume={686},
  number={2},
  pages={801},
  year={2008},
  publisher={IOP Publishing}
}

@article{bromm2003,
  title={Formation of the first supermassive black holes},
  author={Bromm, Volker and Loeb, Abraham},
  journal={The Astrophysical Journal},
  volume={596},
  number={1},
  pages={34},
  year={2003},
  publisher={IOP Publishing}
}

@ARTICLE{wise2008,
       author = {{Wise}, John H. and {Turk}, Matthew J. and {Abel}, Tom},
        title = "{Resolving the Formation of Protogalaxies. II. Central Gravitational Collapse}",
      journal = {\apj},
         year = "2008",
        month = "Aug",
       volume = {682},
       number = {2},
        pages = {745-757},
          doi = {10.1086/588209},
archivePrefix = {arXiv},
       eprint = {0710.1678},
 primaryClass = {astro-ph},
       adsurl = {https://ui.adsabs.harvard.edu/abs/2008ApJ...682..745W}
}

@ARTICLE{regan2009,
       author = {{Regan}, John A. and {Haehnelt}, Martin G.},
        title = "{Pathways to massive black holes and compact star clusters in pre-galactic dark matter haloes with virial temperatures \&gt;\raisebox{-0.5ex}\textasciitilde10000K}",
      journal = {\mnras},
         year = "2009",
        month = "Jun",
       volume = {396},
       number = {1},
        pages = {343-353},
          doi = {10.1111/j.1365-2966.2009.14579.x},
archivePrefix = {arXiv},
       eprint = {0810.2802},
 primaryClass = {astro-ph},
       adsurl = {https://ui.adsabs.harvard.edu/abs/2009MNRAS.396..343R}
}

@ARTICLE{latif2013,
       author = {{Latif}, M.~A. and {Schleicher}, D.~R.~G. and {Schmidt}, W. and
         {Niemeyer}, J.~C.},
        title = "{The characteristic black hole mass resulting from direct collapse in the early Universe}",
      journal = {\mnras},
         year = "2013",
        month = "Dec",
       volume = {436},
       number = {4},
        pages = {2989-2996},
          doi = {10.1093/mnras/stt1786},
archivePrefix = {arXiv},
       eprint = {1309.1097},
 primaryClass = {astro-ph.CO},
       adsurl = {https://ui.adsabs.harvard.edu/abs/2013MNRAS.436.2989L}
}

@ARTICLE{chon2016,
       author = {{Chon}, Sunmyon and {Hirano}, Shingo and {Hosokawa}, Takashi and
         {Yoshida}, Naoki},
        title = "{Cosmological Simulations of Early Black Hole Formation: Halo Mergers, Tidal Disruption, and the Conditions for Direct Collapse}",
      journal = {\apj},
         year = "2016",
        month = "Dec",
       volume = {832},
       number = {2},
          eid = {134},
        pages = {134},
          doi = {10.3847/0004-637X/832/2/134},
archivePrefix = {arXiv},
       eprint = {1603.08923},
 primaryClass = {astro-ph.GA},
       adsurl = {https://ui.adsabs.harvard.edu/abs/2016ApJ...832..134C}
}

@ARTICLE{Wise2019,
       author = {{Wise}, John H. and {Regan}, John A. and {O'Shea}, Brian W. and {Norman}, Michael L. and {Downes}, Turlough P. and {Xu}, Hao},
        title = "{Formation of massive black holes in rapidly growing pre-galactic gas clouds}",
      journal = {\nat},
         year = 2019,
        month = jan,
       volume = {566},
       number = {7742},
        pages = {85-88},
          doi = {10.1038/s41586-019-0873-4},
archivePrefix = {arXiv},
       eprint = {1901.07563},
 primaryClass = {astro-ph.GA},
       adsurl = {https://ui.adsabs.harvard.edu/abs/2019Natur.566...85W}
}

@ARTICLE{mayer2015,
       author = {{Mayer}, Lucio and {Fiacconi}, Davide and {Bonoli}, Silvia and
         {Quinn}, Thomas and {Ro{\v{s}}kar}, Rok and {Shen}, Sijing and
         {Wadsley}, James},
        title = "{Direct Formation of Supermassive Black Holes in Metal-enriched Gas at the Heart of High-redshift Galaxy Mergers}",
      journal = {\apj},
         year = "2015",
        month = "Sep",
       volume = {810},
       number = {1},
          eid = {51},
        pages = {51},
          doi = {10.1088/0004-637X/810/1/51},
archivePrefix = {arXiv},
       eprint = {1411.5683},
 primaryClass = {astro-ph.GA},
       adsurl = {https://ui.adsabs.harvard.edu/abs/2015ApJ...810...51M}
}

@ARTICLE{pezzulli2016,
       author = {{Pezzulli}, Edwige and {Valiante}, Rosa and {Schneider}, Raffaella},
        title = "{Super-Eddington growth of the first black holes}",
      journal = {\mnras},
         year = "2016",
        month = "May",
       volume = {458},
       number = {3},
        pages = {3047-3059},
          doi = {10.1093/mnras/stw505},
archivePrefix = {arXiv},
       eprint = {1603.00475},
 primaryClass = {astro-ph.GA},
       adsurl = {https://ui.adsabs.harvard.edu/abs/2016MNRAS.458.3047P}
}

@article{pezzulli2017b,
  title={The sustainable growth of the first black holes},
  author={Pezzulli, Edwige and Volonteri, Marta and Schneider, Raffaella and Valiante, Rosa},
  journal={Monthly Notices of the Royal Astronomical Society},
  volume={471},
  number={1},
  pages={589--595},
  year={2017},
  publisher={Oxford University Press}
}

@article{valiante2011,
  title={The origin of the dust in high-redshift quasars: the case of SDSS J1148+ 5251},
  author={Valiante, Rosa and Schneider, Raffaella and Salvadori, Stefania and Bianchi, Simone},
  journal={Monthly Notices of the Royal Astronomical Society},
  volume={416},
  number={3},
  pages={1916--1935},
  year={2011},
  publisher={Blackwell Publishing Ltd Oxford, UK}
}

@ARTICLE{debennassuti2017,
       author = {{de Bennassuti}, M. and {Salvadori}, S. and {Schneider}, R. and
         {Valiante}, R. and {Omukai}, K.},
        title = "{Limits on Population III star formation with the most iron-poor stars}",
      journal = {\mnras},
         year = 2017,
        month = feb,
       volume = {465},
       number = {1},
        pages = {926-940},
          doi = {10.1093/mnras/stw2687},
archivePrefix = {arXiv},
       eprint = {1610.05777},
 primaryClass = {astro-ph.GA},
       adsurl = {https://ui.adsabs.harvard.edu/abs/2017MNRAS.465..926D}
}

@ARTICLE{heger2002,
       author = {{Heger}, A. and {Woosley}, S.~E.},
        title = "{The Nucleosynthetic Signature of Population III}",
      journal = {\apj},
         year = 2002,
        month = mar,
       volume = {567},
       number = {1},
        pages = {532-543},
          doi = {10.1086/338487},
archivePrefix = {arXiv},
       eprint = {astro-ph/0107037},
 primaryClass = {astro-ph},
       adsurl = {https://ui.adsabs.harvard.edu/abs/2002ApJ...567..532H}
}

@ARTICLE{omukai2001,
       author = {{Omukai}, Kazuyuki},
        title = "{Primordial Star Formation under Far-Ultraviolet Radiation}",
      journal = {\apj},
         year = "2001",
        month = "Jan",
       volume = {546},
       number = {2},
        pages = {635-651},
          doi = {10.1086/318296},
archivePrefix = {arXiv},
       eprint = {astro-ph/0011446},
 primaryClass = {astro-ph},
       adsurl = {https://ui.adsabs.harvard.edu/abs/2001ApJ...546..635O}
}

@ARTICLE{Xin2025,
       author = {{Xin}, Chengcheng and {Isi}, Maximiliano and {Farr}, Will M. and {Haiman}, Zolt{\'a}n},
        title = "{Identifying Compact Chirping SMBHBs in LSST using Bayesian Analysis}",
      journal = {arXiv e-prints},
         year = 2025,
        month = jun,
          eid = {arXiv:2506.10846},
        pages = {arXiv:2506.10846},
          doi = {10.48550/arXiv.2506.10846},
archivePrefix = {arXiv},
       eprint = {2506.10846},
 primaryClass = {astro-ph.HE},
       adsurl = {https://ui.adsabs.harvard.edu/abs/2025arXiv250610846X}
}

@ARTICLE{Mead2025,
       author = {{Mead}, Jennifer and {Brauer}, Kaley and {Bryan}, Greg L. and {Mac Low}, Mordecai-Mark and {Ji}, Alexander P. and {Wise}, John H. and {Emerick}, Andrew and {Andersson}, Eric P. and {Frebel}, Anna and {C{\^o}t{\'e}}, Benoit},
        title = "{AEOS: Transport of Metals from Minihalos following Population III Stellar Feedback}",
      journal = {\apj},
         year = 2025,
        month = feb,
       volume = {980},
       number = {1},
          eid = {62},
        pages = {62},
          doi = {10.3847/1538-4357/ada3c1},
archivePrefix = {arXiv},
       eprint = {2411.14209},
 primaryClass = {astro-ph.GA},
       adsurl = {https://ui.adsabs.harvard.edu/abs/2025ApJ...980...62M}
}

@ARTICLE{sakurai2020,
       author = {{Sakurai}, Yuya and {Haiman}, Zolt{\'a}n and {Inayoshi}, Kohei},
        title = "{Radiative feedback for supermassive star formation in a massive cloud with H$_{2}$ molecules in an atomic-cooling halo}",
      journal = {\mnras},
         year = 2020,
        month = dec,
       volume = {499},
       number = {4},
        pages = {5960-5971},
          doi = {10.1093/mnras/staa3227},
archivePrefix = {arXiv},
       eprint = {2009.02629},
 primaryClass = {astro-ph.GA},
       adsurl = {https://ui.adsabs.harvard.edu/abs/2020MNRAS.499.5960S}
}

@ARTICLE{toyouchi2023,
       author = {{Toyouchi}, Daisuke and {Inayoshi}, Kohei and {Li}, Wenxiu and {Haiman}, Zolt{\'a}n and {Kuiper}, Rolf},
        title = "{Radiative feedback on supermassive star formation: the massive end of the Population III initial mass function}",
      journal = {\mnras},
         year = 2023,
        month = jan,
       volume = {518},
       number = {2},
        pages = {1601-1616},
          doi = {10.1093/mnras/stac3191},
archivePrefix = {arXiv},
       eprint = {2206.14459},
 primaryClass = {astro-ph.GA},
       adsurl = {https://ui.adsabs.harvard.edu/abs/2023MNRAS.518.1601T}
}

@ARTICLE{khandai2012,
       author = {{Khandai}, Nishikanta and {Feng}, Yu and {DeGraf}, Colin and {Di Matteo}, Tiziana and {Croft}, Rupert A.~C.},
        title = "{The formation of galaxies hosting z {\ensuremath{\sim}} 6 quasars}",
      journal = {\mnras},
         year = 2012,
        month = jul,
       volume = {423},
       number = {3},
        pages = {2397-2406},
          doi = {10.1111/j.1365-2966.2012.21047.x},
archivePrefix = {arXiv},
       eprint = {1111.0692},
 primaryClass = {astro-ph.CO},
       adsurl = {https://ui.adsabs.harvard.edu/abs/2012MNRAS.423.2397K}
}

@ARTICLE{Matsuoka2024,
       author = {{Matsuoka}, Yoshiki and {Izumi}, Takuma and {Onoue}, Masafusa and {Strauss}, Michael A. and {Iwasawa}, Kazushi and {Kashikawa}, Nobunari and {Akiyama}, Masayuki and {Aoki}, Kentaro and {Arita}, Junya and {Imanishi}, Masatoshi and {Ishimoto}, Rikako and {Kawaguchi}, Toshihiro and {Kohno}, Kotaro and {Lee}, Chien-Hsiu and {Nagao}, Tohru and {Silverman}, John D. and {Toba}, Yoshiki},
        title = "{Discovery of Merging Twin Quasars at z = 6.05}",
      journal = {\apjl},
         year = 2024,
        month = apr,
       volume = {965},
       number = {1},
          eid = {L4},
        pages = {L4},
          doi = {10.3847/2041-8213/ad35c7},
archivePrefix = {arXiv},
       eprint = {2405.02465},
 primaryClass = {astro-ph.GA},
       adsurl = {https://ui.adsabs.harvard.edu/abs/2024ApJ...965L...4M}
}

@ARTICLE{scoggins2025,
       author = {{Scoggins}, Matthew T. and {Haiman}, Zoltan and {Pacucci}, Fabio},
        title = "{Heavy black hole seed survivors in dwarf galaxies: a case study of Leo I}",
      journal = {arXiv e-prints},
         year = 2025,
        month = nov,
          eid = {arXiv:2511.04736},
        pages = {arXiv:2511.04736},
          doi = {10.48550/arXiv.2511.04736},
archivePrefix = {arXiv},
       eprint = {2511.04736},
 primaryClass = {astro-ph.GA},
       adsurl = {https://ui.adsabs.harvard.edu/abs/2025arXiv251104736S}
}

@ARTICLE{Wolcott-Green2017,
       author = {{Wolcott-Green}, J. and {Haiman}, Z. and {Bryan}, G.~L.},
        title = "{Beyond J$_{crit}$: a critical curve for suppression of H$_{2}$-cooling in protogalaxies}",
      journal = {\mnras},
         year = 2017,
        month = aug,
       volume = {469},
       number = {3},
        pages = {3329-3336},
          doi = {10.1093/mnras/stx167},
archivePrefix = {arXiv},
       eprint = {1609.02142},
 primaryClass = {astro-ph.GA},
       adsurl = {https://ui.adsabs.harvard.edu/abs/2017MNRAS.469.3329W}
}

@ARTICLE{Regan2020b,
       author = {{Regan}, John A. and {Wise}, John H. and {O'Shea}, Brian W. and {Norman}, Michael L.},
        title = "{The emergence of the first star-free atomic cooling haloes in the Universe}",
      journal = {\mnras},
         year = 2020,
        month = feb,
       volume = {492},
       number = {2},
        pages = {3021-3031},
          doi = {10.1093/mnras/staa035},
archivePrefix = {arXiv},
       eprint = {1908.02823},
 primaryClass = {astro-ph.GA},
       adsurl = {https://ui.adsabs.harvard.edu/abs/2020MNRAS.492.3021R}
}

@ARTICLE{chon2020,
       author = {{Chon}, Sunmyon and {Omukai}, Kazuyuki},
        title = "{Supermassive star formation via super competitive accretion in slightly metal-enriched clouds}",
      journal = {\mnras},
         year = 2020,
        month = may,
       volume = {494},
       number = {2},
        pages = {2851-2860},
          doi = {10.1093/mnras/staa863},
archivePrefix = {arXiv},
       eprint = {2001.06491},
 primaryClass = {astro-ph.GA},
       adsurl = {https://ui.adsabs.harvard.edu/abs/2020MNRAS.494.2851C}
}

@ARTICLE{bondi1952,
       author = {{Bondi}, H.},
        title = "{On spherically symmetrical accretion}",
      journal = {\mnras},
         year = 1952,
        month = jan,
       volume = {112},
        pages = {195},
          doi = {10.1093/mnras/112.2.195},
       adsurl = {https://ui.adsabs.harvard.edu/abs/1952MNRAS.112..195B}
}

@ARTICLE{hoyle1941,
       author = {{Hoyle}, F. and {Lyttleton}, R.~A.},
        title = "{On the accretion theory of stellar evolution}",
      journal = {\mnras},
         year = 1941,
        month = jan,
       volume = {101},
        pages = {227},
          doi = {10.1093/mnras/101.4.227},
       adsurl = {https://ui.adsabs.harvard.edu/abs/1941MNRAS.101..227H}
}

@ARTICLE{ahn2009,
       author = {{Ahn}, Kyungjin and {Shapiro}, Paul R. and {Iliev}, Ilian T. and
         {Mellema}, Garrelt and {Pen}, Ue-Li},
        title = "{The Inhomogeneous Background Of H$_{2}$-Dissociating Radiation During Cosmic Reionization}",
      journal = {\apj},
         year = 2009,
        month = apr,
       volume = {695},
       number = {2},
        pages = {1430-1445},
          doi = {10.1088/0004-637X/695/2/1430},
archivePrefix = {arXiv},
       eprint = {0807.2254},
 primaryClass = {astro-ph},
       adsurl = {https://ui.adsabs.harvard.edu/abs/2009ApJ...695.1430A}
}

@ARTICLE{habouzit2016,
       author = {{Habouzit}, M{\'e}lanie and {Volonteri}, Marta and {Latif}, Muhammad and
         {Dubois}, Yohan and {Peirani}, S{\'e}bastien},
        title = "{On the number density of `direct collapse' black hole seeds}",
      journal = {\mnras},
         year = "2016",
        month = "Nov",
       volume = {463},
       number = {1},
        pages = {529-540},
          doi = {10.1093/mnras/stw1924},
archivePrefix = {arXiv},
       eprint = {1601.00557},
 primaryClass = {astro-ph.GA},
       adsurl = {https://ui.adsabs.harvard.edu/abs/2016MNRAS.463..529H}
}

@ARTICLE{valiante2018observability,
       author = {{Valiante}, Rosa and {Schneider}, Raffaella and {Zappacosta}, Luca and
         {Graziani}, Luca and {Pezzulli}, Edwige and {Volonteri}, Marta},
        title = "{Chasing the observational signatures of seed black holes at z \&gt; 7: candidate observability}",
      journal = {\mnras},
         year = "2018",
        month = "May",
       volume = {476},
       number = {1},
        pages = {407-420},
          doi = {10.1093/mnras/sty213},
       adsurl = {https://ui.adsabs.harvard.edu/abs/2018MNRAS.476..407V}
}

@ARTICLE{valiante2018statistics,
       author = {{Valiante}, Rosa and {Schneider}, Raffaella and {Graziani}, Luca and
         {Zappacosta}, Luca},
        title = "{Chasing the observational signatures of seed black holes at z \&gt; 7: candidate statistics}",
      journal = {\mnras},
         year = "2018",
        month = "Mar",
       volume = {474},
       number = {3},
        pages = {3825-3834},
          doi = {10.1093/mnras/stx3028},
archivePrefix = {arXiv},
       eprint = {1801.08165},
 primaryClass = {astro-ph.GA},
       adsurl = {https://ui.adsabs.harvard.edu/abs/2018MNRAS.474.3825V}
}

@ARTICLE{wolcottgreen2017,
       author = {{Wolcott-Green}, J. and {Haiman}, Z. and {Bryan}, G.~L.},
        title = "{Beyond J$_{crit}$: a critical curve for suppression of H$_{2}$-cooling in protogalaxies}",
      journal = {\mnras},
         year = "2017",
        month = "Aug",
       volume = {469},
       number = {3},
        pages = {3329-3336},
          doi = {10.1093/mnras/stx167},
archivePrefix = {arXiv},
       eprint = {1609.02142},
 primaryClass = {astro-ph.GA},
       adsurl = {https://ui.adsabs.harvard.edu/abs/2017MNRAS.469.3329W}
}

@ARTICLE{Juodzbalis2025,
       author = {{Juod{\v{z}}balis}, Ignas and {Marconcini}, Cosimo and {D'Eugenio}, Francesco and {Maiolino}, Roberto and {Marconi}, Alessandro and {{\"U}bler}, Hannah and {Scholtz}, Jan and {Ji}, Xihan and {Arribas}, Santiago and {Bennett}, Jake S. and {Bromm}, Volker and {Bunker}, Andrew J. and {Carniani}, Stefano and {Charlot}, St{\'e}phane and {Cresci}, Giovanni and {Dayal}, Pratika and {Egami}, Eiichi and {Fabian}, Andrew and {Inayoshi}, Kohei and {Isobe}, Yuki and {Ivey}, Lucy and {Jones}, Gareth C. and {Koudmani}, Sophie and {Laporte}, Nicolas and {Liu}, Boyuan and {Lyu}, Jianwei and {Mazzolari}, Giovanni and {Monty}, Stephanie and {Parlanti}, Eleonora and {P{\'e}rez-Gonz{\'a}lez}, Pablo G. and {Perna}, Michele and {Robertson}, Brant and {Schneider}, Raffaella and {Sijacki}, Debora and {Tacchella}, Sandro and {Trinca}, Alessandro and {Valiante}, Rosa and {Volonteri}, Marta and {Witstok}, Joris and {Zhang}, Saiyang},
        title = "{A direct black hole mass measurement in a Little Red Dot at the Epoch of Reionization}",
      journal = {arXiv e-prints},
         year = 2025,
        month = aug,
          eid = {arXiv:2508.21748},
        pages = {arXiv:2508.21748},
          doi = {10.48550/arXiv.2508.21748},
archivePrefix = {arXiv},
       eprint = {2508.21748},
 primaryClass = {astro-ph.GA},
       adsurl = {https://ui.adsabs.harvard.edu/abs/2025arXiv250821748J}
}

@ARTICLE{Lin2024,
       author = {{Lin}, Xiaojing and {Wang}, Feige and {Fan}, Xiaohui and {Cai}, Zheng and {Champagne}, Jaclyn B. and {Sun}, Fengwu and {Volonteri}, Marta and {Yang}, Jinyi and {Hennawi}, Joseph F. and {Ba{\~n}ados}, Eduardo and {Barth}, Aaron and {Eilers}, Anna-Christina and {Farina}, Emanuele Paolo and {Liu}, Weizhe and {Jin}, Xiangyu and {Jun}, Hyunsung D. and {Lupi}, Alessandro and {Kakiichi}, Koki and {Mazzucchelli}, Chiara and {Onoue}, Masafusa and {Pan}, Zhiwei and {Pizzati}, Elia and {Rojas-Ruiz}, Sof{\'\i}a and {Schindler}, Jan-Torge and {Trakhtenbrot}, Benny and {Shen}, Yue and {Trebitsch}, Maxime and {Zhuang}, Ming-Yang and {Endsley}, Ryan and {Meyer}, Romain A. and {Li}, Zihao and {Li}, Mingyu and {Pudoka}, Maria and {Tee}, Wei Leong and {Wu}, Yunjing and {Zhang}, Haowen},
        title = "{A SPectroscopic Survey of Biased Halos In the Reionization Era (ASPIRE): Broad-line AGN at z = 4‑5 Revealed by JWST/NIRCam WFSS}",
      journal = {\apj},
         year = 2024,
        month = oct,
       volume = {974},
       number = {1},
          eid = {147},
        pages = {147},
          doi = {10.3847/1538-4357/ad6565},
archivePrefix = {arXiv},
       eprint = {2407.17570},
 primaryClass = {astro-ph.GA},
       adsurl = {https://ui.adsabs.harvard.edu/abs/2024ApJ...974..147L}
}

@ARTICLE{vanWassenhove2010,
       author = {{van Wassenhove}, S. and {Volonteri}, M. and {Walker}, M.~G. and {Gair}, J.~R.},
        title = "{Massive black holes lurking in Milky Way satellites}",
      journal = {\mnras},
         year = 2010,
        month = oct,
       volume = {408},
       number = {2},
        pages = {1139-1146},
          doi = {10.1111/j.1365-2966.2010.17189.x},
archivePrefix = {arXiv},
       eprint = {1001.5451},
 primaryClass = {astro-ph.CO},
       adsurl = {https://ui.adsabs.harvard.edu/abs/2010MNRAS.408.1139V}
}

@ARTICLE{wolcottgreen2020suppression,
       author = {{Wolcott-Green}, Jemma and {Haiman}, Zolt{\'a}n and {Bryan}, Greg L.},
        title = "{Suppression of H$_{2}$-cooling in protogalaxies aided by trapped Ly{\ensuremath{\alpha}} cooling radiation}",
      journal = {\mnras},
         year = 2021,
        month = jan,
       volume = {500},
       number = {1},
        pages = {138-144},
          doi = {10.1093/mnras/staa3057},
archivePrefix = {arXiv},
       eprint = {2001.05498},
 primaryClass = {astro-ph.GA},
       adsurl = {https://ui.adsabs.harvard.edu/abs/2021MNRAS.500..138W}
}

@article{shang2010supermassive,
  title={Supermassive black hole formation by direct collapse: keeping protogalactic gas H2 free in dark matter haloes with virial temperatures T vir> rsim 104 K},
  author={Shang, Cien and Bryan, Greg L and Haiman, Zoltan},
  journal={Monthly Notices of the Royal Astronomical Society},
  volume={402},
  number={2},
  pages={1249--1262},
  year={2010},
  publisher={Blackwell Publishing Ltd Oxford, UK}
}

@article{larson1998early,
  title={Early star formation and the evolution of the stellar initial mass function in galaxies},
  author={Larson, Richard B},
  journal={Monthly Notices of the Royal Astronomical Society},
  volume={301},
  number={2},
  pages={569--581},
  year={1998},
  publisher={Wiley Online Library}
}

@article{sugimura2014critical,
  title={The critical radiation intensity for direct collapse black hole formation: dependence on the radiation spectral shape},
  author={Sugimura, Kazuyuki and Omukai, Kazuyuki and Inoue, Akio K},
  journal={Monthly Notices of the Royal Astronomical Society},
  volume={445},
  number={1},
  pages={544--553},
  year={2014},
  publisher={The Royal Astronomical Society}
}

@article{dijkstra2008fluctuations,
  title={Fluctuations in the high-redshift Lyman--Werner background: close halo pairs as the origin of supermassive black holes},
  author={Dijkstra, Mark and Haiman, Zolt{\'a}n and Mesinger, Andrei and Wyithe, J Stuart B},
  journal={Monthly Notices of the Royal Astronomical Society},
  volume={391},
  number={4},
  pages={1961--1972},
  year={2008},
  publisher={Blackwell Publishing Ltd Oxford, UK}
}

@article{begelman2006formation,
  title={Formation of supermassive black holes by direct collapse in pre-galactic haloes},
  author={Begelman, Mitchell C and Volonteri, Marta and Rees, Martin J},
  journal={Monthly Notices of the Royal Astronomical Society},
  volume={370},
  number={1},
  pages={289--298},
  year={2006},
  publisher={Blackwell Publishing Ltd Oxford, UK}
}

@ARTICLE{dimatteo2012,
       author = {{Di Matteo}, T. and {Khandai}, N. and {DeGraf}, C. and {Feng}, Y. and
         {Croft}, R.~A.~C. and {Lopez}, J. and {Springel}, V.},
        title = "{Cold Flows and the First Quasars}",
      journal = {\apjl},
         year = 2012,
        month = feb,
       volume = {745},
       number = {2},
          eid = {L29},
        pages = {L29},
          doi = {10.1088/2041-8205/745/2/L29},
archivePrefix = {arXiv},
       eprint = {1107.1253},
 primaryClass = {astro-ph.CO},
       adsurl = {https://ui.adsabs.harvard.edu/abs/2012ApJ...745L..29D}
}

@ARTICLE{schaye2015,
       author = {{Schaye}, Joop and {Crain}, Robert A. and {Bower}, Richard G. and
         {Furlong}, Michelle and {Schaller}, Matthieu and {Theuns}, Tom and
         {Dalla Vecchia}, Claudio and {Frenk}, Carlos S. and {McCarthy}, I.~G. and
         {Helly}, John C. and {Jenkins}, Adrian and {Rosas-Guevara}, Y.~M. and
         {White}, Simon D.~M. and {Baes}, Maarten and {Booth}, C.~M. and
         {Camps}, Peter and {Navarro}, Julio F. and {Qu}, Yan and
         {Rahmati}, Alireza and {Sawala}, Till and {Thomas}, Peter A. and
         {Trayford}, James},
        title = "{The EAGLE project: simulating the evolution and assembly of galaxies and their environments}",
      journal = {\mnras},
         year = 2015,
        month = jan,
       volume = {446},
       number = {1},
        pages = {521-554},
          doi = {10.1093/mnras/stu2058},
archivePrefix = {arXiv},
       eprint = {1407.7040},
 primaryClass = {astro-ph.GA},
       adsurl = {https://ui.adsabs.harvard.edu/abs/2015MNRAS.446..521S}
}

@ARTICLE{sassano2021,
       author = {{Sassano}, Federica and {Schneider}, Raffaella and {Valiante}, Rosa and {Inayoshi}, Kohei and {Chon}, Sunmyon and {Omukai}, Kazuyuki and {Mayer}, Lucio and {Capelo}, Pedro R.},
        title = "{Light, medium-weight, or heavy? The nature of the first supermassive black hole seeds}",
      journal = {\mnras},
         year = 2021,
        month = sep,
       volume = {506},
       number = {1},
        pages = {613-632},
          doi = {10.1093/mnras/stab1737},
archivePrefix = {arXiv},
       eprint = {2106.08330},
 primaryClass = {astro-ph.GA},
       adsurl = {https://ui.adsabs.harvard.edu/abs/2021MNRAS.506..613S}
}

@ARTICLE{lupi2021,
       author = {{Lupi}, Alessandro and {Haiman}, Zolt{\'a}n and {Volonteri}, Marta},
        title = "{Forming massive seed black holes in high-redshift quasar host progenitors}",
      journal = {\mnras},
         year = 2021,
        month = may,
       volume = {503},
       number = {4},
        pages = {5046-5060},
          doi = {10.1093/mnras/stab692},
archivePrefix = {arXiv},
       eprint = {2102.05051},
 primaryClass = {astro-ph.GA},
       adsurl = {https://ui.adsabs.harvard.edu/abs/2021MNRAS.503.5046L}
}

@ARTICLE{sheth2001,
       author = {{Sheth}, Ravi K. and {Mo}, H.~J. and {Tormen}, Giuseppe},
        title = "{Ellipsoidal collapse and an improved model for the number and spatial distribution of dark matter haloes}",
      journal = {\mnras},
         year = 2001,
        month = may,
       volume = {323},
       number = {1},
        pages = {1-12},
          doi = {10.1046/j.1365-8711.2001.04006.x},
archivePrefix = {arXiv},
       eprint = {astro-ph/9907024},
 primaryClass = {astro-ph},
       adsurl = {https://ui.adsabs.harvard.edu/abs/2001MNRAS.323....1S}
}

@ARTICLE{lodato2006,
       author = {{Lodato}, Giuseppe and {Natarajan}, Priyamvada},
        title = "{Supermassive black hole formation during the assembly of pre-galactic discs}",
      journal = {\mnras},
         year = 2006,
        month = oct,
       volume = {371},
       number = {4},
        pages = {1813-1823},
          doi = {10.1111/j.1365-2966.2006.10801.x},
archivePrefix = {arXiv},
       eprint = {astro-ph/0606159},
 primaryClass = {astro-ph},
       adsurl = {https://ui.adsabs.harvard.edu/abs/2006MNRAS.371.1813L}
}

@ARTICLE{wang2008,
       author = {{Wang}, Peng and {Abel}, Tom},
        title = "{Dynamical Treatment of Virialization Heating in Galaxy Formation}",
      journal = {\apj},
         year = 2008,
        month = jan,
       volume = {672},
       number = {2},
        pages = {752-756},
          doi = {10.1086/523623},
       adsurl = {https://ui.adsabs.harvard.edu/abs/2008ApJ...672..752W}
}

@ARTICLE{haardt1996,
       author = {{Haardt}, Francesco and {Madau}, Piero},
        title = "{Radiative Transfer in a Clumpy Universe. II. The Ultraviolet Extragalactic Background}",
      journal = {\apj},
         year = 1996,
        month = apr,
       volume = {461},
        pages = {20},
          doi = {10.1086/177035},
archivePrefix = {arXiv},
       eprint = {astro-ph/9509093},
 primaryClass = {astro-ph},
       adsurl = {https://ui.adsabs.harvard.edu/abs/1996ApJ...461...20H}
}

@ARTICLE{chon2022,
       author = {{Chon}, Sunmyon and {Ono}, Haruka and {Omukai}, Kazuyuki and {Schneider}, Raffaella},
        title = "{Impact of the cosmic background radiation on the initial mass function of metal-poor stars}",
      journal = {\mnras},
         year = 2022,
        month = aug,
       volume = {514},
       number = {3},
        pages = {4639-4654},
          doi = {10.1093/mnras/stac1549},
archivePrefix = {arXiv},
       eprint = {2205.15328},
 primaryClass = {astro-ph.GA},
       adsurl = {https://ui.adsabs.harvard.edu/abs/2022MNRAS.514.4639C}
}

@ARTICLE{trinca2022,
       author = {{Trinca}, Alessandro and {Schneider}, Raffaella and {Valiante}, Rosa and {Graziani}, Luca and {Zappacosta}, Luca and {Shankar}, Francesco},
        title = "{The low-end of the black hole mass function at cosmic dawn}",
      journal = {\mnras},
         year = 2022,
        month = mar,
       volume = {511},
       number = {1},
        pages = {616-640},
          doi = {10.1093/mnras/stac062},
archivePrefix = {arXiv},
       eprint = {2201.02630},
 primaryClass = {astro-ph.GA},
       adsurl = {https://ui.adsabs.harvard.edu/abs/2022MNRAS.511..616T}
}

@ARTICLE{valiante2021,
       author = {{Valiante}, Rosa and {Colpi}, Monica and {Schneider}, Raffaella and {Mangiagli}, Alberto and {Bonetti}, Matteo and {Cerini}, Giulia and {Fairhurst}, Stephen and {Haardt}, Francesco and {Mills}, Cameron and {Sesana}, Alberto},
        title = "{Unveiling early black hole growth with multifrequency gravitational wave observations}",
      journal = {\mnras},
         year = 2021,
        month = jan,
       volume = {500},
       number = {3},
        pages = {4095-4109},
          doi = {10.1093/mnras/staa3395},
archivePrefix = {arXiv},
       eprint = {2010.15096},
 primaryClass = {astro-ph.GA},
       adsurl = {https://ui.adsabs.harvard.edu/abs/2021MNRAS.500.4095V}
}

@ARTICLE{fan2019,
       author = {{Fan}, Xiaohui and {Wang}, Feige and {Yang}, Jinyi and {Keeton}, Charles R. and {Yue}, Minghao and {Zabludoff}, Ann and {Bian}, Fuyan and {Bonaglia}, Marco and {Georgiev}, Iskren Y. and {Hennawi}, Joseph F. and {Li}, Jiangtao and {McGreer}, Ian D. and {Naidu}, Rohan and {Pacucci}, Fabio and {Rabien}, Sebastian and {Thompson}, David and {Venemans}, Bram and {Walter}, Fabian and {Wang}, Ran and {Wu}, Xue-Bing},
        title = "{The Discovery of a Gravitationally Lensed Quasar at z = 6.51}",
      journal = {\apjl},
         year = 2019,
        month = jan,
       volume = {870},
       number = {2},
          eid = {L11},
        pages = {L11},
          doi = {10.3847/2041-8213/aaeffe},
archivePrefix = {arXiv},
       eprint = {1810.11924},
 primaryClass = {astro-ph.GA},
       adsurl = {https://ui.adsabs.harvard.edu/abs/2019ApJ...870L..11F}
}

@ARTICLE{sassano2023,
       author = {{Sassano}, Federica and {Capelo}, Pedro R. and {Mayer}, Lucio and {Schneider}, Raffaella and {Valiante}, Rosa},
        title = "{Super-critical accretion of medium-weight seed black holes in gaseous proto-galactic nuclei}",
      journal = {\mnras},
         year = 2023,
        month = feb,
       volume = {519},
       number = {2},
        pages = {1837-1855},
          doi = {10.1093/mnras/stac3608},
archivePrefix = {arXiv},
       eprint = {2204.10330},
 primaryClass = {astro-ph.GA},
       adsurl = {https://ui.adsabs.harvard.edu/abs/2023MNRAS.519.1837S}
}

@ARTICLE{Naidu2025,
       author = {{Naidu}, Rohan P. and {Matthee}, Jorryt and {Katz}, Harley and {de Graaff}, Anna and {Oesch}, Pascal and {Smith}, Aaron and {Greene}, Jenny E. and {Brammer}, Gabriel and {Weibel}, Andrea and {Hviding}, Raphael and {Chisholm}, John and {Labb\textbackslash'e}, Ivo and {Simcoe}, Robert A. and {Witten}, Callum and {Atek}, Hakim and {Baggen}, Josephine F.~W. and {Belli}, Sirio and {Bezanson}, Rachel and {Boogaard}, Leindert A. and {Bose}, Sownak and {Covelo-Paz}, Alba and {Dayal}, Pratika and {Fudamoto}, Yoshinobu and {Furtak}, Lukas J. and {Giovinazzo}, Emma and {Goulding}, Andy and {Gronke}, Max and {Heintz}, Kasper E. and {Hirschmann}, Michaela and {Illingworth}, Garth and {Inoue}, Akio K. and {Johnson}, Benjamin D. and {Leja}, Joel and {Leonova}, Ecaterina and {McConachie}, Ian and {Maseda}, Michael V. and {Natarajan}, Priyamvada and {Nelson}, Erica and {Setton}, David J. and {Shivaei}, Irene and {Sobral}, David and {Stefanon}, Mauro and {Tacchella}, Sandro and {Toft}, Sune and {Torralba}, Alberto and {van Dokkum}, Pieter and {van der Wel}, Arjen and {Volonteri}, Marta and {Walter}, Fabian and {Wang}, Bingjie and {Watson}, Darach},
        title = "{A ``Black Hole Star'' Reveals the Remarkable Gas-Enshrouded Hearts of the Little Red Dots}",
      journal = {arXiv e-prints},
         year = 2025,
        month = mar,
          eid = {arXiv:2503.16596},
        pages = {arXiv:2503.16596},
          doi = {10.48550/arXiv.2503.16596},
archivePrefix = {arXiv},
       eprint = {2503.16596},
 primaryClass = {astro-ph.GA},
       adsurl = {https://ui.adsabs.harvard.edu/abs/2025arXiv250316596N}
}

@ARTICLE{Maiolino2025,
       author = {{Maiolino}, Roberto and {Uebler}, Hannah and {D'Eugenio}, Francesco and {Scholtz}, Jan and {Juodzbalis}, Ignas and {Ji}, Xihan and {Perna}, Michele and {Bromm}, Volker and {Dayal}, Pratika and {Koudmani}, Sophie and {Liu}, Boyuan and {Schneider}, Raffaella and {Sijacki}, Debora and {Valiante}, Rosa and {Trinca}, Alessandro and {Zhang}, Saiyang and {Volonteri}, Marta and {Inayoshi}, Kohei and {Carniani}, Stefano and {Nakajima}, Kimihiko and {Isobe}, Yuki and {Witstok}, Joris and {Jones}, Gareth C. and {Tacchella}, Sandro and {Arribas}, Santiago and {Bunker}, Andrew and {Cataldi}, Elisa and {Charlot}, Stephane and {Cresci}, Giovanni and {Curti}, Mirko and {Fabian}, Andrew C. and {Katz}, Harley and {Kumari}, Nimisha and {Laporte}, Nicolas and {Mazzolari}, Giovanni and {Robertson}, Brant and {Sun}, Fengwu and {Rodriguez Del Pino}, Bruno and {Venturi}, Giacomo},
        title = "{A black hole in a near-pristine galaxy 700 million years after the Big Bang}",
      journal = {arXiv e-prints},
         year = 2025,
        month = may,
          eid = {arXiv:2505.22567},
        pages = {arXiv:2505.22567},
          doi = {10.48550/arXiv.2505.22567},
archivePrefix = {arXiv},
       eprint = {2505.22567},
 primaryClass = {astro-ph.GA},
       adsurl = {https://ui.adsabs.harvard.edu/abs/2025arXiv250522567M}
}

@ARTICLE{Obrennan2025,
       author = {{O'Brennan}, Hannah and {Regan}, John A. and {Brennan}, John and {McCaffrey}, Joe and {Wise}, John H. and {Visbal}, Eli and {Trinca}, Alessandro and {Norman}, Michael L.},
        title = "{Predicting the number density of heavy seed massive black holes due to an intense Lyman-Werner field}",
      journal = {arXiv e-prints},
         year = 2025,
        month = feb,
          eid = {arXiv:2502.00574},
        pages = {arXiv:2502.00574},
          doi = {10.48550/arXiv.2502.00574},
archivePrefix = {arXiv},
       eprint = {2502.00574},
 primaryClass = {astro-ph.CO},
       adsurl = {https://ui.adsabs.harvard.edu/abs/2025arXiv250200574O}
}

@ARTICLE{massonneau2023,
       author = {{Massonneau}, Warren and {Dubois}, Yohan and {Volonteri}, Marta and {Beckmann}, Ricarda S.},
        title = "{How the super-Eddington regime affects black hole spin evolution in high-redshift galaxies}",
      journal = {\aap},
         year = 2023,
        month = jan,
       volume = {669},
          eid = {A143},
        pages = {A143},
          doi = {10.1051/0004-6361/202244874},
archivePrefix = {arXiv},
       eprint = {2209.01369},
 primaryClass = {astro-ph.GA},
       adsurl = {https://ui.adsabs.harvard.edu/abs/2023A&A...669A.143M}
}

@ARTICLE{degraaff2025,
       author = {{de Graaff}, Anna and {Rix}, Hans-Walter and {Naidu}, Rohan P. and {Labbe}, Ivo and {Wang}, Bingjie and {Leja}, Joel and {Matthee}, Jorryt and {Katz}, Harley and {Greene}, Jenny E. and {Hviding}, Raphael E. and {Baggen}, Josephine and {Bezanson}, Rachel and {Boogaard}, Leindert A. and {Brammer}, Gabriel and {Dayal}, Pratika and {van Dokkum}, Pieter and {Goulding}, Andy D. and {Hirschmann}, Michaela and {Maseda}, Michael V. and {McConachie}, Ian and {Miller}, Tim B. and {Nelson}, Erica and {Oesch}, Pascal A. and {Setton}, David J. and {Shivaei}, Irene and {Weibel}, Andrea and {Whitaker}, Katherine E. and {Williams}, Christina C.},
        title = "{A remarkable Ruby: Absorption in dense gas, rather than evolved stars, drives the extreme Balmer break of a Little Red Dot at $z=3.5$}",
      journal = {arXiv e-prints},
         year = 2025,
        month = mar,
          eid = {arXiv:2503.16600},
        pages = {arXiv:2503.16600},
          doi = {10.48550/arXiv.2503.16600},
archivePrefix = {arXiv},
       eprint = {2503.16600},
 primaryClass = {astro-ph.GA},
       adsurl = {https://ui.adsabs.harvard.edu/abs/2025arXiv250316600D}
}

@ARTICLE{Metha2026,
       author = {{Mehta}, Daxal H. and {Regan}, John A. and {Prole}, Lewis},
        title = "{The growth of light seed black holes in the early Universe}",
      journal = {Nature Astronomy},
         year = 2026,
        month = jan,
          doi = {10.1038/s41550-025-02767-5},
archivePrefix = {arXiv},
       eprint = {2601.14395},
 primaryClass = {astro-ph.GA},
       adsurl = {https://ui.adsabs.harvard.edu/abs/2026NatAs.tmp...21M}
}

@ARTICLE{Mehta2024,
       author = {{Mehta}, Daxal and {Regan}, John A. and {Prole}, Lewis},
        title = "{Growth of Light-Seed Black Holes in Gas-Rich Galaxies at High Redshift}",
      journal = {The Open Journal of Astrophysics},
         year = 2024,
        month = nov,
       volume = {7},
          eid = {107},
        pages = {107},
          doi = {10.33232/001c.126629},
archivePrefix = {arXiv},
       eprint = {2409.08326},
 primaryClass = {astro-ph.GA},
       adsurl = {https://ui.adsabs.harvard.edu/abs/2024OJAp....7E.107M}
}

@ARTICLE{Shi2024b,
       author = {{Shi}, Yanlong and {Kremer}, Kyle and {Hopkins}, Philip F.},
        title = "{Feedback-regulated seed black hole growth in star-forming molecular clouds and galactic nuclei}",
      journal = {\aap},
         year = 2024,
        month = nov,
       volume = {691},
          eid = {A24},
        pages = {A24},
          doi = {10.1051/0004-6361/202450964},
archivePrefix = {arXiv},
       eprint = {2405.12164},
 primaryClass = {astro-ph.GA},
       adsurl = {https://ui.adsabs.harvard.edu/abs/2024A&A...691A..24S}
}

@ARTICLE{Shi2026,
       author = {{Shi}, Yanlong and {Murray}, Norman},
        title = "{The in-situ growth of stellar-mass ``light'' seed black holes in nuclear star clusters}",
      journal = {arXiv e-prints},
         year = 2026,
        month = mar,
          eid = {arXiv:2603.10581},
        pages = {arXiv:2603.10581},
          doi = {10.48550/arXiv.2603.10581},
archivePrefix = {arXiv},
       eprint = {2603.10581},
 primaryClass = {astro-ph.GA},
       adsurl = {https://ui.adsabs.harvard.edu/abs/2026arXiv260310581S}
}

@ARTICLE{Prole2024b,
       author = {{Prole}, Lewis R. and {Regan}, John A. and {Glover}, Simon C.~O. and {Klessen}, Ralf S. and {Priestley}, Felix D. and {Clark}, Paul C.},
        title = "{Heavy black hole seed formation in high-z atomic cooling halos}",
      journal = {\aap},
         year = 2024,
        month = may,
       volume = {685},
          eid = {A31},
        pages = {A31},
          doi = {10.1051/0004-6361/202348903},
archivePrefix = {arXiv},
       eprint = {2312.06769},
 primaryClass = {astro-ph.GA},
       adsurl = {https://ui.adsabs.harvard.edu/abs/2024A&A...685A..31P}
}

@ARTICLE{Yanagisawa2026,
       author = {{Yanagisawa}, Hiroto and {Ouchi}, Masami and {Golubchik}, Miriam and {Oguri}, Masamune and {Fujimoto}, Seiji and {Kokorev}, Vasily and {Brammer}, Gabriel and {Sun}, Fengwu and {Nakane}, Minami and {Harikane}, Yuichi and {Umeda}, Hiroya and {Akins}, Hollis B. and {Atek}, Hakim and {Bauer}, Franz E. and {Brada{\v{c}}}, Maru{\v{s}}a and {Chisholm}, John and {Coe}, Dan and {Diego}, Jose M. and {Ferguson}, Henry C. and {Finkelstein}, Steven L. and {Furtak}, Lukas J. and {Inayoshi}, Kohei and {Koekemoer}, Anton M. and {Matthee}, Jorryt and {Naidu}, Rohan P. and {Ono}, Yoshiaki and {Pan}, Richard and {Richard}, Johan and {Robbins}, Luke and {Willott}, Chris and {Zitrin}, Adi and {Amor{\'\i}n}, Ricardo O. and {Bradley}, Larry D. and {Bromm}, Volker and {Conselice}, Christopher J. and {Dayal}, Pratika and {Kartaltepe}, Jeyhan S. and {Lopes}, Paulo A.~A. and {Lucas}, Ray A. and {Magdis}, Georgios E. and {Martis}, Nicholas S. and {Papovich}, Casey and {Schaerer}, Daniel and {Valentino}, Francesco and {Vanzella}, Eros and {Allingham}, Joseph F.~V. and {Grogin}, Norman A. and {Gonz{\'a}lez-Otero}, Mauro and {Ricotti}, Massimo and {Windhorst}, Rogier A.},
        title = "{VENUS: Two Faint Little Red Dots Separated by $\sim70\,\mathrm{pc}$ Hidden in a Single Lensed Galaxy at $z\sim7$}",
      journal = {arXiv e-prints},
         year = 2026,
        month = jan,
          eid = {arXiv:2601.06015},
        pages = {arXiv:2601.06015},
          doi = {10.48550/arXiv.2601.06015},
archivePrefix = {arXiv},
       eprint = {2601.06015},
 primaryClass = {astro-ph.GA},
       adsurl = {https://ui.adsabs.harvard.edu/abs/2026arXiv260106015Y}
}

@ARTICLE{Champagne2025,
       author = {{Champagne}, Jaclyn B. and {Wang}, Feige and {Yang}, Jinyi and {Fan}, Xiaohui and {Hennawi}, Joseph F. and {Sun}, Fengwu and {Ba{\~n}ados}, Eduardo and {Bosman}, Sarah E.~I. and {Costa}, Tiago and {Habouzit}, Melanie and {Jin}, Xiangyu and {Jun}, Hyunsung D. and {Li}, Mingyu and {Liu}, Weizhe and {Loiacono}, Federica and {Lupi}, Alessandro and {Mazzucchelli}, Chiara and {Pudoka}, Maria and {Rojas-Ruiz}, Sof{\'\i}a and {Tee}, Wei Leong and {Trebitsch}, Maxime and {Zhang}, Haowen and {Zhuang}, Ming-Yang and {Zou}, Siwei},
        title = "{A Quasar-anchored Protocluster at z = 6.6 in the ASPIRE Survey. II. An Environmental Analysis of Galaxy Properties in an Overdense Structure}",
      journal = {\apj},
         year = 2025,
        month = mar,
       volume = {981},
       number = {2},
          eid = {114},
        pages = {114},
          doi = {10.3847/1538-4357/adb1bc},
archivePrefix = {arXiv},
       eprint = {2410.03827},
 primaryClass = {astro-ph.GA},
       adsurl = {https://ui.adsabs.harvard.edu/abs/2025ApJ...981..114C}
}

@ARTICLE{Shi2023,
       author = {{Shi}, Yanlong and {Kremer}, Kyle and {Grudi{\'c}}, Michael Y. and {Gerling-Dunsmore}, Hannalore J. and {Hopkins}, Philip F.},
        title = "{Hyper-Eddington black hole growth in star-forming molecular clouds and galactic nuclei: can it happen?}",
      journal = {\mnras},
         year = 2023,
        month = jan,
       volume = {518},
       number = {3},
        pages = {3606-3621},
          doi = {10.1093/mnras/stac3245},
archivePrefix = {arXiv},
       eprint = {2208.05025},
 primaryClass = {astro-ph.GA},
       adsurl = {https://ui.adsabs.harvard.edu/abs/2023MNRAS.518.3606S}
}

@ARTICLE{Kocevski2024,
       author = {{Kocevski}, Dale D. and {Finkelstein}, Steven L. and {Barro}, Guillermo and {Taylor}, Anthony J. and {Calabr{\`o}}, Antonello and {Laloux}, Brivael and {Buchner}, Johannes and {Trump}, Jonathan R. and {Leung}, Gene C.~K. and {Yang}, Guang and {Dickinson}, Mark and {P{\'e}rez-Gonz{\'a}lez}, Pablo G. and {Pacucci}, Fabio and {Inayoshi}, Kohei and {Somerville}, Rachel S. and {McGrath}, Elizabeth J. and {Akins}, Hollis B. and {Bagley}, Micaela B. and {Bowler}, Rebecca A.~A. and {Bisigello}, Laura and {Carnall}, Adam and {Casey}, Caitlin M. and {Cheng}, Yingjie and {Cleri}, Nikko J. and {Costantin}, Luca and {Cullen}, Fergus and {Davis}, Kelcey and {Donnan}, Callum T. and {Dunlop}, James S. and {Ellis}, Richard S. and {Ferguson}, Henry C. and {Fujimoto}, Seiji and {Fontana}, Adriano and {Giavalisco}, Mauro and {Grazian}, Andrea and {Grogin}, Norman A. and {Hathi}, Nimish P. and {Hirschmann}, Michaela and {Huertas-Company}, Marc and {Holwerda}, Benne W. and {Illingworth}, Garth and {Juneau}, St{\'e}phanie and {Kartaltepe}, Jeyhan S. and {Koekemoer}, Anton M. and {Li}, Wenxiu and {Lucas}, Ray A. and {Magee}, Dan and {Mason}, Charlotte and {McLeod}, Derek J. and {McLure}, Ross J. and {Napolitano}, Lorenzo and {Papovich}, Casey and {Pirzkal}, Nor and {Rodighiero}, Giulia and {Santini}, Paola and {Wilkins}, Stephen M. and {Yung}, L.~Y. Aaron},
        title = "{The Rise of Faint, Red Active Galactic Nuclei at z > 4: A Sample of Little Red Dots in the JWST Extragalactic Legacy Fields}",
      journal = {\apj},
         year = 2025,
        month = jun,
       volume = {986},
       number = {2},
          eid = {126},
        pages = {126},
          doi = {10.3847/1538-4357/adbc7d},
archivePrefix = {arXiv},
       eprint = {2404.03576},
 primaryClass = {astro-ph.GA},
       adsurl = {https://ui.adsabs.harvard.edu/abs/2025ApJ...986..126K}
}

@ARTICLE{Maiolino2024bhs,
       author = {{Maiolino}, Roberto and {Scholtz}, Jan and {Curtis-Lake}, Emma and {Carniani}, Stefano and {Baker}, William and {de Graaff}, Anna and {Tacchella}, Sandro and {{\"U}bler}, Hannah and {D'Eugenio}, Francesco and {Witstok}, Joris and {Curti}, Mirko and {Arribas}, Santiago and {Bunker}, Andrew J. and {Charlot}, St{\'e}phane and {Chevallard}, Jacopo and {Eisenstein}, Daniel J. and {Egami}, Eiichi and {Ji}, Zhiyuan and {Jones}, Gareth C. and {Lyu}, Jianwei and {Rawle}, Tim and {Robertson}, Brant and {Rujopakarn}, Wiphu and {Perna}, Michele and {Sun}, Fengwu and {Venturi}, Giacomo and {Williams}, Christina C. and {Willott}, Chris},
        title = "{JADES: The diverse population of infant black holes at 4 < z < 11: Merging, tiny, poor, but mighty}",
      journal = {\aap},
         year = 2024,
        month = nov,
       volume = {691},
          eid = {A145},
        pages = {A145},
          doi = {10.1051/0004-6361/202347640},
archivePrefix = {arXiv},
       eprint = {2308.01230},
 primaryClass = {astro-ph.GA},
       adsurl = {https://ui.adsabs.harvard.edu/abs/2024A&A...691A.145M}
}

@ARTICLE{Greene2024,
       author = {{Greene}, Jenny E. and {Labbe}, Ivo and {Goulding}, Andy D. and {Furtak}, Lukas J. and {Chemerynska}, Iryna and {Kokorev}, Vasily and {Dayal}, Pratika and {Volonteri}, Marta and {Williams}, Christina C. and {Wang}, Bingjie and {Setton}, David J. and {Burgasser}, Adam J. and {Bezanson}, Rachel and {Atek}, Hakim and {Brammer}, Gabriel and {Cutler}, Sam E. and {Feldmann}, Robert and {Fujimoto}, Seiji and {Glazebrook}, Karl and {de Graaff}, Anna and {Khullar}, Gourav and {Leja}, Joel and {Marchesini}, Danilo and {Maseda}, Michael V. and {Matthee}, Jorryt and {Miller}, Tim B. and {Naidu}, Rohan P. and {Nanayakkara}, Themiya and {Oesch}, Pascal A. and {Pan}, Richard and {Papovich}, Casey and {Price}, Sedona H. and {van Dokkum}, Pieter and {Weaver}, John R. and {Whitaker}, Katherine E. and {Zitrin}, Adi},
        title = "{UNCOVER Spectroscopy Confirms the Surprising Ubiquity of Active Galactic Nuclei in Red Sources at z > 5}",
      journal = {\apj},
         year = 2024,
        month = mar,
       volume = {964},
       number = {1},
          eid = {39},
        pages = {39},
          doi = {10.3847/1538-4357/ad1e5f},
archivePrefix = {arXiv},
       eprint = {2309.05714},
 primaryClass = {astro-ph.GA},
       adsurl = {https://ui.adsabs.harvard.edu/abs/2024ApJ...964...39G}
}

@ARTICLE{furtak2024,
       author = {{Furtak}, Lukas J. and {Labb{\'e}}, Ivo and {Zitrin}, Adi and {Greene}, Jenny E. and {Dayal}, Pratika and {Chemerynska}, Iryna and {Kokorev}, Vasily and {Miller}, Tim B. and {Goulding}, Andy D. and {de Graaff}, Anna and {Bezanson}, Rachel and {Brammer}, Gabriel B. and {Cutler}, Sam E. and {Leja}, Joel and {Pan}, Richard and {Price}, Sedona H. and {Wang}, Bingjie and {Weaver}, John R. and {Whitaker}, Katherine E. and {Atek}, Hakim and {Bogd{\'a}n}, {\'A}kos and {Charlot}, St{\'e}phane and {Curtis-Lake}, Emma and {van Dokkum}, Pieter and {Endsley}, Ryan and {Feldmann}, Robert and {Fudamoto}, Yoshinobu and {Fujimoto}, Seiji and {Glazebrook}, Karl and {Juneau}, St{\'e}phanie and {Marchesini}, Danilo and {Maseda}, Micheal V. and {Nelson}, Erica and {Oesch}, Pascal A. and {Plat}, Ad{\`e}le and {Setton}, David J. and {Stark}, Daniel P. and {Williams}, Christina C.},
        title = "{A high black-hole-to-host mass ratio in a lensed AGN in the early Universe}",
      journal = {\nat},
         year = 2024,
        month = apr,
       volume = {628},
       number = {8006},
        pages = {57-61},
          doi = {10.1038/s41586-024-07184-8},
archivePrefix = {arXiv},
       eprint = {2308.05735},
 primaryClass = {astro-ph.GA},
       adsurl = {https://ui.adsabs.harvard.edu/abs/2024Natur.628...57F}
}

@ARTICLE{kokorev2025,
       author = {{Kokorev}, Vasily and {Caputi}, Karina I. and {Greene}, Jenny E. and {Dayal}, Pratika and {Trebitsch}, Maxime and {Cutler}, Sam E. and {Fujimoto}, Seiji and {Labb{\'e}}, Ivo and {Miller}, Tim B. and {Iani}, Edoardo and {Navarro-Carrera}, Rafael and {Rinaldi}, Pierluigi},
        title = "{A Census of Photometrically Selected Little Red Dots at 4 < z < 9 in JWST Blank Fields}",
      journal = {\apj},
         year = 2024,
        month = jun,
       volume = {968},
       number = {1},
          eid = {38},
        pages = {38},
          doi = {10.3847/1538-4357/ad4265},
archivePrefix = {arXiv},
       eprint = {2401.09981},
 primaryClass = {astro-ph.GA},
       adsurl = {https://ui.adsabs.harvard.edu/abs/2024ApJ...968...38K}
}

@ARTICLE{yue2024,
       author = {{Yue}, Minghao and {Eilers}, Anna-Christina and {Simcoe}, Robert A. and {Mackenzie}, Ruari and {Matthee}, Jorryt and {Kashino}, Daichi and {Bordoloi}, Rongmon and {Lilly}, Simon J. and {Naidu}, Rohan P.},
        title = "{EIGER. V. Characterizing the Host Galaxies of Luminous Quasars at z {\ensuremath{\gtrsim}} 6}",
      journal = {\apj},
         year = 2024,
        month = may,
       volume = {966},
       number = {2},
          eid = {176},
        pages = {176},
          doi = {10.3847/1538-4357/ad3914},
archivePrefix = {arXiv},
       eprint = {2309.04614},
 primaryClass = {astro-ph.GA},
       adsurl = {https://ui.adsabs.harvard.edu/abs/2024ApJ...966..176Y}
}

@ARTICLE{Pizzati2024,
       author = {{Pizzati}, Elia and {Hennawi}, Joseph F. and {Schaye}, Joop and {Schaller}, Matthieu and {Eilers}, Anna-Christina and {Wang}, Feige and {Frenk}, Carlos S. and {Elbers}, Willem and {Helly}, John C. and {Mackenzie}, Ruari and {Matthee}, Jorryt and {Bordoloi}, Rongmon and {Kashino}, Daichi and {Naidu}, Rohan P. and {Yue}, Minghao},
        title = "{A unified model for the clustering of quasars and galaxies at z {\ensuremath{\approx}} 6}",
      journal = {\mnras},
         year = 2024,
        month = nov,
       volume = {534},
       number = {4},
        pages = {3155-3175},
          doi = {10.1093/mnras/stae2307},
archivePrefix = {arXiv},
       eprint = {2403.12140},
 primaryClass = {astro-ph.GA},
       adsurl = {https://ui.adsabs.harvard.edu/abs/2024MNRAS.534.3155P}
}

@ARTICLE{Chiaki2023,
       author = {{Chiaki}, Gen and {Chon}, Sunmyon and {Omukai}, Kazuyuki and {Trinca}, Alessandro and {Schneider}, Raffaella and {Valiante}, Rosa},
        title = "{Direct-collapse black hole formation induced by internal radiation of host haloes}",
      journal = {\mnras},
         year = 2023,
        month = may,
       volume = {521},
       number = {2},
        pages = {2845-2859},
          doi = {10.1093/mnras/stad689},
archivePrefix = {arXiv},
       eprint = {2303.01762},
 primaryClass = {astro-ph.GA},
       adsurl = {https://ui.adsabs.harvard.edu/abs/2023MNRAS.521.2845C}
}

@ARTICLE{ventura2024,
       author = {{Ventura}, Emanuele M. and {Qin}, Yuxiang and {Balu}, Sreedhar and {Wyithe}, J. Stuart B.},
        title = "{Semi-analytic modelling of Pop. III star formation and metallicity evolution - I. Impact on the UV luminosity functions at z = 9-16}",
      journal = {\mnras},
         year = 2024,
        month = mar,
       volume = {529},
       number = {1},
        pages = {628-646},
          doi = {10.1093/mnras/stae567},
archivePrefix = {arXiv},
       eprint = {2401.07396},
 primaryClass = {astro-ph.GA},
       adsurl = {https://ui.adsabs.harvard.edu/abs/2024MNRAS.529..628V}
}

%%%%%%%%%%%%%%%%% APPENDICES %%%%%%%%%%%%%%%%%%%%%

\appendix

\section{The Lyman-Werner background in CAT}
\label{sec:AppendixA}
In the \textsc{cat} model, a cosmic UV background is progressively built up through the emission from stars and BHs within the simulated overdensity. This allows the model to capture the radiative feedback from Lyman-Werner photons, which can suppress $\rm H_2$ formation, inhibit gas cooling, and regulate star formation in pristine mini-halos, potentially leading to direct collapse once the halo enters the atomic-cooling regime.
To follow the average UV background across the simulated overdensity, in \textsc{cat} we calculate the cumulative flux at the observed frequency $\nu_{\rm obs}$ and redshift $z_{\rm obs}$ as \citep[in units of $\mathrm{erg \, s^{-1} \, cm^{-2} \, Hz^{-1} \, sr^{-1}}$, 
see e.g.]{haardt1996}
\begin{equation}
J(\nu_{\rm obs}, z_{\rm obs}) = \frac{(1+z_{\rm obs})^{3}}{4\pi}
\int_{z_{\rm obs}}^{z_{\rm max}} 
dz \, c \left| \frac{dt}{dz} \right| 
\, \epsilon(\nu_z, z) \, e^{-\tau_{\rm H_2}(\nu_{\rm obs}, z_{\rm obs}, z)},
\label{eq:LWflux}
\end{equation}
where $\tau_{\rm H_2}$ is the $\rm H_2$ optical depth in the LW band and $\epsilon(\nu_z, z)$ represents the comoving emissivity, i.e. the monochromatic luminosity per unit comoving volume in the LW band, which is computed summing over all the emitting sources at each given redshift $z$. We refer the reader to \citet{valiante2016} for a detailed description of the emissivities and spectral energy distributions adopted in the model for PopIII and PopII stars, as well as for accreting BHs.
The redshift $z_{\rm max}$ represents the maximum redshift from which an LW photon emitted at $z > z_{\rm obs}$ can reach the observer at $z_{\rm obs}$ before being redshifted outside the LW band, into a hydrogen Lyman resonance line, which, in a dark-screen approximation, can be defined as $
(1+z_{\rm max})/(1+z_{\rm obs}) = \nu_i / \nu_{\rm obs}$,
where $\nu_i$ is the first Lyman-line frequency above the observed one \citep[see e.g.][]{Haiman2000}.
The value of the optical depth $\tau_{\rm H_2}$ depends in general on the H$_2$ number density, the line profile, and the probability that the molecule dissociates after a transition. The intergalactic absorption averaged over the LW band is computed using the modulation factor given by the fitting formula \citep{ahn2009}:
\begin{equation}
e^{-\tau_{\rm H_2}} =
\begin{cases}
1.7\, \exp\left[-\left(\frac{r_{\rm c,Mpc}}{116.29\,\alpha}\right)^{0.68}\right] - 0.7, & \text{if } r_{\rm c,Mpc}/\alpha \leq 97.39,\\
0, & \text{if } r_{\rm c,Mpc}/\alpha > 97.39.
\end{cases}
\label{eq:tauH2}
\end{equation}
Here $r_{\rm c,Mpc}$ is the comoving distance between the emitting source at redshift $z$ and the observer at $z_{\rm obs}$,
\begin{equation}
r_{\rm c,Mpc} = - \int_{z_{\rm obs}}^{z} \frac{c\, dz'}{H(z')},
\label{eq:rc}
\end{equation}
with $H(z) = H_0 \left[\Omega_{\rm M}(1+z)^3 + \Omega_\Lambda\right]^{1/2}$. while $\alpha$ represents a scaling factor defined as
\begin{equation}
\alpha = 
\left(\frac{h}{0.7}\right)^{-1}
\left(\frac{\Omega_{\rm M}}{0.27}\right)^{-1/2}
\left(\frac{1+z}{21}\right)^{-1/2}.
\end{equation}
These assumptions result in an average attenuation of the UV flux that increases with the comoving ratio $r_{\rm c,Mpc}/\alpha$, approaching zero when $r_{\rm c,Mpc} = 97.39\,\alpha$, i.e. at the maximum distance from which an observer can receive LW photons emitted by a source at redshift $z$.

In the updated version of \textsc{cat} presented in this work, in addition to the global average background, we also track the local incident LW flux on each system that is a potential candidate for DCBH formation, to study how spatial fluctuations in the LW background affect the distribution and number of DCBH formation sites. This is done by computing the cumulative LW flux irradiating each galaxy from all other active systems in the overdensity (hosting stellar populations and/or AGNs), whose spatial distribution is known from the underlying N-body merger trees:
\begin{equation}
J_{\rm LW,\,local} \simeq \sum_{i=1}^{N_{\rm sources}} 
\frac{L_{{\rm LW},i}}{16 \, \pi^2 \, r_{{\rm ph},i}^{2}} ,
\end{equation}
where $L_{{\rm LW},i}$ is the total LW luminosity emitted by each irradiating source ($ \rm erg \ s^{-1} \ Hz^{-1}$), $r_{{\rm ph},i}$ is the physical separation between the source and the target galaxy, and we can approximate $e^{-\tau_{\rm H_2}} \simeq 1$ given the small comoving size of the simulated overdensity ($\sim 2~\rm cMpc$).  

As discussed in Section \ref{sec:SeedDistribution}, accurately tracking the local incident flux is crucial to identify galaxies where the LW flux exceeds the threshold for direct collapse, particularly at early times when the average flux across the overdensity is still low. We continue to compute the local flux on each galaxy until the average LW background reaches the assumed DCBH formation threshold $J_{\rm crit}$. Beyond this point, local flux enhancements no longer affect BH seeding, and the LW flux evolution is then tracked using the global average background.

\section{Black hole and stellar mass distributions of heavy-seed descendants}
\label{sec:AppendixB}

\begin{figure}
\centering
\includegraphics[width=\linewidth]{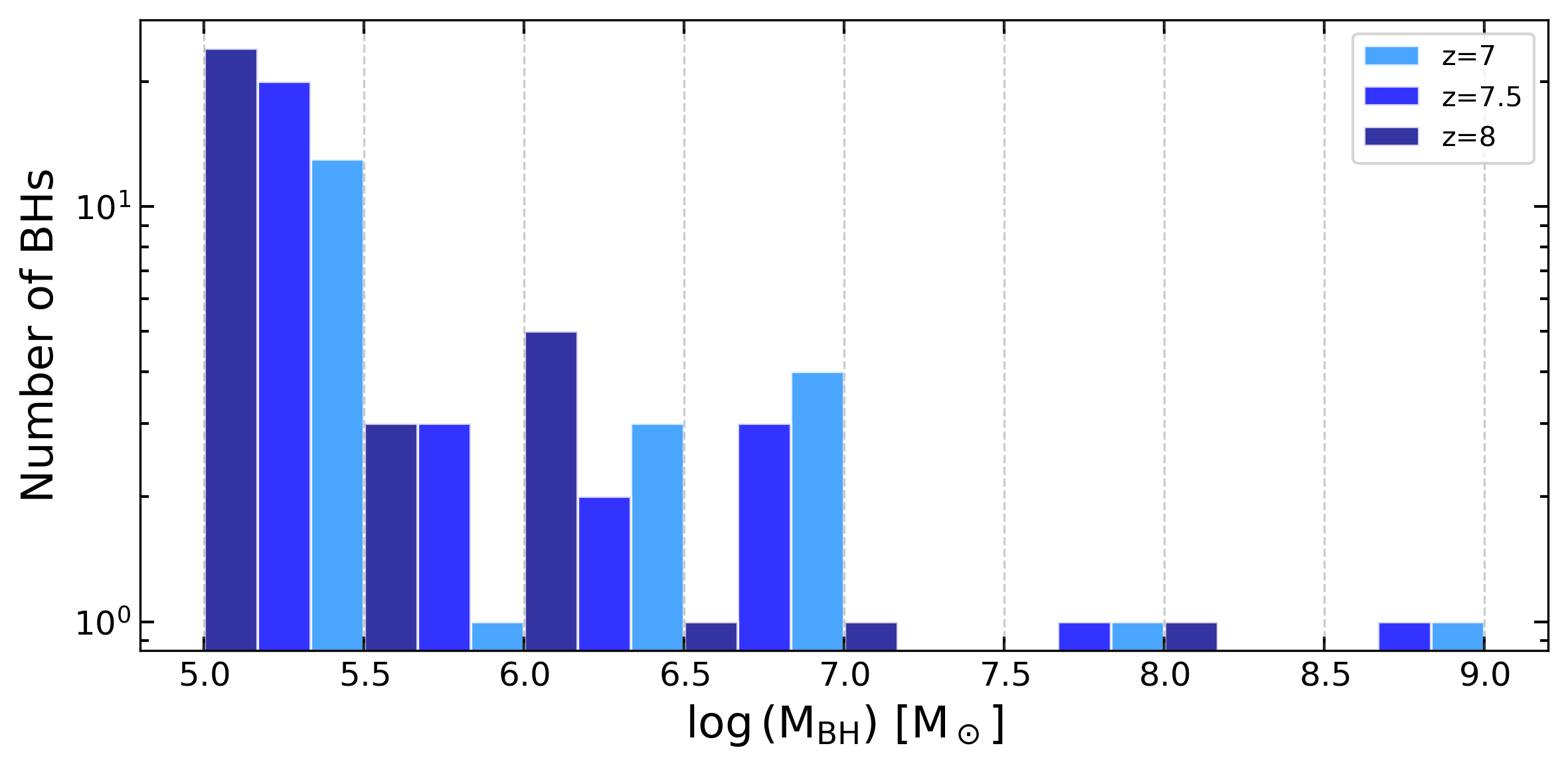}
\includegraphics[width=\linewidth]{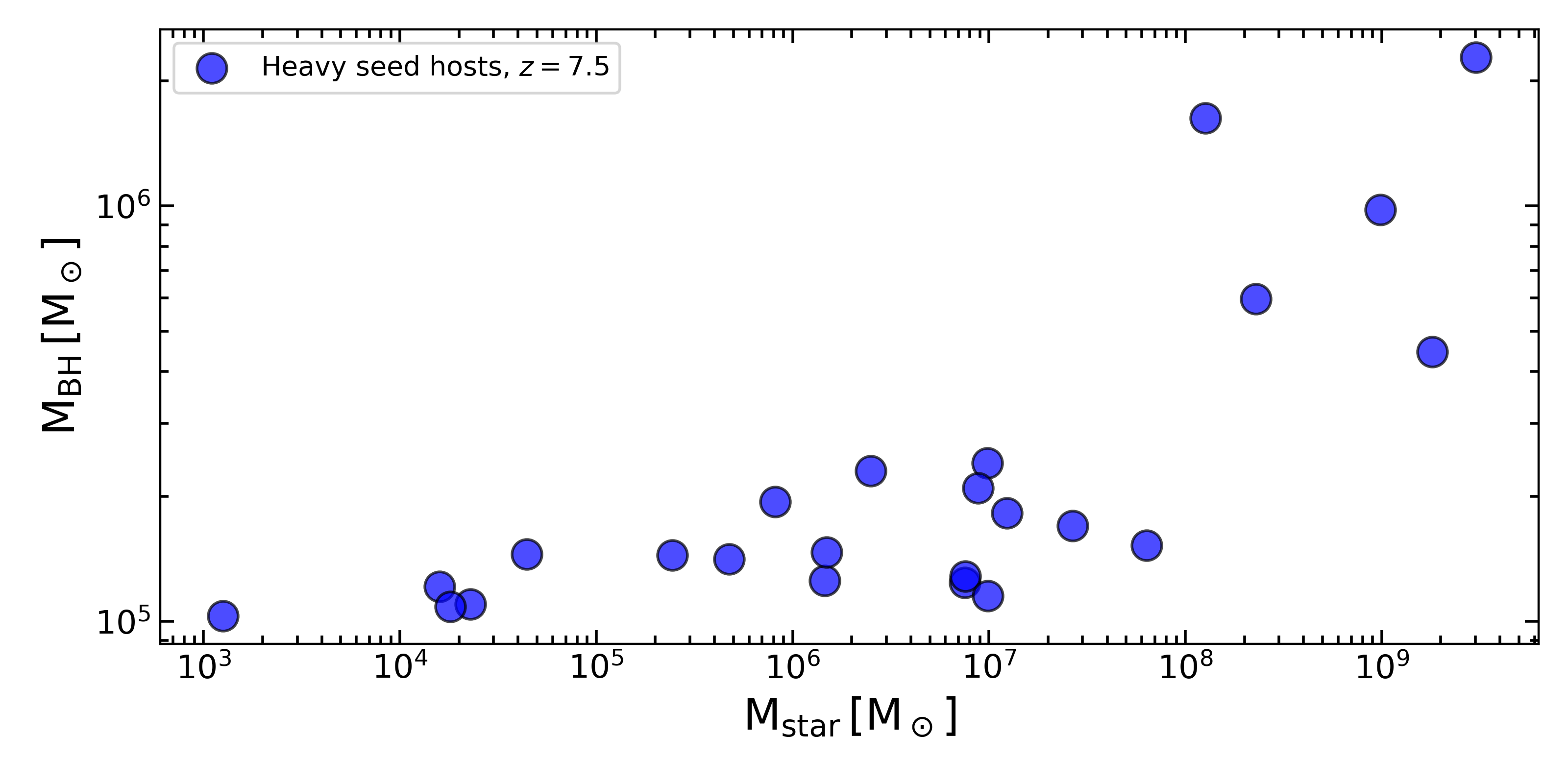}
\caption{\textit{Upper panel}: Distribution of BH masses in the simulated overdensity at $z=8$, $z=7.5$ and $7$, shown in bins of $0.5 \rm \ dex$. \textit{Lower panel}: BH mass vs stellar mass of galaxies hosting heavy-seed descendants with $\Mbh < 10^{6.5} \ \msun$ at $z=7.5$. The same systems, to which we restricted our observability analysis, are the ones highlighted in red in Fig. \ref{fig:MapBHz7}.}
\label{fig:BHMF}
\end{figure}

In the upper panel of Fig.~\ref{fig:BHMF} we present the mass distribution of all SMBHs in the simulated overdensity at $z = 8$, 7.5, and 7, binned in intervals of 0.5 dex. By $z=8$, only three descendants of heavy seeds have grown to masses above $\Mbh > 10^{6.5} \ \msun$. These objects reside in the three most massive galaxy progenitors, where sustained gas inflows allow for a more continuous accretion history. Between $z=8$ and $z=7$, the redshift interval over which we perform our observational analysis, only a small number of additional systems experience efficient growth, bringing the total number above this mass threshold to six by $z=7$.

Since our focus is on the observability of the population of relatively ungrown systems, which are expected to undergo short and intermittent active accretion phases, and in order to mitigate biases associated with the presence of multiple massive galaxy progenitors in our single merger-tree realization, we restrict our observability analysis to heavy-seed descendants with $\Mbh \leq 10^{6.5} \ \msun$.

The lower panel of Fig.~\ref{fig:BHMF} shows the BH mass-stellar mass relation for the selected galaxies at $z = 7.5$, corresponding to the sources highlighted in red in Fig.~\ref{fig:MapBHz7}. This population spans a wide range of evolutionary stages. A small fraction of systems exhibit very low stellar masses, $\Mstar \sim 10^{3}$-$10^{5} \ \msun$, likely reflecting cases in which the formation of the central DCBH and the associated feedback strongly depleted the host galaxy gas reservoir, effectively suppressing the growth of both components. The majority of systems populate the range $\Mstar \sim 10^{5}$-$10^{8} \ \msun$, where the galaxy has assembled a substantial stellar component while BH growth remains limited, plausibly occurring only in short accretion episodes that result in, at most, a modest increase over the initial seed mass. Finally, a few systems reside in more massive galaxies, with $\Mstar > 10^{8} \ \msun$, where the higher gas availability and deeper gravitational potential wells enable a more sustained growth of the central SMBHs, that in all cases reach masses $\Mbh > 4 \times 10^{5} \ \msun$.

Overall, this diversity highlights how the emerging spectral properties of these systems during active accretion phases are expected to vary significantly, with some sources exhibiting clear AGN signatures, while in others the emission from the surrounding host galaxy might still dominate the total observed spectrum.

The vast range of host-galaxy stellar masses also shows that the BH-to-stellar mass ratio does not necessarily provide a strong diagnostic of the underlying MBH seed formation channel. In fact, as discussed in Sec. \ref{sec:discussion}, shallower surveys are likely to be biased toward the detection of more massive DCBH descendants, that are hosted in brighter and more evolved galaxies, characterized by relatively standard values of $\Mbh/\Mstar \sim 10^{-3}$.

% Don't change these lines
\bsp	% typesetting comment
\label{lastpage}
\end{document}